\documentclass[10pt,twocolumn,letterpaper]{article}

\usepackage[pagenumbers]{iccv} %

\usepackage{amsmath}
\usepackage{amssymb}
\usepackage{makecell}
\usepackage{microtype}
\usepackage{multirow}
\usepackage{enumitem}
\usepackage{bbm}
\newcommand{\redcross}{\textcolor{red}{\texttimes}}
\newcommand{\greenmark}{\textcolor{mydarkgreen}{\checkmark}}
\newcommand{\orangemark}{\textcolor{orange}{\checkmark}}

\newcommand{\videoLength}{L}
\newcommand{\videoInput}{\mathbf{I}}
\newcommand{\videoOutput}{\hat{\mathbf{I}}}
\newcommand{\videoClear}{\mathbf{I}^{c}}
\newcommand{\videoWeather}{\mathbf{I}^{w}}

\newcommand{\diffusionNoise}{\mathbb{\epsilon}}
\newcommand{\diffusionTime}{\tau}

\newcommand{\diffusionCond}{\mathbf{c}}
\newcommand{\diffusionModel}{\mathbf{f}}
\newcommand{\diffusionModelParams}{\theta}

\newcommand{\latent}{\mathbf{z}}
\newcommand{\latentClear}{\mathbf{z}^{c}}
\newcommand{\latentWeather}{\mathbf{z}^{w}}

\newcommand{\vaeEncoder}{\mathcal{E}}
\newcommand{\vaeDecoder}{\mathcal{D}}
\newcommand{\dataDistribution}{p_{\text{data}}}

\newcommand{\strength}{\mathbf{s}}
\newcommand{\strengthMap}{\mathbf{S}}

\newcommand{\simulation}{c \rightarrow w}
\newcommand{\removal}{w \rightarrow c}
\newcommand{\simulationModel}{\mathbf{f}_{\diffusionModelParams}^{\simulation}}
\newcommand{\removalModel}{\mathbf{f}_{\diffusionModelParams}^{\removal}}
\newcommand{\loss}{\mathcal{L}}
\newcommand{\lossSimulation}{\loss^{\simulation}}
\newcommand{\lossRemoval}{\loss^{\removal}}

\newcommand{\INFO}[1]{}
\newcommand{\TODO}[1]{}
\newcommand{\CL}[1]{}
\newcommand{\ZW}[1]{}
\newcommand{\ZG}[1]{}
\newcommand{\SW}[1]{}
\newcommand{\RF}[1]{}
\newcommand{\YZ}[1]{}
\newcommand{\SF}[1]{}

\newcommand{\ourmodel}{\textsc{WeatherWeaver}\xspace}
\newcommand{\parahead}[1]{\vspace{1mm}\noindent\textbf{#1}\hspace{2mm}}

\definecolor{iccvblue}{rgb}{0.21,0.49,0.74}
\usepackage[pagebackref,breaklinks,colorlinks,allcolors=iccvblue]{hyperref}

\def\paperID{5663} %
\def\confName{ICCV}
\def\confYear{2025}

\title{Controllable Weather Synthesis and Removal with Video Diffusion Models}

\author{
Chih-Hao Lin$^{1,2}$,
Zian Wang$^{1,3,4}$,
Ruofan Liang$^{1,3,4}$,
Yuxuan Zhang$^{1}$,\\
Sanja Fidler$^{1,2,3}$, 
Shenlong Wang$^{2}$,
Zan Gojcic$^{1}$\\
$^1$NVIDIA 
\quad $^2$University of Illinois Urbana-Champaign
\quad $^3$University of Toronto 
\quad $^4$Vector Institute \\
\large{\href{https://research.nvidia.com/labs/toronto-ai/WeatherWeaver/}{Project Website}}
}

\begin{document}

\newcommand{\todocite}[1]{\textcolor{blue}{Citation needed []}}
\newcommand{\shenlongsay}[1]{\textcolor{blue}{[{\it Shenlong: #1}]}}
\newcommand{\jiabin}[1]{\textcolor{cyan}{[{\it Jia-Bin: #1}]}}

\newcommand{\mfigure}[2]{\includegraphics[width=#1\linewidth]{#2}}
\newcommand{\mpage}[2]
{
\begin{minipage}{#1\linewidth}\centering
#2
\end{minipage}
}

\newcolumntype{L}[1]{>{\raggedright\let\newline\\\arraybackslash\hspace{0pt}}m{#1}}
\newcolumntype{C}[1]{>{\centering\let\newline\\\arraybackslash\hspace{0pt}}m{#1}}
\newcolumntype{R}[1]{>{\raggedleft\let\newline\\\arraybackslash\hspace{0pt}}m{#1}}

\newcommand{\xpar}[1]{\noindent\textbf{#1}\ \ }
\newcommand{\vpar}[1]{\vspace{3mm}\noindent\textbf{#1}\ \ }

\newcommand{\ignorethis}[1]{}
\newcommand{\norm}[1]{\lVert#1\rVert}
\newcommand{\fcseven}{$\mbox{fc}_7$}

\newcommand{\topic}[1]
{
\vspace{1mm}\noindent\textbf{#1}
}

\def\naive{na\"{\i}ve\xspace}
\def\Naive{Na\"{\i}ve\xspace}

\makeatletter
\DeclareRobustCommand\onedot{\futurelet\@let@token\@onedot}
\def\@onedot{\ifx\@let@token.\else.\null\fi\xspace}

\def\iid{\emph{i.i.d}\onedot}
\def\eg{\emph{e.g}\onedot} \def\Eg{\emph{E.g}\onedot}
\def\ie{\emph{i.e}\onedot} \def\Ie{\emph{I.e}\onedot}
\def\cf{\emph{c.f}\onedot} \def\Cf{\emph{C.f}\onedot}
\def\etc{\emph{etc}\onedot} \def\vs{\emph{vs}\onedot}
\def\wrt{w.r.t\onedot} \def\dof{d.o.f\onedot}
\def\etal{et al\onedot}
\makeatother

\definecolor{citecolor}{RGB}{34,139,34}
\definecolor{mydarkblue}{rgb}{0,0.08,1}
\definecolor{mydarkgreen}{rgb}{0.02,0.6,0.02}
\definecolor{mydarkred}{rgb}{0.8,0.02,0.02}
\definecolor{mydarkorange}{rgb}{0.40,0.2,0.02}
\definecolor{mypurple}{RGB}{111,0,255}
\definecolor{myred}{rgb}{1.0,0.0,0.0}
\definecolor{mygold}{rgb}{0.75,0.6,0.12}
\definecolor{myblue}{rgb}{0,0.2,0.8}
\definecolor{mydarkgray}{rgb}{0.66,0.66,0.66}

\newcommand{\myparagraph}[1]{\vspace{-6pt}\paragraph{#1}}

\newcommand{\bbR}{{\mathbb{R}}}
\newcommand{\bK}{\mathbf{K}}
\newcommand{\bX}{\mathbf{X}}
\newcommand{\bY}{\mathbf{Y}}
\newcommand{\bk}{\mathbf{k}}
\newcommand{\bx}{\mathbf{x}}
\newcommand{\by}{\mathbf{y}}
\newcommand{\bhy}{\hat{\mathbf{y}}}
\newcommand{\bty}{\tilde{\mathbf{y}}}
\newcommand{\bG}{\mathbf{G}}
\newcommand{\bI}{\mathbf{I}}
\newcommand{\bg}{\mathbf{g}}
\newcommand{\bS}{\mathbf{S}}
\newcommand{\bs}{\mathbf{s}}
\newcommand{\bM}{\mathbf{M}}
\newcommand{\bw}{\mathbf{w}}
\newcommand{\eye}{\mathbf{I}}
\newcommand{\bU}{\mathbf{U}}
\newcommand{\bV}{\mathbf{V}}
\newcommand{\bW}{\mathbf{W}}
\newcommand{\bn}{\mathbf{n}}
\newcommand{\bv}{\mathbf{v}}
\newcommand{\bq}{\mathbf{q}}
\newcommand{\bR}{\mathbf{R}}
\newcommand{\bi}{\mathbf{i}}
\newcommand{\bj}{\mathbf{j}}
\newcommand{\bp}{\mathbf{p}}
\newcommand{\bt}{\mathbf{t}}
\newcommand{\bJ}{\mathbf{J}}
\newcommand{\bu}{\mathbf{u}}
\newcommand{\bB}{\mathbf{B}}
\newcommand{\bD}{\mathbf{D}}
\newcommand{\bz}{\mathbf{z}}
\newcommand{\bP}{\mathbf{P}}
\newcommand{\bC}{\mathbf{C}}
\newcommand{\bA}{\mathbf{A}}
\newcommand{\bZ}{\mathbf{Z}}
\newcommand{\bff}{\mathbf{f}}
\newcommand{\bF}{\mathbf{F}}
\newcommand{\bo}{\mathbf{o}}
\newcommand{\bc}{\mathbf{c}}
\newcommand{\bT}{\mathbf{T}}
\newcommand{\bQ}{\mathbf{Q}}
\newcommand{\bL}{\mathbf{L}}
\newcommand{\bl}{\mathbf{l}}
\newcommand{\ba}{\mathbf{a}}
\newcommand{\bE}{\mathbf{E}}
\newcommand{\be}{\mathbf{e}}
\newcommand{\bH}{\mathbf{H}}
\newcommand{\bd}{\mathbf{d}}
\newcommand{\br}{\mathbf{r}}
\newcommand{\bb}{\mathbf{b}}
\newcommand{\bh}{\mathbf{h}}

\newcommand{\btheta}{\bm{\theta}}
\newcommand{\bhh}{\hat{\mathbf{h}}}
\newcommand{\ci}{{\cal I}}
\newcommand{\ct}{{\cal T}}
\newcommand{\co}{{\cal O}}
\newcommand{\ck}{{\cal K}}
\newcommand{\cu}{{\cal U}}
\newcommand{\cS}{{\cal S}}
\newcommand{\cQ}{{\cal Q}}
\newcommand{\cT}{{\cal S}}
\newcommand{\cC}{{\cal C}}
\newcommand{\cE}{{\cal E}}
\newcommand{\cF}{{\cal F}}
\newcommand{\cL}{{\cal L}}
\newcommand{\X}{{\cal{X}}}
\newcommand{\Y}{{\cal Y}}
\newcommand{\cH}{{\cal H}}
\newcommand{\cP}{{\cal P}}
\newcommand{\cN}{{\cal N}}
\newcommand{\cU}{{\cal U}}
\newcommand{\cV}{{\cal V}}
\newcommand{\cX}{{\cal X}}
\newcommand{\cY}{{\cal Y}}
\newcommand{\graph}{{\cal H}}
\newcommand{\bayes}{{\cal B}}
\newcommand{\cx}{{\cal X}}
\newcommand{\cg}{{\cal G}}
\newcommand{\cm}{{\cal M}}
\newcommand{\cM}{{\cal M}}
\newcommand{\cG}{{\cal G}}
\newcommand{\cR}{\cal{R}}
\newcommand{\R}{\cal{R}}
\newcommand{\eig}{\mathrm{eig}}

\newcommand{\bbS}{\mathbb{S}}

\newcommand{\D}{{\cal D}}
\newcommand{\bfp}{{\bf p}}
\newcommand{\bfd}{{\bf d}}

\newcommand{\cv}{{\cal V}}
\newcommand{\ce}{{\cal E}}
\newcommand{\cy}{{\cal Y}}
\newcommand{\cz}{{\cal Z}}
\newcommand{\cb}{{\cal B}}
\newcommand{\cq}{{\cal Q}}
\newcommand{\cd}{{\cal D}}
\newcommand{\bcf}{{\cal F}}
\newcommand{\cI}{\mathcal{I}}

\newcommand{\ut}{^{(t)}}
\newcommand{\up}{^{(t-1)}}

\newcommand{\bpi}{\boldsymbol{\pi}}
\newcommand{\bphi}{\boldsymbol{\phi}}
\newcommand{\bPhi}{\boldsymbol{\Phi}}
\newcommand{\bmu}{\boldsymbol{\mu}}
\newcommand{\bSigma}{\boldsymbol{\Sigma}}
\newcommand{\bGamma}{\boldsymbol{\Gamma}}
\newcommand{\bbeta}{\boldsymbol{\beta}}
\newcommand{\bomega}{\boldsymbol{\omega}}
\newcommand{\blambda}{\boldsymbol{\lambda}}
\newcommand{\bkappa}{\boldsymbol{\kappa}}
\newcommand{\btau}{\boldsymbol{\tau}}
\newcommand{\balpha}{\boldsymbol{\alpha}}
\def\bgamma{\boldsymbol\gamma}

\newcommand{\prox}{{\mathrm{prox}}}

\newcommand{\pardev}[2]{\frac{\partial #1}{\partial #2}}
\newcommand{\dev}[2]{\frac{d #1}{d #2}}
\newcommand{\dw}{\delta\bw}
\newcommand{\lab}{\mathcal{L}}
\newcommand{\unlab}{\mathcal{U}}
\newcommand{\ind}{1{\hskip -2.5 pt}\hbox{I}}
\newcommand{\ff}[2]{   \cf_{\prec (#1 \rightarrow #2)}}
\newcommand{\vv}[2]{   \cv_{\prec (#1 \rightarrow #2)}}
\newcommand{\dd}[2]{   \delta_{#1 \rightarrow #2}}
\newcommand{\ld}[2]{   \lambda_{#1 \rightarrow #2}}
\newcommand{\en}[2]{  \bD(#1|| #2)}
\newcommand{\ex}[3]{  \bE_{#1 \sim #2}\left[ #3\right]} 
\newcommand{\exd}[2]{  \bE_{#1 }\left[ #2\right]}

\newcommand{\se}[1]{\mathfrak{se}(#1)}
\newcommand{\SE}[1]{\mathbb{SE}(#1)}
\newcommand{\so}[1]{\mathfrak{so}(#1)}
\newcommand{\SO}[1]{\mathbb{SO}(#1)}

\newcommand{\poselow}{\xi}
\newcommand{\pose}{\bm{\poselow}}
\newcommand{\linpose}{\pose^\ell}
\newcommand{\cbpose}{\pose^c}
\newcommand{\rateparam}{v_i}
\newcommand{\bapose}{\bm{\poselow}_i}
\newcommand{\trackingpose}{\bm{\poselow}}
\newcommand{\rotlow}{\omega}
\newcommand{\rot}{\bm{\rotlow}}
\newcommand{\translow}{v}
\newcommand{\trans}{\bm{\translow}}
\newcommand{\hnorm}[1]{\left\lVert#1\right\rVert_{\gamma}}
\newcommand{\lnorm}[1]{\left\lVert#1\right\rVert}
\newcommand{\barate}{v_i}
\newcommand{\trackingrate}{v}
\newcommand{\imgpt}{\mathbf{u}_{i,k,j}}
\newcommand{\mappt}{\mathbf{X}_{j}}
\newcommand{\timet}[1]{\bar{t}_{#1}}
\newcommand{\mf}[1]{\text{MF}_{#1}}
\newcommand{\kmf}[1]{\text{KMF}_{#1}}
\newcommand{\Exp}{\text{Exp}}
\newcommand{\Log}{\text{Log}}

\newcommand{\shiftleft}[2]{\makebox[0pt][r]{\makebox[#1][l]{#2}}}
\newcommand{\shiftright}[2]{\makebox[#1][r]{\makebox[0pt][l]{#2}}}

\twocolumn[{%
\maketitle
\vspace{-1cm}
\renewcommand\twocolumn[1][]{#1}%
\begin{center}
    \centering
\vspace{-4mm}
\includegraphics[width=1.0\textwidth]{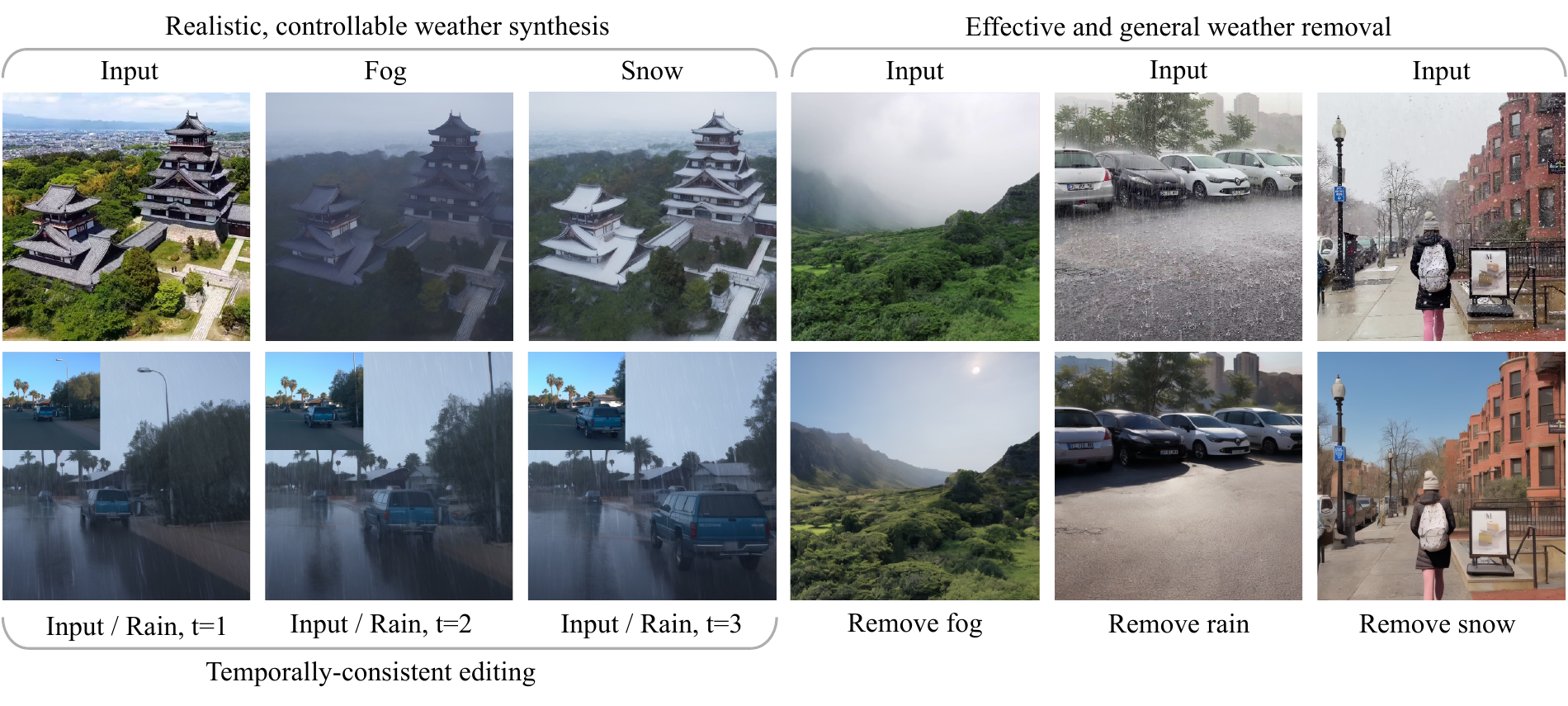}
\vspace{-8mm}
\captionof{figure}{
We introduce \ourmodel{}, a generative editing method for synthesizing and removing weather effects. Given an input video, it 
creates corresponding videos with diverse weather condition (rain, snow, fog, clouds)
and precise control over the intensity (left), 
removes weather from real footage (right). The results are photorealistic, temporally consistent, and faithfully preserve the original scene.
}    
\label{fig:teaser}
\end{center}
}]

\begin{abstract}
Generating realistic and controllable weather effects in videos is valuable for many applications. 
Physics-based weather simulation requires precise reconstructions that are hard to scale to in-the-wild videos, while current video editing often lacks realism and control.
In this work, we introduce \ourmodel{}, a video diffusion model that synthesizes diverse weather effects---including rain, snow, fog, and clouds---directly into any input video without the need for 3D modeling.
Our model provides precise control over weather effect intensity and supports blending various weather types, ensuring both realism and adaptability.
To overcome the scarcity of paired training data, we propose a novel data strategy combining synthetic videos, generative image editing, and auto-labeled real-world videos. 
Extensive evaluations show that our method outperforms state-of-the-art methods
in weather simulation and removal, providing high-quality, physically plausible, and scene-identity-preserving results over various real-world videos.
\end{abstract}

\vspace{-2mm}
\section{Introduction}
\vspace{-2mm}

Simulating photorealistic weather effects in videos, such as rain, snow, fog, or clouds, is a challenging yet essential task in computer vision and graphics. High-quality weather simulations enable a range of creative applications in film production, AR/VR, and video games. Moreover, controllable weather simulation is invaluable for training and evaluating perception systems in safety-critical domains such as autonomous driving and robotics, where robust performance under diverse weather conditions is crucial. 

Comprehensive weather simulation must capture both transient effects—such as falling rain, swirling snow, or drifting fog—and persistent or accumulative changes, such as 
snow buildup on the ground or water puddles after rain. 
In modern graphics engines, transient effects are often handled using particle-based simulations~\cite{feldman2002modeling, garg2006photorealistic, stomakhin2013material}, while persistent changes are approximated by modifying scene asset materials~\cite{unrealengine}. However, these methods rely on detailed, simulation-ready 3D models, limiting their applicability to synthetic environments. Recent work has attempted to adapt such pipelines to real-world videos by reconstructing scenes through methods like NeRF~\cite{mildenhall2020nerf} or 3DGS~\cite{kerbl3Dgaussians}, but imperfect reconstructions frequently introduce blending artifacts and unnatural shading~\cite{Li2023ClimateNeRF}. 

Instead of employing a two-stage \emph{reconstruct-then-simulate} approach, we formulate weather simulation in real-world videos as a video-to-video translation task, leveraging the recent success of large video generative models in video editing. Nevertheless, straightforward adaptations of general video editing methods fail to deliver the necessary realism—particularly for transient phenomena—and lack precise control over the weather type and intensity (Fig.~\ref{fig:qual_forward}). Two main challenges contribute to this: \textbf{(i)} acquiring high-quality paired data (videos of the same scene under different weather conditions) is difficult to scale in real-world settings, and \textbf{(ii)} directly translating from one weather condition to another (e.g., rainy to snowy) is inherently complex, as it requires removing one weather effect while adding another.

To overcome these challenges, we draw inspiration from modern graphics engines, which treat weather simulation as an added effect applied to an existing scene consisting of geometry, materials, and lighting. Concretely, we split our pipeline into two video diffusion models: a \textsc{weather removal model} that translates a real-world video into a “canonical,” weather-free video\footnote{Note that \emph{canonical weather} representation is not strictly defined. In this work, we use the term to refer to a clear sunny or overcast sky.}, and a \textsc{weather synthesis model} that adds weather effects to a “canonical” video with precise control over both intensity and type of weather. 
This split offers two main advantages. First, the \textsc{weather removal model} can serve as a pseudo-labeling engine, producing paired data with realistically looking weather effects. Second, confining the \textsc{weather synthesis model} to solely adding the weather effects simplifies its task. 

High-quality paired video training data is crucial to ensure both realism and scene preservation for the proposed models. 
However, acquiring real-world paired videos of the same dynamic scene is challenging. To address this, we introduce a new data strategy and train our models on a carefully curated combination of three data sources (see Table~\ref{tab:datasets}). 
First, we render a synthetic video dataset using standard graphics engines and fully modeled 3D environments, allowing precise control over weather attributes but yielding a synthetic appearance. Second, we generate paired image data via large image generative models (e.g., SDXL~\cite{podell2023sdxl}) by leveraging Prompt-to-Prompt~\cite{hertz2022prompt} method. This strategy yields more realistic outputs, albeit with lack of precise control and limitation to image data. Finally,  we use these datasets to train the \textsc{weather removal model} and apply it to automatically convert real-world videos with weather effects to their ``canonical'' clear-day video, thus creating a large dataset of highly realistic video pairs. For training the \textsc{weather synthesis model}, we use all three sources of data.

Our resulting framework, \ourmodel{}, outperforms state-of-the-art methods by producing high-quality, controllable weather effects in real-world videos with precise control of intensity and type of weather. In summary, our contributions are: 
\begin{itemize} 
\item A controllable weather synthesis model that adds diverse weather effects to real-world videos, offering precise control over both intensity and type.
\item A weather removal model that effectively handles both transient (\eg rain, snow) and persistent (\eg clouds, rain puddle, snow coverage) weather effects.
\item A data curation strategy that combines synthetic data, generative models outputs, and auto-labeled real-world videos, thus improving realism and diversity of the paired data. 
\end{itemize}

\begin{figure*}[t]
\centering
\vspace{-5mm}
\includegraphics[width=0.9\linewidth]{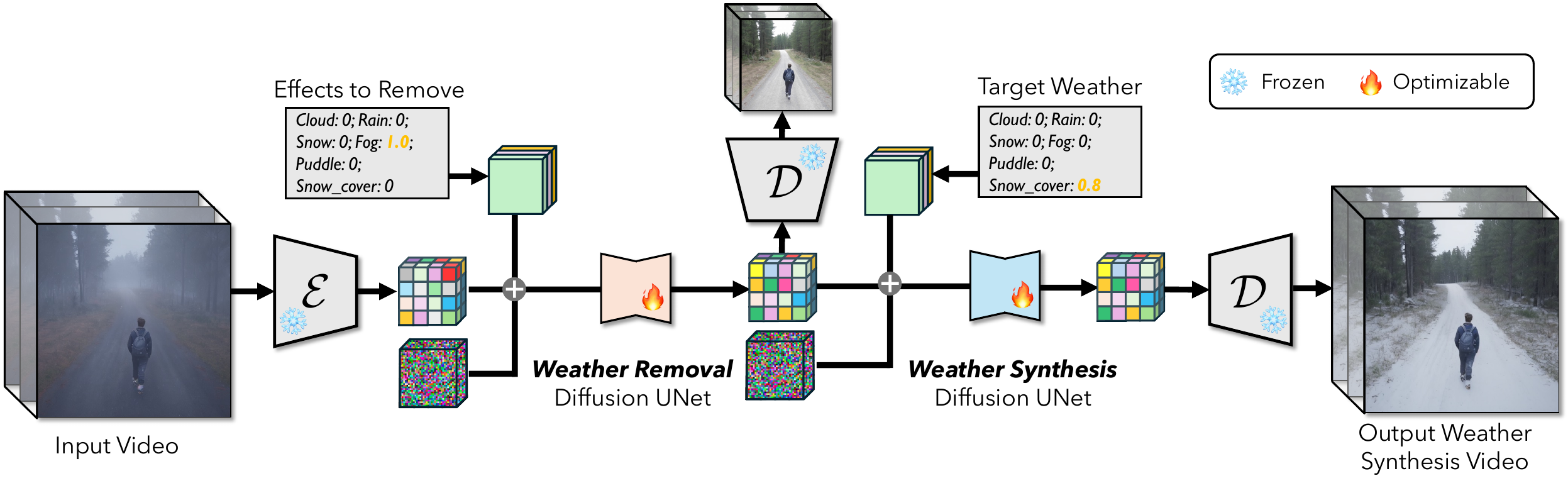}
\vspace{-5mm}
\caption{\textbf{Model Overview.} 
Our controllable weather simulation framework includes two complementary models for both weather removal and weather synthesis. These models can be used both independently and combined for weather editing tasks. %
}
\vspace{-5mm}
\label{fig:framework}
\end{figure*}

\vspace{-2mm}
\section{Related Work}
\vspace{-2mm}

\parahead{Video Editing}\label{sec:realted_video_editing}
Image editing with generative priors has been extensively studied \cite{avrahami2022blended, hertz2022prompt, meng2021sdedit, tumanyan2023plug}. However, directly applying image diffusion models in a frame-wise manner to video often leads to temporal inconsistencies. To mitigate flicker and jitter artifacts, recent methods~\cite{ceylan2023pix2video, khachatryan2023text2video, zhang2023controlvideo} inverts the initial latent code and employs cross-attention control to enforce frame consistency. Similarly, \cite{qi2023fatezero, geyer2023tokenflow} fuse attention maps or diffusion features from the source video with those from the generated video, thereby preserving fine details and ensuring content consistency with source frames. 
Other approaches \cite{esser2023structure, liang2024flowvid, feng2024ccedit, DiffusionRenderer} incorporate structural constraints or auxiliary information—such as depth maps, optical flow or G-buffers—to align generated frames with the original geometry and motion. Alternatively, some methods~\cite{hsu2024autovfx, haque2023instruct} build 3D representations from source videos and apply a diffusion prior for 3D editing to ensure consistency. 

Given sufficient computational budget, an alternative line of work explored one-shot fine-tuning to personalize the model to target video \cite{wu2023tune, molad2023dreamix, shin2024edit}. 
Our work builds on a pretrained video diffusion model, but eliminates the need for per-video fine-tuning and provides more precise control.

\parahead{Weather Synthesis}\label{sec:related_weather_simulation}
serves as a valuable augmentation to existing data and benefits perception tasks under adversarial weather conditions~\cite{schmalfuss2023distracting, volk2019towards, tremblay2021rain, von2019simulating}.
ClimateGAN~\cite{schmidt2021climategan, cosne2020using} generates flood images from depth information; \cite{hahner2019semantic} synthesize controllable fog based on depth and semantics. These methods focus on specific weather effects for static images. 
Similarly, \cite{schmidt2019visualizing} uses CycleGAN~\cite{zhu2017unpaired} for image editing on a climate dataset. 
In contrast, \ourmodel{} is a general framework that synthesizes and controls various weather effects, including transient effects (\eg rain, snow) in videos.

An alternative line of works synthesizes weather effects in 3D representations with graphics techniques~\cite{sulsky1995application}. 
\cite{feldman2002modeling, gissler2020implicit, stomakhin2013material} simulate snow particles and their interaction with objects and wind. These methods are typically limited to synthetic environments. ClimateNeRF~\cite{Li2023ClimateNeRF} and subsequent works~\cite{dai2025rainygs, fiebelman2025let} extend classic weather simulation by inserting physical entities into neural 3D reconstructions~\cite{muller2022instant, kerbl3Dgaussians}, but they require accurate geometry that is challenging to acquire from sparse capture. 
\ourmodel{} leverages a data-driven video diffusion model, bypassing the need for geometry reconstruction and enabling realistic effects on diverse and dynamic videos. 

\parahead{Weather Removal}\label{sec:related_weather_removal}
is a long-standing problem for robust computer vision systems.
Early methods targeted specific weather effects, such as deraining~\cite{yang2017deep, qian2018attentive, quan2021removing, xiao2022image}, dehazing \cite{cai2016dehazenet, li2017aod, liu2019griddehazenet, wu2021contrastive}, and desnowing~\cite{liu2018desnownet, chen2020jstasr, chen2023msp}, using specialized architectures tailored to each weather type. 
Recent approaches unify weather removal under a single model. 
All-in-One~\cite{li2020all} handles fog, rain, and snow with a unified CNN model.
\cite{valanarasu2022transweather, sun2024restoring, zhu2024mwformer} used transformer architectures with dedicated attention mechanisms to further improve restoration quality across diverse weather effects. 
ViWS-Net~\cite{yang2023video} introduced a video weather removal framework that incorporates temporal information for enhanced video restoration. 
Recent works explored using generative modelsfor weather removal~\cite{ye2023adverse, ozdenizci2023, chen2024teaching}. WeatherDiffusion~\cite{ozdenizci2023} uses patch-based diffusion denoising to effectively remove weather artifacts while preserving image details.
Prior works and benchmarks in weather removal primarily focus on transient effects like fog, rain, and snow, neglecting persistent weather effects such as cloud, puddle, and snow coverage.

\vspace{-2mm}
\section{Preliminary: Video Diffusion Model}
\vspace{-2mm}

Diffusion models generate samples from a data distribution $\dataDistribution(\videoInput)$ by iteratively refining noisy inputs through a denoising process~\cite{sohl2015deep, ho2020denoising, dhariwal2021diffusion}. 
In the context of videos, video diffusion models (VDMs) typically operate in a compressed latent space to reduce computational complexity~\cite{blattmann2023stable}. An input video $\videoInput \in \mathbb{R}^{\videoLength \times H \times W \times 3}$, with $\videoLength$ frames at resolution $H \times W$, is encoded into a latent representation $\latent = \vaeEncoder(\videoInput) \in \mathbb{R}^{l \times h \times w \times C}$ using a pre-trained VAE encoder $\vaeEncoder$. The diffusion process is then applied within this latent space. 

During training, noisy versions of the latent representation $\latent_\diffusionTime$ are generated by adding Gaussian noise $\diffusionNoise$ to the original latent $\latent_0$ using a predefined noise schedule~\cite{Karras2022edm} $\latent_\diffusionTime = \alpha_\diffusionTime \latent_0 + \sigma_\diffusionTime \diffusionNoise$ at timestep $\diffusionTime$. 
The diffusion model is trained to reverse this process using a denoising score matching objective~\cite{Karras2022edm}
$\|\diffusionModel_{\diffusionModelParams} \left(\latent_\diffusionTime; \diffusionCond, \diffusionTime \right) - \latent_0\|_2^2$ where $\diffusionCond$ denotes optional conditioning information. 
Once trained, the model generates new video samples by iteratively denoising Gaussian noise. 
The final output video $\videoOutput$ is reconstructed by decoding the denoised latent with the VAE decoder $\vaeDecoder$.

Our method is designed to be model-agnostic and can be applied to any video diffusion model. In this work, we build on Stable Video Diffusion~\cite{blattmann2023stable}, which compresses the spatial dimensions of the video by a factor of 8 while preserving the temporal resolution, using a latent dimension of $C=4$. 

\vspace{-2mm}
\section{Method}\label{sec:method}
\vspace{-2mm}

\begin{figure*}[t]
\centering
\vspace{-6mm}
\includegraphics[width=1.0\textwidth]{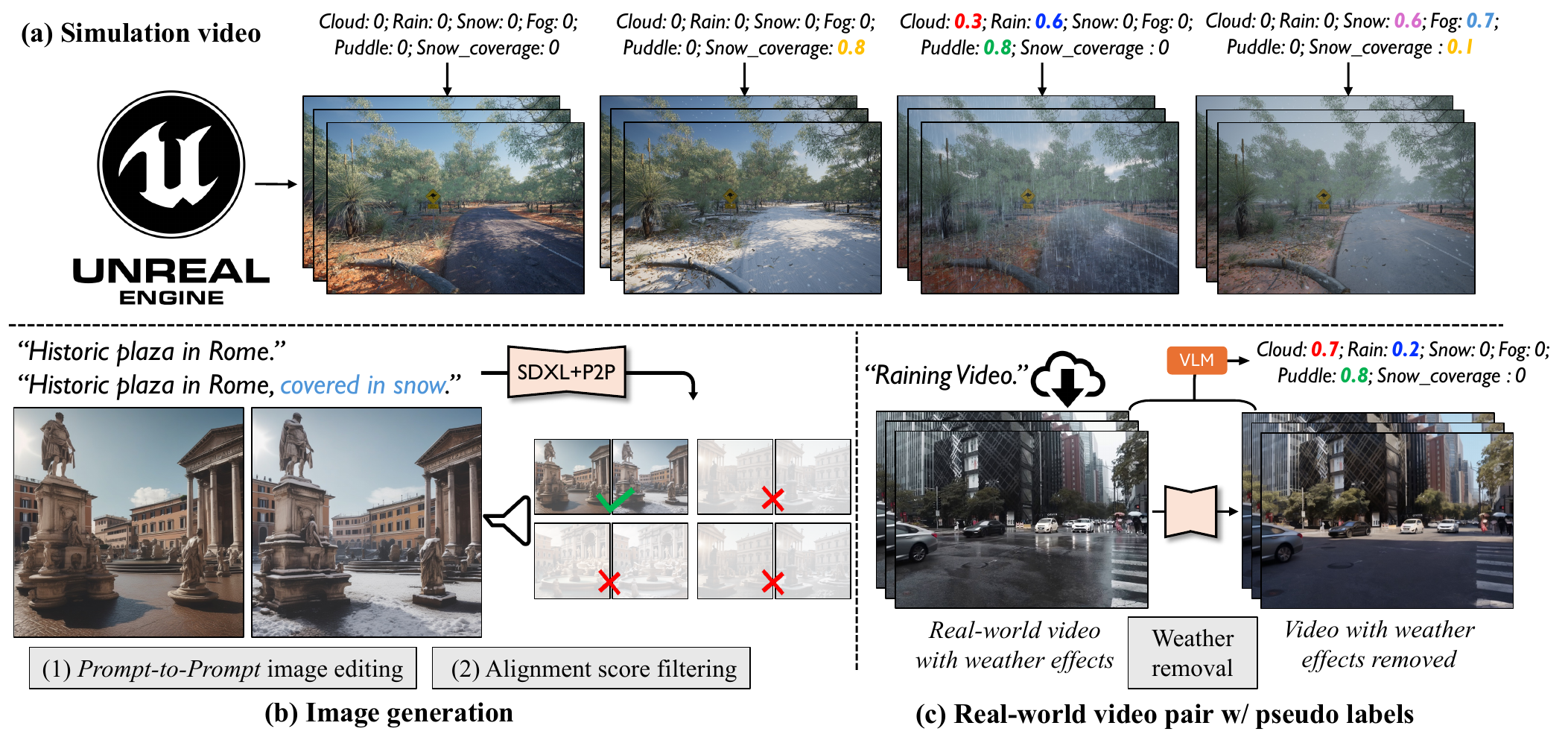}
\vspace{-9mm}
\caption{\textbf{Data Strategy.} 
We collect paired image and video data from (a) simulation engine, (b) text-to-image generative models with Prompt-to-prompt~\cite{hertz2022prompt}, and (c) auto-labeling real-world online videos. 
}
\vspace{-5mm}
\label{fig:data_strategy}
\end{figure*}

\begin{table}[t]
\centering
\resizebox{1.0\linewidth}{!}{%
\begin{tabular}{lcccccc}
\toprule
\textbf{Dataset}  & \textbf{Size}  & \textbf{\makecell{Weather \\ Controllability}} & \textbf{\makecell{Temporal \\ Consistency}} & \textbf{Realism} & \textbf{\makecell{Scene \\ Diversity}} & \textbf{\makecell{Trajectory \\ Diversity}} \\
\midrule

Simulation  & 2080k & \greenmark & \greenmark & \orangemark & \redcross & \greenmark \\

Generation  & 1147k & \orangemark & \redcross & \orangemark & \greenmark & \redcross \\

Real videos & 460k & \redcross & \greenmark & \greenmark & \orangemark & \orangemark \\

\bottomrule
\end{tabular}
}
\vspace{-3mm}
\caption{\textbf{Dataset Statistics.} We collect the weather data from three heterogeneous data sources, and mark each properties as high (\greenmark), moderate (\orangemark), and low/none (\redcross). The data size is the number of image pairs (with and without weather effects).}
\vspace{-6mm}
\label{tab:datasets}
\end{table}

We formulate weather simulation in real-world videos as a video-to-video translation task using two complementary and controllable video diffusion models. The \textsc{weather removal model} removes existing weather effects to generate a clear day video, while the \textsc{weather synthesis model} adds weather effects to the clear day video with precise control over both type and intensity.

To train these models, we decompose weather into its fundamental components (Sec.~\ref{sec:network}), curate a diverse multi-source dataset (Sec.~\ref{sec:data_collection}), and propose a staged training strategy (Sec.~\ref{sec:training}). The overall pipeline is shown in Fig.~\ref{fig:framework}.

\vspace{-1mm}
\subsection{Model Design}\label{sec:network}
\vspace{-1mm}
Our method is designed to flexibly represent and control individual weather effects. Both the weather removal and synthesis are formulated as conditional video generation task and use the same network architecture. 

\parahead{Representing Weather Effects} 
To enable precise control over weather type and intensity, we decompose weather into six distinct effects: 1) cloud, 2) fog, 3) rain, 4) snow, 5) puddle, and 6) snow coverage (i.e., persistent snow accumulation on the ground and objects). Each effect is parameterized by a continuous strength value $s \in \mathbb{R}^{+}$, where higher values indicate stronger manifestations (e.g., denser fog or heavier rain). The overall weather condition for a video is thus represented by the vector
\[
\strength = (s_{\text{cloud}}, s_{\text{fog}}, s_{\text{rain}}, s_{\text{snow}}, s_{\text{puddle}}, s_{\text{snow\_coverage}}) \in \mathbb{R}^6.
\]
This parametric representation precisely captures weather variations and offers intuitive control over both the type and intensity of effects applied to the input video. By combining individual conditions, our model can synthesize a wide array of realistic weather conditions (Fig.~\ref{fig:qual_forward},~\ref{fig:controllable}).

\parahead{Weather Synthesis}  
Given an input video $\videoClear$ and a conditioning signal $\mathbf{s}$, our \textsc{weather synthesis model} outputs the synthesized video with desired weather effects $\mathbf{\hat{I}}^{w}$. 
We formulate weather synthesis as a conditional video generation task, and aim to approximate weather synthesis in a data-driven manner, allowing the model to operate on arbitrary input videos without relying on explicit 3D geometry.

Our \textsc{weather synthesis model} $\simulationModel$ is initialized with the pre-trained weights of Stable Video Diffusion and operates in the VAE latent space. Specifically, for each data sample $(\videoClear, \videoWeather, \mathbf{s})$, we encode both the input video $\videoClear$ and the corresponding weather-affected video $\videoWeather$ into the latent space using the VAE encoder:
$$
\latent_0^c = \vaeEncoder(\videoClear) \in \mathbb{R}^{l \times h \times w \times C}, 
\latent_0^w = \vaeEncoder(\videoWeather) \in \mathbb{R}^{l \times h \times w \times C} 
$$
To represent the strength of the weather effect, we construct a condition map $\strengthMap$ by expanding the condition vectors across spatial and temporal dimensions $\strengthMap = \mathbbm{1} \otimes \strength \in \mathbb{R}^{l \times h \times w \times 6}$, where $\mathbbm{1} \in \mathbb{R}^{l\times h \times w}$ denotes an all-one tensor.

During training, noisy video latents are obtained by adding Gaussian noise following the predefined noise schedule $\latent_\diffusionTime^w = \alpha_\diffusionTime \latent_0^w + \sigma_\diffusionTime \diffusionNoise$. 
In each denoising step, the noisy latent $\latent_\diffusionTime^w$, the video latent $\latent_0^c$, and the weather strength map $\strengthMap$ are concatenated as input into the UNet denoising function $\simulationModel$. 
To handle the concatenated input conditions, we add zero-initialized extra channels to the first convolution layer of the UNet.  
The model is optimized using the denoising score matching objective~\cite{Karras2022edm}: 
\begin{equation}\label{eq:loss_simulation}
    \lossSimulation = \|\simulationModel(\latentWeather_\diffusionTime; \latentClear_0, \strengthMap, \diffusionTime) - \latentWeather_0\|_2^2
\end{equation}

\parahead{Weather Removal} is similarly formulated as a conditional video generation task, sharing the same architecture as the \textsc{weather synthesis model}.
Given an input video with weather effects $\videoWeather$, and weather strengths $\strength$ indicating the effects to remove, the \textsc{weather removal model} generates the corresponding clear-day video $\mathbf{\hat{I}}^{c}$. 

During training, Gaussian noise is added to the clear-day video latent $\latentClear_0$ to create noisy latent $\latentClear_\diffusionTime$. 
The noisy latent is concatenated with the input video latent $\latentWeather_0$ and the weather strength map $\strengthMap$ to form the input for the UNet denoising function $\removalModel$. The training objective is defined as:
\begin{equation}\label{eq:loss_removal}
    \lossRemoval = \|\removalModel(\latentClear_\diffusionTime, \latentWeather_0, \strengthMap, \diffusionTime) - \latentClear_0\|_2^2
\end{equation}

At inference time, both weather synthesis and removal models produce photorealistic edited videos by iteratively denoising Gaussian noise with learned denoising functions.

\vspace{-1mm}
\subsection{Data Collection}\label{sec:data_collection}
\vspace{-1mm}
High-quality paired video data $(\videoClear, \videoWeather, \strength)$ is essential for training our models, where $\videoClear$ denotes clear-day videos without weather effects, $\videoWeather$ the corresponding videos with weather effects, and $\strength$ represents the strength of these effects. Collecting such data in real-world scenarios is challenging, and existing public datasets~\cite{blattmann2023stable, bain2021frozen, schuhmann2022laion} do not meet these specific requirements. 
To bridge this gap, we propose a data collection strategy that leverages three complementary sources: \textit{Simulation}, \textit{Generation}, and auto-labeled \textit{Real-World Videos}. Table~\ref{tab:datasets} summarizes the key properties of these sources, and Fig.~\ref{fig:data_strategy} shows examples of the collected data.

\parahead{Simulation}
To obtain paired video data with precise weather control, we use synthetic environments in Unreal Engine~\cite{unrealengine}. Specifically, we select four large-scale, artist-generated outdoor scenes consisting of city streets, wild forests, towns, and rural areas and simulate six weather effects at varying intensities. To mimic real-world conditions, we also randomly combine these individual effects.

We generate diverse camera trajectories by sampling an initial pose and then randomly selecting subsequent poses within defined spatial bounds, using collision detection to avoid asset intersections. Lighting was varied by randomly sampling environment maps covering different times of day.

By automating this workflow via Unreal Engine scripting, we produced 20.8k video pairs, each comprising 100 frames with labeled ground truth weather effects.

\parahead{Generation}
High-quality synthetic assets are costly to obtain and often lack scene diversity. In contrast, generative models can synthesize a rich variety of data and scale with compute. To make use this resource, we follow \citet{brooks2023instructpix2pix} and use Prompt-to-Prompt~\cite{hertz2022prompt} in combination with SDXL~\cite{podell2023sdxl} to generate paired images—with and without weather effects—while maintaining structural consistency. 

Specifically, we use large language models~\cite{OpenAI_ChatGPT, brown2020language} to generate 61k scene descriptions (\eg ``A coastal road bordered by palm trees'') and 10 pairs of weather-related captions for each of the six weather effects (\eg ``on a sunny day'' versus ``on a snowy day''). 
These paired captions enable us to generate image pairs through Prompt-to-Prompt. 
To synthesize varying weather intensities, we adjust the cross-attention weights for weather-related tokens (e.g., “snowy”) and use these weights as strength labels (see ~\cite{hertz2022prompt} for further details).

We observed that the generative model often fails to adhere to the provided prompts. To address this, we filter the generated samples by measuring the consistency between image pairs and their corresponding caption pairs in the CLIP embedding space~\cite{radford2021learning}, following the approach in~\cite{brooks2023instructpix2pix, gal2022stylegan}. We then select the top 4\% of samples based on their consistency scores. For each selected sample, we generate 5 image pairs with varying effect strengths, resulting in a total of 1,147k high-quality paired images that capture diverse weather variations across numerous scenes.

Although this pipeline produces image pairs rather than video pairs, the diversity provided by these images significantly benefits our model. Extending attention-based techniques to text-to-video generation~\cite{blattmann2023stable, hong2022cogvideo, yang2024cogvideox} is promising but demands considerably more resources and less scalable. Hence, we leave video-based data generation for future work.

\parahead{Real-world Videos} offer high diversity and realism, yet obtaining paired examples with and without weather effects remains challenging. To address this, we introduce an auto-labeling strategy that leverages the abundance of photorealistic weather videos available online to generate additional training data for our \textsc{weather synthesis model}. 

Specifically, we collect online videos capturing significant weather events such as heavy rainstorms and snowfall. We then use our pre-trained \textsc{weather removal model} and generate corresponding weather-free versions, effectively transforming the input videos into clear-day sequences (see Fig.\ref{fig:data_strategy}). To label the weather effect strengths we use a vision-language model (VLM) \cite{Qwen2-VL} with in-context learning. By providing the VLM with simulation data examples and their corresponding strength labels, we instruct them to estimate weather effect strengths for the collected real-world videos.

In total, we collected and processed 4.6k video pairs (100 frames per video) that capture the realistic appearance and dynamic variations of diverse weather conditions.

\vspace{-1mm}
\subsection{Training Strategy}\label{sec:training}
\vspace{-1mm}
We use a multi-stage training strategy to combine the strengths of different data sources. 
We first train the \textsc{weather removal model} $\removalModel$ using a combination of simulation and generation data. 
Since the generation dataset contains only images, we perform image-video co-training by treating each image as a single-frame video. 
Once trained, we use the model and auto-label real-world videos by generating corresponding videos with weather effects removed. 

For \textsc{weather synthesis model} $\simulationModel$, we start by training on both simulation and generation data, enabling the model to learn precise control over weather effects. Finally, we jointly train $\simulationModel$ on all three data sources of simulation, generation, and auto-labeled real-world video data.

\begin{figure*}[t]
    \centering\setlength{\tabcolsep}{0.1em}
    \vspace{-7mm}
    \resizebox{1.0\textwidth}{!}{%
    \begin{tabular}{@{}lccccc@{}}
    
    & Input & TokenFlow~\cite{geyer2023tokenflow} & WeatherDiffusion~\cite{ozdenizci2023} & Histoformer~\cite{sun2024restoring} & Ours \\[0.2em]

    \raisebox{2.5\normalbaselineskip}[0pt][0pt]{\rotatebox[origin=c]{90}{Fog}} & \frame{\includegraphics[width=0.2\textwidth]{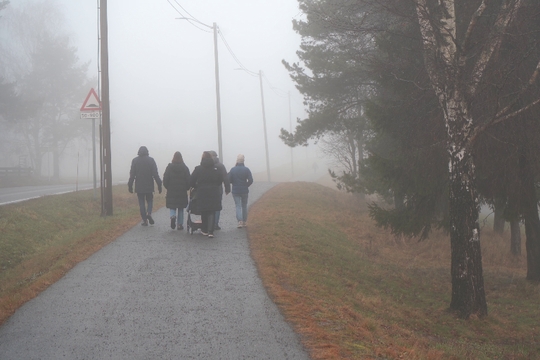}} & 
    \frame{\includegraphics[width=0.2\textwidth]{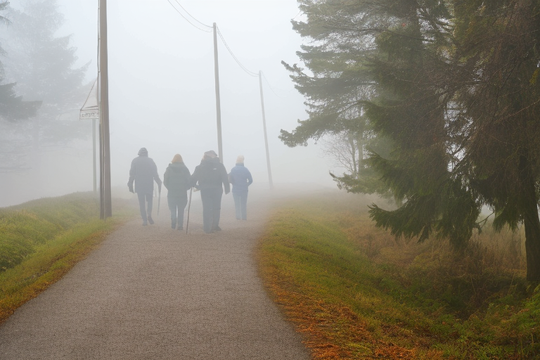}} &
    \frame{\includegraphics[width=0.2\textwidth]{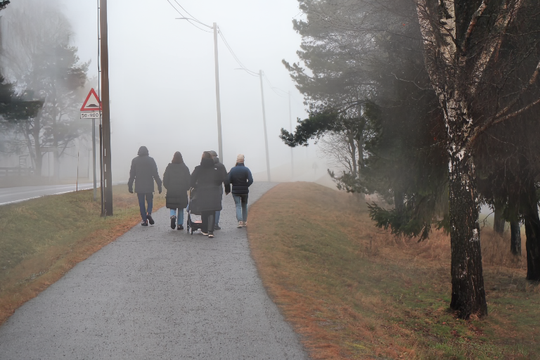}} & 
    \frame{\includegraphics[width=0.2\textwidth]{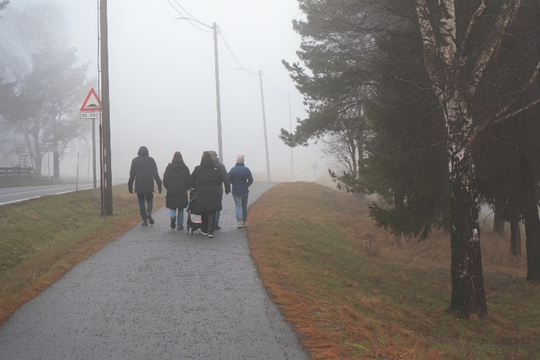}} & 
    \frame{\includegraphics[width=0.2\textwidth]{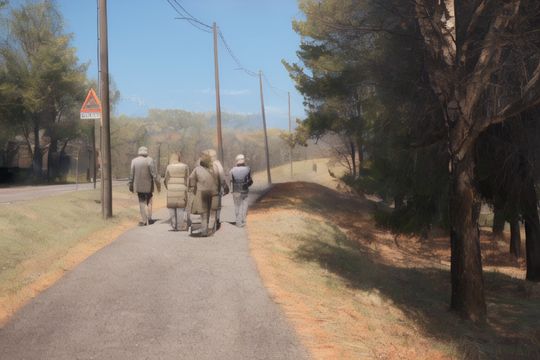}} \\

    \raisebox{2.5\normalbaselineskip}[0pt][0pt]{\rotatebox[origin=c]{90}{Rain}} & \frame{\includegraphics[width=0.2\textwidth]{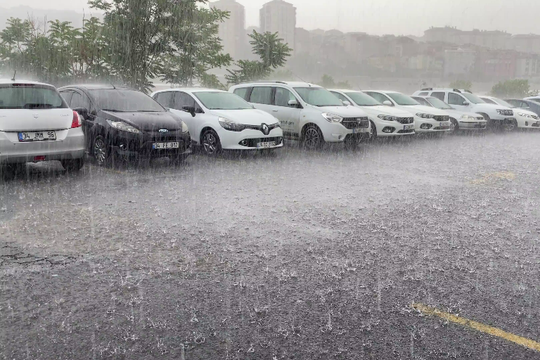}} & 
    \frame{\includegraphics[width=0.2\textwidth]{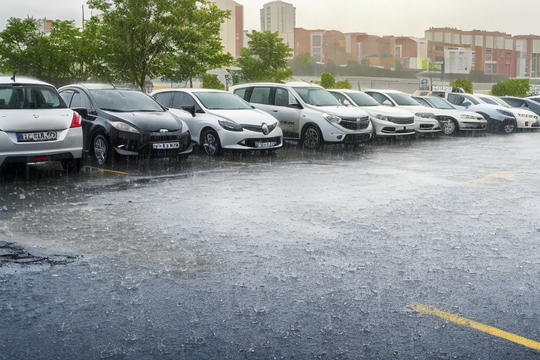}} &
    \frame{\includegraphics[width=0.2\textwidth]{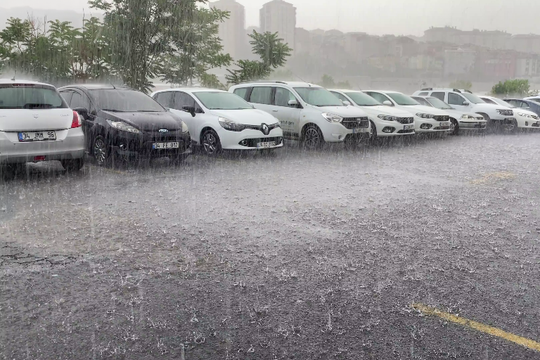}} & 
    \frame{\includegraphics[width=0.2\textwidth]{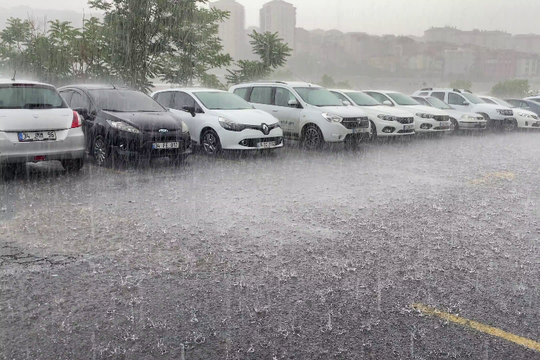}} & 
    \frame{\includegraphics[width=0.2\textwidth]{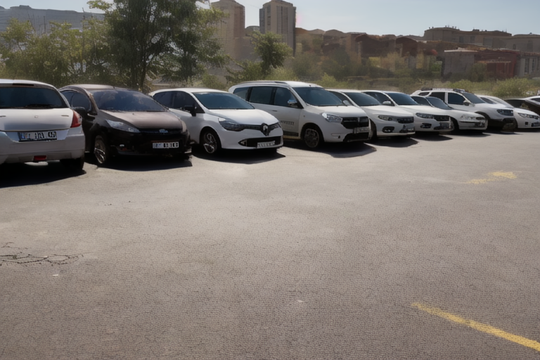}} \\

    \raisebox{2.5\normalbaselineskip}[0pt][0pt]{\rotatebox[origin=c]{90}{Snow + Fog}} & \frame{\includegraphics[width=0.2\textwidth]{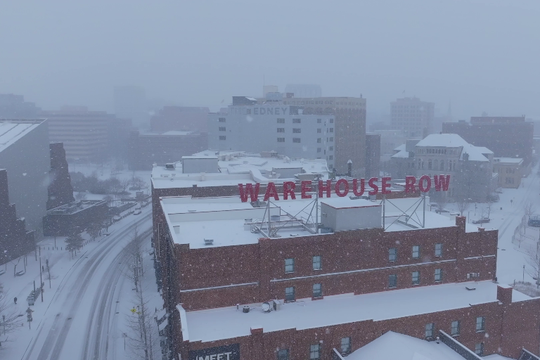}} & 
    \frame{\includegraphics[width=0.2\textwidth]{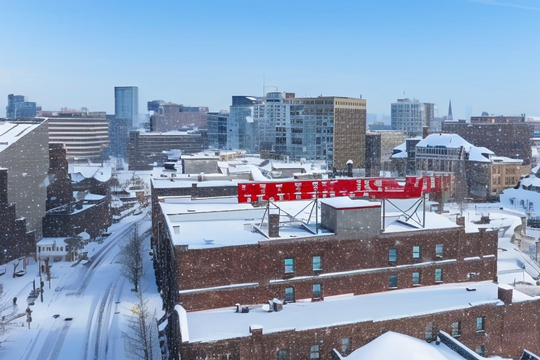}} &
    \frame{\includegraphics[width=0.2\textwidth]{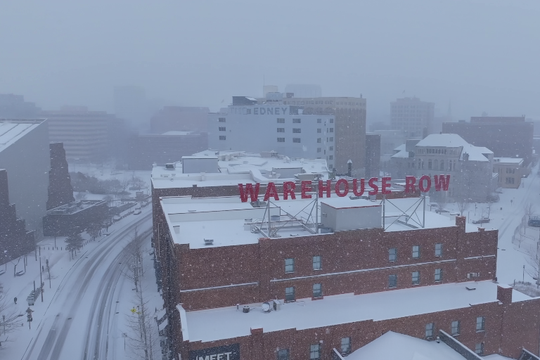}} & 
    \frame{\includegraphics[width=0.2\textwidth]{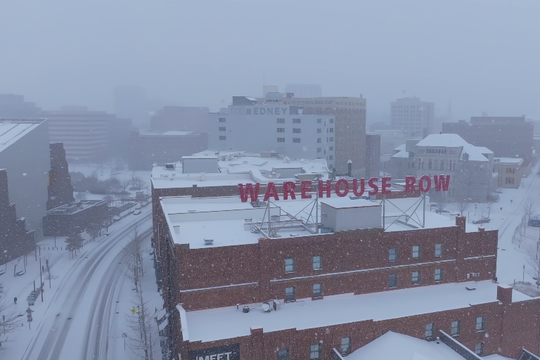}} & 
    \frame{\includegraphics[width=0.2\textwidth]{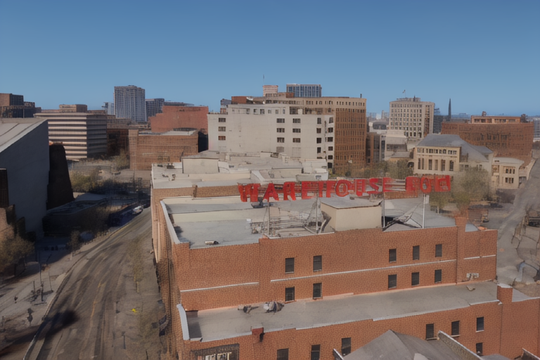}} \\

    \end{tabular}%
    }
    \vspace{-4mm}
    \caption{{Qualitative comparison with state-of-the-art methods on weather removal.}}
    \vspace{-4mm}
    \label{fig:qual_inverse}
\end{figure*}

\begin{figure*}[t]
    \centering\setlength{\tabcolsep}{0.1em}
    \resizebox{1.0\textwidth}{!}{%
    \begin{tabular}{@{}lccccc@{}}
    
    & Input & AnyV2V~\cite{ku2024anyv2v} & TokenFlow~\cite{geyer2023tokenflow} & FRESCO~\cite{yang2024fresco} & Ours \\[0.2em]

    \raisebox{2.0\normalbaselineskip}[0pt][0pt]{\rotatebox[origin=c]{90}{Fog}} & \frame{\includegraphics[width=0.2\textwidth]{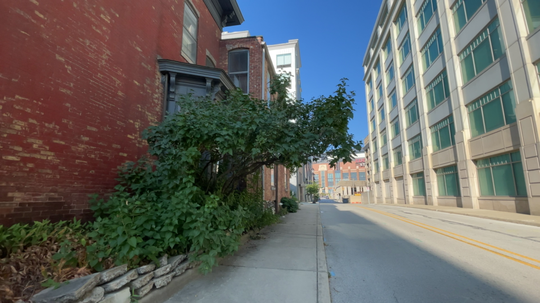}} & 
    \frame{\includegraphics[width=0.2\textwidth]{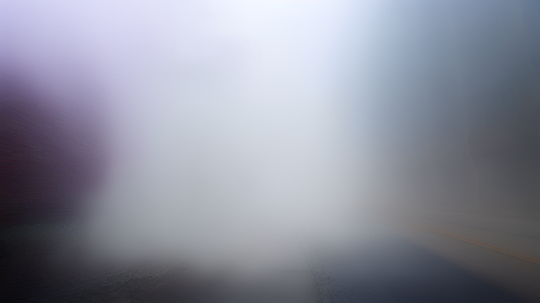}} &
    \frame{\includegraphics[width=0.2\textwidth]{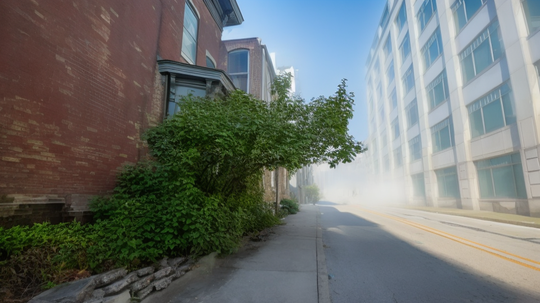}} & 
    \frame{\includegraphics[width=0.2\textwidth]{figures/images/qual_forward/dl3dv-adb95/fog/tokenflow/00011_tokenflow.png}} & 
    \frame{\includegraphics[width=0.2\textwidth]{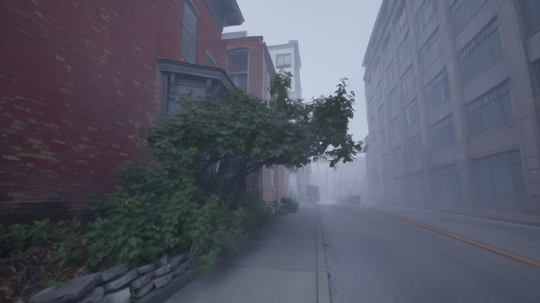}} \\

    \raisebox{2.5\normalbaselineskip}[0pt][0pt]{\rotatebox[origin=c]{90}{Rain}} & \frame{\includegraphics[width=0.2\textwidth]{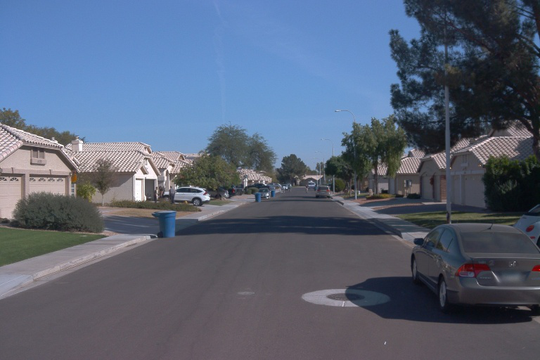}} & 
    \frame{\includegraphics[width=0.2\textwidth]{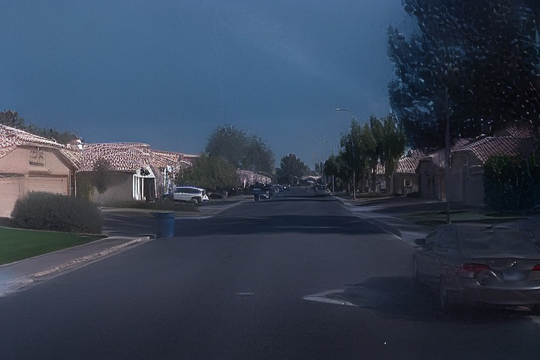}} &
    \frame{\includegraphics[width=0.2\textwidth]{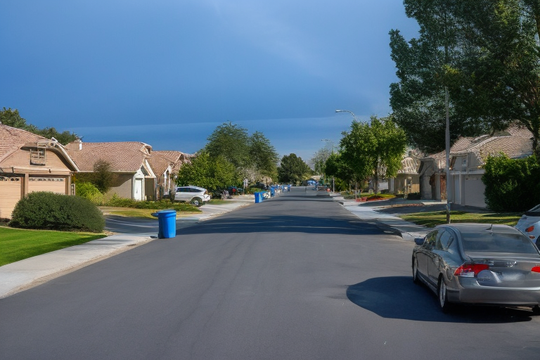}} & 
    \frame{\includegraphics[width=0.2\textwidth]{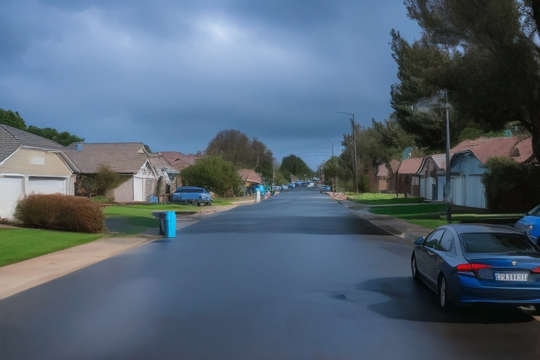}} & 
    \frame{\includegraphics[width=0.2\textwidth]{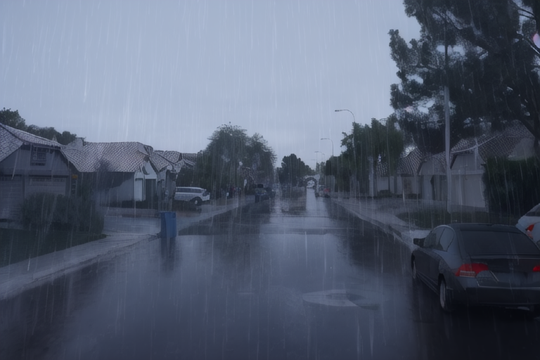}} \\

    \raisebox{2.5\normalbaselineskip}[0pt][0pt]{\rotatebox[origin=c]{90}{Snow}} & \frame{\includegraphics[width=0.2\textwidth]{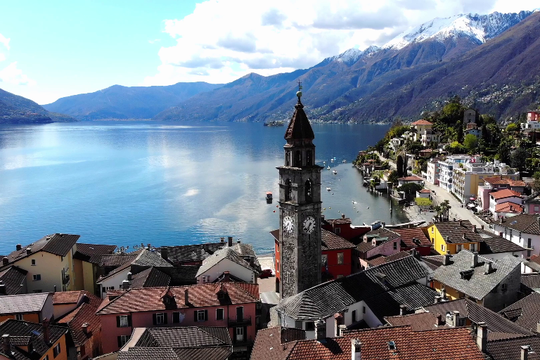}} & 
    \frame{\includegraphics[width=0.2\textwidth]{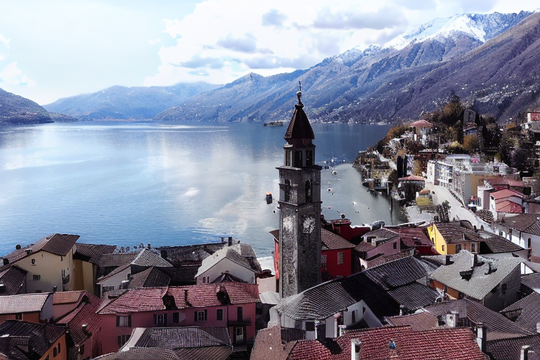}} &
    \frame{\includegraphics[width=0.2\textwidth]{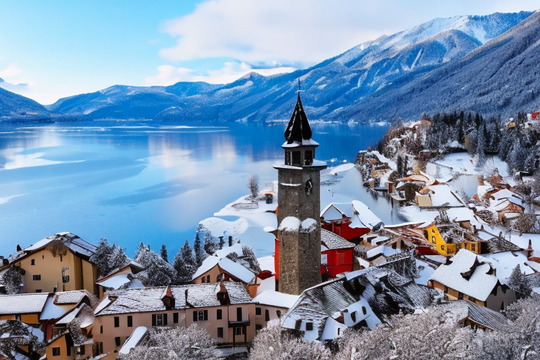}} & 
    \frame{\includegraphics[width=0.2\textwidth]{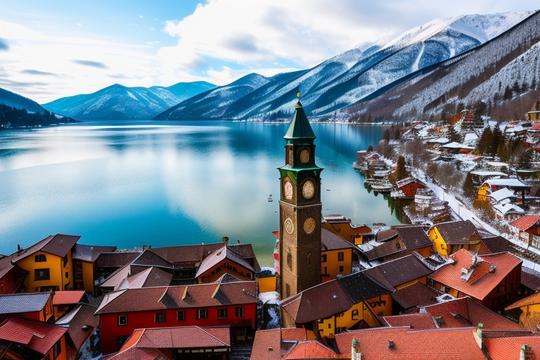}} & 
    \frame{\includegraphics[width=0.2\textwidth]{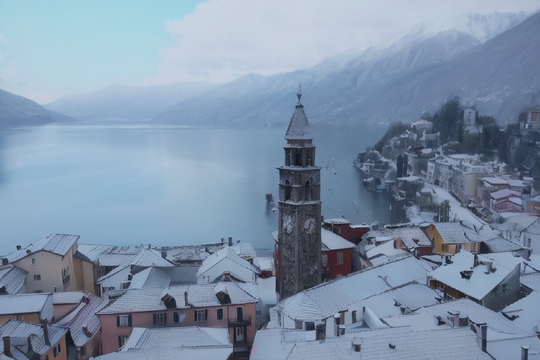}} \\

    \end{tabular}%
    }
    \vspace{-4mm}    
    \caption{{Qualitative comparison with state-of-the-art methods on weather synthesis.}
    }
    \vspace{-5mm}    
    \label{fig:qual_forward}
\end{figure*}

\vspace{-2mm}
\section{Experiments}\label{sec:experiments}
\vspace{-2mm}
We extensively evaluate our method on real-world video sequences and compare with state-of-the-art. 
Both qualitative and quantitative results demonstrate the effectiveness of our approach for weather synthesis, removal, and downstream applications. Video results are included in the Supplement.

\parahead{Datasets}
To evaluate generalization and ensure a fair comparison with baselines, we collect test video sequences from three distinct, non-overlapping sources: 
driving sequences from the Waymo Open Dataset~\cite{sun2020scalability}, outdoor scenes from DL3DV~\cite{ling2024dl3dv}, and casual in-the-wild videos from Pexels~\cite{pexels}. 
In total, we use 40 videos for weather synthesis and 55 videos (with fog, rain, or snow) for weather removal evaluation.

\parahead{Baselines}
We compare our method with diffusion-based video editing approaches, including Text2Live~\cite{bar2022text2live}, AnyV2V~\cite{ku2024anyv2v}, TokenFlow~\cite{geyer2023tokenflow}, and FRESCO~\cite{yang2024fresco}. These works rely on text input for guidance. 
To enable scalable evaluation and reduce human bias, we use state-of-the-art VLM~\cite{Qwen2-VL} to generate synthesis/removal prompts from the first frame of each input sequence. 
We also compare with specialized methods for weather removal, including WeatherDiffusion~\cite{ozdenizci2023} and Histoformer~\cite{sun2024restoring}. Finally, we perform qualitative comparison with ClimateNeRF~\cite{Li2023ClimateNeRF} on weather synthesis. 

\parahead{Evaluation Metrics} 
For weather synthesis, all methods generate three effects (fog, rain, snow) for each input video. 
Our method uses a fixed effect strength of 1.0 to generate the results. 
To measure how well the output aligns with target effects, we use VLM~\cite{Qwen2-VL} to estimate alignment scores (denoted as Align.~VLM) based on weather descriptions, and measure the average cosine similarity of edited frames using CLIP~\cite{radford2021learning} (denoted as Align.~CLIP). 
Following prior works~\cite{wu2023tune, cong2023flatten}, we also adopt PickScore~\cite{kirstain2023pick}, which estimates alignment with human preferences. 
Temporal consistency is evaluated using VBench++~\cite{huang2023vbench, huang2024vbench++}, which computes CLIP feature similarity across frames and evaluate motion smoothness using motion priors from video model~\cite{li2023amt}. 
Structure preservation is measured using the DINO Structure score (DINO Struct.), following~\cite{tumanyan2022splicing, parmar2024one}, with all scores multiplied by 100. 
Finally, we evaluate the perceptual quality of generated videos with a user study.

\vspace{-1mm}
\subsection{Quantitative Evaluation}\label{sec:quan_eval}
\vspace{-1mm}
Table~\ref{tab:quant} shows the quantitative comparison of weather synthesis and removal tasks compared with four baseline methods. 
Our method consistently outperforms all baselines in terms of Align.~VLM, Align.~CLIP, and PickScore, demonstrating its effectiveness in synthesizing diverse weather conditions and removing existing weather effects. 
For structure preservation (DINO Struct.), our method ranks second best in synthesis and third best in removal, suggesting that while videos are modified with weather change, the overall structure is preserved well. 
While WeatherDiffusion~\cite{ozdenizci2023} and Histoformer~\cite{sun2024restoring} achieve higher structure preservation scores, their outputs often fail to remove weather effects, resulting in videos that are nearly identical to the inputs. This limitation is reflected in their lower alignment scores, PickScores, and the qualitative results shown in Fig.~\ref{fig:qual_inverse}. 
The supplementary video shows that our method also demonstrates good temporal consistency and motion smoothness.

\parahead{User Study}
We conducted a user study to assess the perceptual quality of our method's video outputs. 
Participants were shown the reference input video alongside two edited video results--one generated by our method and the other by a baseline model, with the order randomized. 
For each sample pair, we invited 11 users to perform binary selection from the video pairs, and used majority voting to determine the preferred video for each comparison. 
For the task of weather synthesis, users are instructed to select the video with more realistic weather effects. For weather removal, users select the videos with least visible weather effects. 
We repeat the full user study three times, and report the average percentage of samples where our method is preferred over baselines in Table~\ref{tab:user_study}. 
We also provide the standard deviation across the three experiments. 

Additionally, following recent research on using VLMs as perceptual evaluators~\cite{wu2023gpteval3d}, we randomly extract a single frame of each video and conduct the same evaluation on \textit{image} pairs using Qwen2.5-VL-72B~\cite{Qwen2-VL} as the perceptual evaluator. 
Our method is consistently preferred by both human and VLM evaluators on both weather synthesis and removal tasks. 

\vspace{-1mm}
\subsection{Qualitative Evaluation}
\vspace{-1mm}
Fig.~\ref{fig:qual_forward} compares our weather synthesis results with state-of-the-art video editing models~\cite{ku2024anyv2v, geyer2023tokenflow, yang2024fresco}. 
Our method effectively adapts lighting conditions for different weather, such as removing shadows and dimming lake reflections to simulate cloudy shading. 
Compared to baselines, our method introduces realistic weather elements that prior methods cannot handle, including reflective puddles, snow-covered roofs, falling snow and rain. 
Our approach preserves the overall structure by only modifying weather-related regions, while previous methods often change shapes, colors, and hallucinate new contents.

\begin{table}[t]
\centering
\resizebox{1.0\linewidth}{!}{%
\begin{tabular}{lcccccc}
\multicolumn{7}{c}{\large Weather Synthesis} \\
\toprule
Method & \makecell{Align. \\ VLM $\uparrow$} & \makecell{Align. \\ CLIP $\uparrow$} & PickScore $\uparrow$ & \makecell{Temporal \\ Consistency $\uparrow$} & \makecell{Motion \\ Smooth. $\uparrow$} & \makecell{DINO \\ Struct. $\downarrow$} \\
\midrule
Text2Live~\cite{bar2022text2live}   & 70.45 & \textbf{0.22} & 20.41 & \textbf{0.96} & \textbf{0.99} & 3.86 \\
AnyV2V~\cite{ku2024anyv2v}          & 65.62 & 0.18 & 20.11 & 0.95 & 0.98 & 3.98 \\
TokenFlow~\cite{geyer2023tokenflow} & 62.38 & 0.17 & 19.89 & \textbf{0.96} & 0.97 & \textbf{1.93} \\
FRESCO~\cite{yang2024fresco}        & 70.23 & 0.18 & 19.81 & 0.95 & 0.98 & 2.42 \\
Ours                                & \textbf{77.29} & \textbf{0.22} & \textbf{20.75} & \textbf{0.96} & \textbf{0.99} & 2.30 \\
\bottomrule
\\

\multicolumn{7}{c}{\large Weather Removal} \\
\toprule
Method & \makecell{Align. \\ VLM $\uparrow$} & \makecell{Align. \\ CLIP $\uparrow$} & PickScore $\uparrow$ & \makecell{Temporal \\ Consistency $\uparrow$} & \makecell{Motion \\ Smooth. $\uparrow$} & \makecell{DINO \\ Struct. $\downarrow$} \\
\midrule
TokenFlow~\cite{geyer2023tokenflow}   & 66.39 & 0.15 & 19.07 & \textbf{0.98} & 0.98 & 2.20 \\
FRESCO~\cite{yang2024fresco}          & 60.98 & 0.16 & 18.94 & 0.97 & 0.98 & 2.71 \\
WeatherDiffusion~\cite{ozdenizci2023} & 22.79 & 0.15 & 18.82 & \textbf{0.98} & \textbf{0.99} & 0.26 \\
Histoformer~\cite{sun2024restoring}   & 13.30 & 0.15 & 18.81 & \textbf{0.98} & \textbf{0.99} & \textbf{0.05} \\
Ours                                  & \textbf{71.61} & \textbf{0.17} & \textbf{19.10} & \textbf{0.98} & \textbf{0.99} & 2.09 \\
\bottomrule

\end{tabular}
}
\vspace{-3mm}
\caption{{Quantitative evaluation for weather synthesis and removal.}}
\label{tab:quant}
\vspace{-4mm}
\end{table}

\begin{table}[t]
\centering
\resizebox{1.0\linewidth}{!}{%
\begin{tabular}{lcccccc}
\multicolumn{7}{c}{\large Weather Synthesis} \\
\toprule
\multirow{2}{*}{Baselines} & \multicolumn{3}{c}{Human Evaluator} & \multicolumn{3}{c}{VLM Evaluator} \\
\cmidrule(lr){2-4} \cmidrule(lr){5-7} 
& Fog &	Rain &	Snow & Fog &	Rain &	Snow \\
\midrule
AnyV2V~\cite{ku2024anyv2v}          & $85\% \pm 24\%$  & $86 \% \pm 18 \%$ & $82 \% \pm 19\%$  & $80\%$ & $70\%$ & $58\%$ \\
FRESCO~\cite{yang2024fresco}        & $60\% \pm 17\%$  & $76 \% \pm 4 \%$  & $78 \% \pm 23\%$  & $60\%$ & $50\%$ & $53\%$ \\
Text2Live~\cite{bar2022text2live}   & $89\% \pm 4\%$   & $88 \% \pm 10 \%$ & $76 \% \pm 19\%$  & $80\%$ & $80\%$ & $73\%$ \\
TokenFlow~\cite{geyer2023tokenflow} & $59\% \pm 10\%$  & $66 \% \pm 10 \%$ & $67 \% \pm 10\%$  & $58\%$ & $55\%$ & $50\%$ \\
\bottomrule
\\
\multicolumn{7}{c}{\large Weather Removal} \\
\toprule
\multirow{2}{*}{Baselines} & \multicolumn{3}{c}{Human Evaluator} & \multicolumn{3}{c}{VLM Evaluator} \\
\cmidrule(lr){2-4} \cmidrule(lr){5-7} 
& Fog &	Rain &	Snow & Fog &	Rain &	Snow \\
\midrule
AnyV2V~\cite{ku2024anyv2v}            & $74 \% \pm 6\%$  & $62\% \pm 21\%$  & $70\% \pm  7\%$  & $63\%$ & $75\%$ & $63\%$ \\
FRESCO~\cite{yang2024fresco}          & $59 \% \pm 6\%$  & $71\% \pm 15\%$  & $67\% \pm 22\%$  & $88\%$ & $65\%$ & $67\%$ \\
Text2Live~\cite{bar2022text2live}     & $85 \% \pm 17\%$ & $94\% \pm 11\%$  & $93\% \pm 12\%$  & $75\%$ & $90\%$ & $92\%$ \\
TokenFlow~\cite{geyer2023tokenflow}   & $52 \% \pm 6\%$  & $65\% \pm 18\%$  & $75\% \pm 17\%$  & $50\%$ & $60\%$ & $58\%$ \\
Histoformer~\cite{sun2024restoring}   & $82 \% \pm 6\%$  & $80\% \pm 14\%$  & $82\% \pm 16\%$  & $75\%$ & $65\%$ & $75\%$ \\
WeatherDiffusion~\cite{ozdenizci2023} & $89 \% \pm 11\%$ & $87\% \pm 14\%$  & $87\% \pm 14\%$  & $100\%$ & $60\%$ & $75\%$ \\
\bottomrule
\end{tabular}
}
\vspace{-3mm}
\caption{\textbf{User study.} Evaluated by human and VLM evaluators, we report the percentage of samples where Ours is preferred over baselines. A preference $> 50\%$ indicates Ours outperforming baselines. 
} 
\label{tab:user_study}
\vspace{-4mm}
\end{table}

\begin{figure}[t]
    \centering\setlength{\tabcolsep}{0.1em}
    \resizebox{1.0\linewidth}{!}{%
    \begin{tabular}{@{}ccc@{}}

    \footnotesize{Fog: $s_{\text{fog}} = 0.5$} & \footnotesize{Fog: $s_{\text{fog}} = 0.8$} & \footnotesize{Fog: $s_{\text{fog}} = 1.0$}  \\
     
    \frame{\includegraphics[width=0.4\linewidth]{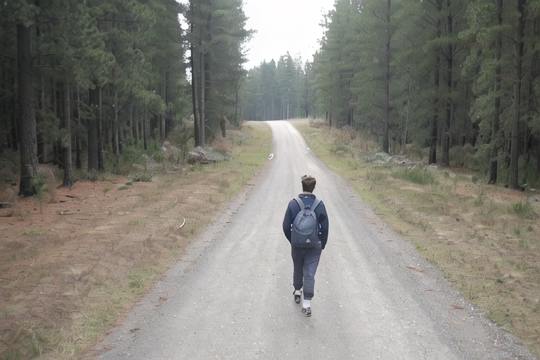}} &
    \frame{\includegraphics[width=0.4\linewidth]{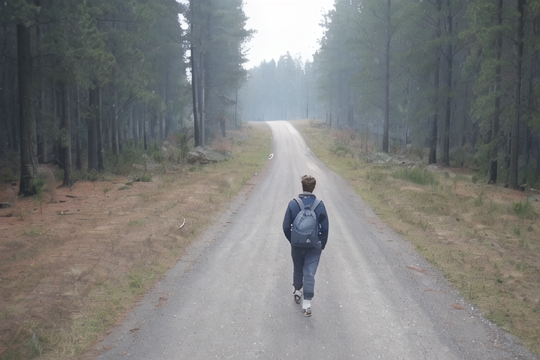}} &
    \frame{\includegraphics[width=0.4\linewidth]{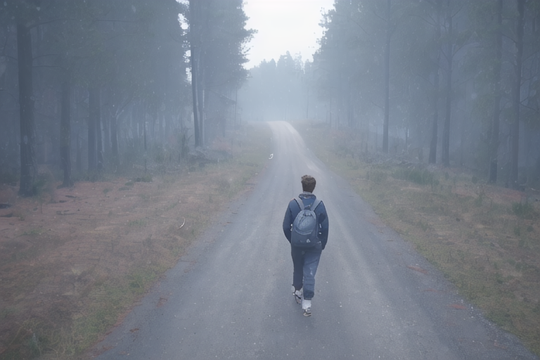}}  \\

    \footnotesize{Puddle: $s_{\text{puddle}} = 0.2$} & \footnotesize{Puddle: $s_{\text{puddle}} = 0.5$} & \footnotesize{Puddle: $s_{\text{puddle}} = 1.0$}  \\
     
    \frame{\includegraphics[width=0.4\linewidth]{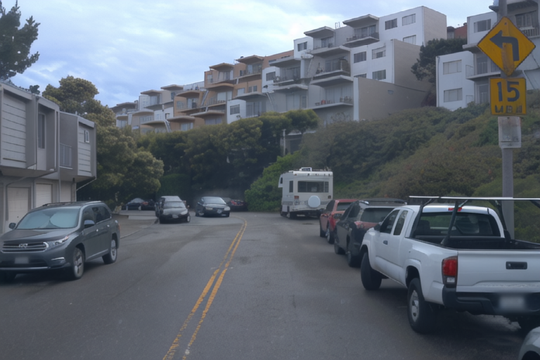}} &
    \frame{\includegraphics[width=0.4\linewidth]{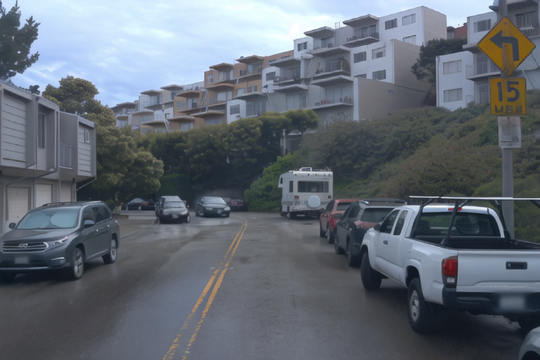}} &
    \frame{\includegraphics[width=0.4\linewidth]{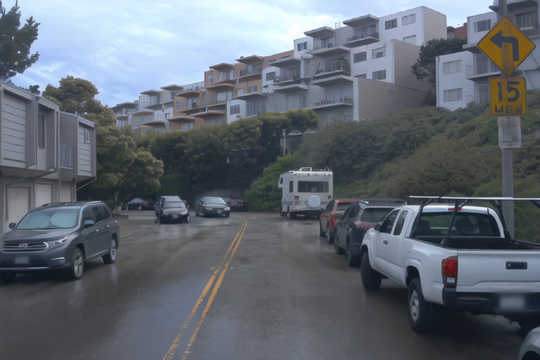}}  \\
    
    \end{tabular}%
    }
    \vspace{-4mm}
    \caption{{Controlling the strength of weather effects.}
    }
    \vspace{-3mm}
    \label{fig:slider}
\end{figure}

\begin{figure*}[t]
    \vspace{-5mm}
    \centering\setlength{\tabcolsep}{0.1em}
    \resizebox{1.0\textwidth}{!}{%
    \begin{tabular}{@{}lcccccc@{}}
    
     & Input & + cloud & + rain & + puddle & - rain & - cloud \\[0.2em]

    \raisebox{2.5\normalbaselineskip}[0pt][0pt]{\rotatebox[origin=c]{90}{Rainy day}} & 
    \frame{\includegraphics[width=0.2\textwidth]{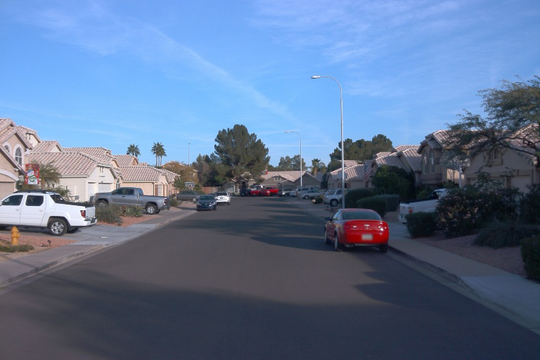}} & 
    \frame{\includegraphics[width=0.2\textwidth]{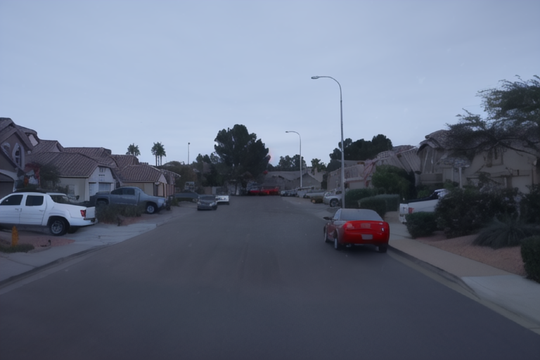}} &
    \frame{\includegraphics[width=0.2\textwidth]{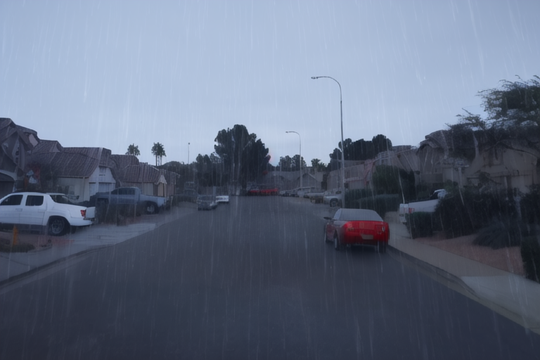}} & 
    \frame{\includegraphics[width=0.2\textwidth]{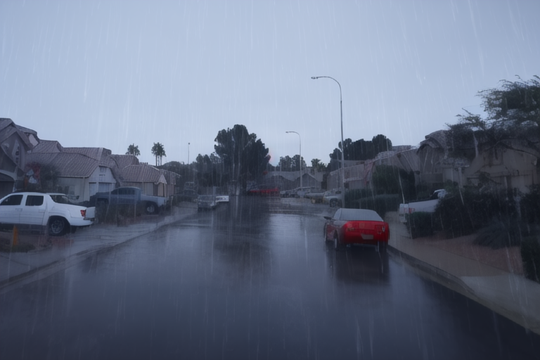}} & 
    \frame{\includegraphics[width=0.2\textwidth]{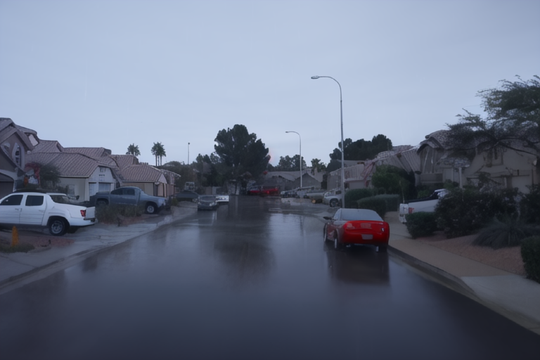}} &
    \frame{\includegraphics[width=0.2\textwidth]{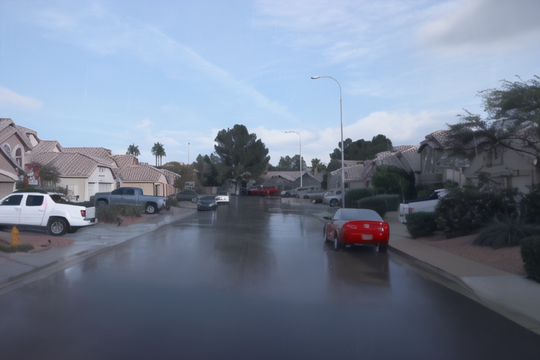}} \\

    & Input & + cloud & + snowflake & + fog + snow & - fog - snowflake & - cloud \\[0.2em]

    \raisebox{2.5\normalbaselineskip}[0pt][0pt]{\rotatebox[origin=c]{90}{Snowy day}} & 
    \frame{\includegraphics[width=0.2\textwidth]{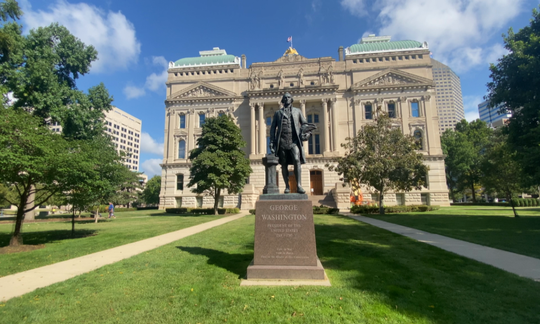}} & 
    \frame{\includegraphics[width=0.2\textwidth]{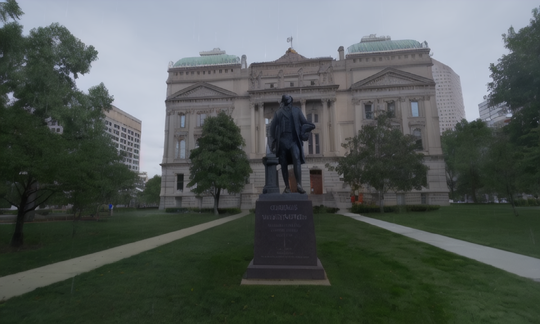}} &
    \frame{\includegraphics[width=0.2\textwidth]{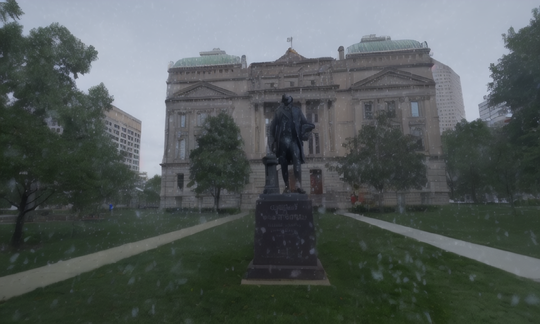}} & 
    \frame{\includegraphics[width=0.2\textwidth]{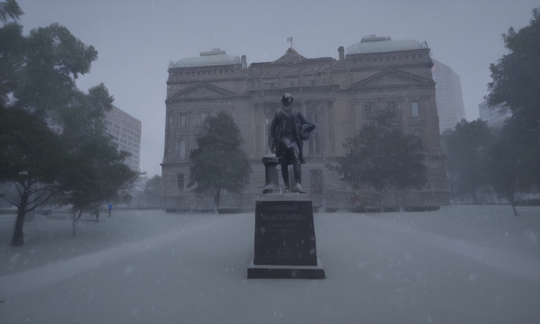}} & 
    \frame{\includegraphics[width=0.2\textwidth]{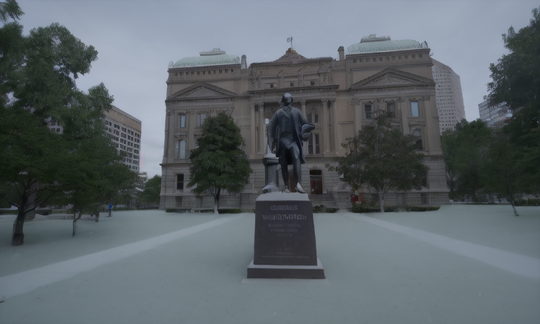}} &
    \frame{\includegraphics[width=0.2\textwidth]{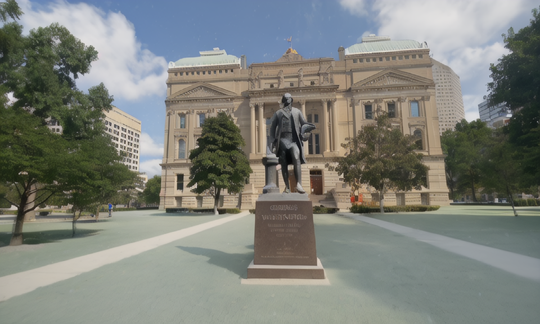}} \\

    \end{tabular}%
    }
    \vspace{-4mm}
    \caption{
    \textbf{Weather Editing with Multiple Effects.} 
    Our method allows sequential application and combination of multiple effects. 
    From left to right, we control the weather effect strengths and simulate how weather changes during rainy/snowy days.
    }
    \vspace{-5mm}
    \label{fig:controllable}
\end{figure*}

\begin{figure}[t]
    \centering\setlength{\tabcolsep}{0.1em}
    \resizebox{1.0\linewidth}{!}{%
    \begin{tabular}{@{}ccc@{}}

    \footnotesize{Input} & \footnotesize{Image model} & \footnotesize{No simulation data}  \\
     
    \frame{\includegraphics[width=0.4\linewidth]{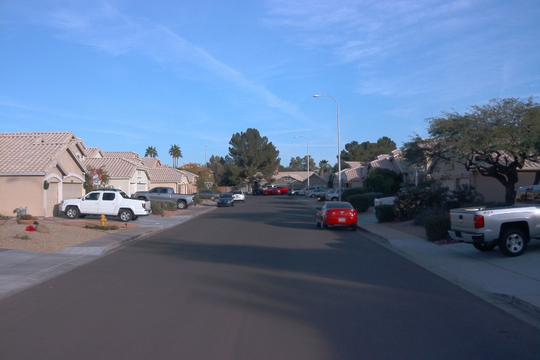}} &
    \frame{\includegraphics[width=0.4\linewidth]{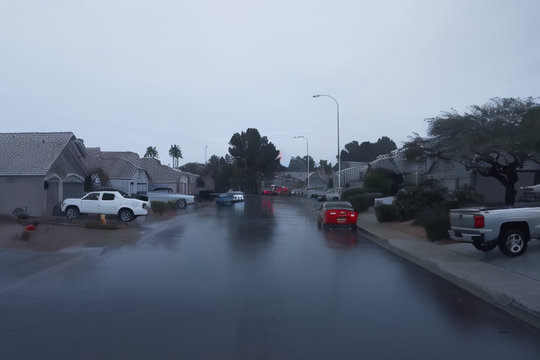}} &
    \frame{\includegraphics[width=0.4\linewidth]{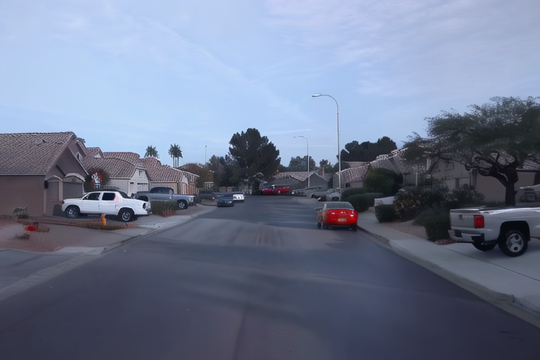}}  \\

    \footnotesize{No generation data} & \footnotesize{No real data} & \footnotesize{Full model}  \\

    \frame{\includegraphics[width=0.4\linewidth]{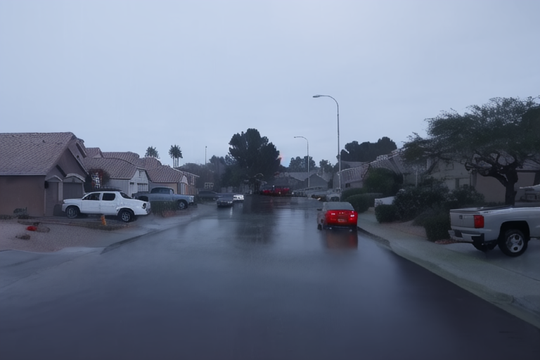}} &
    \frame{\includegraphics[width=0.4\linewidth]{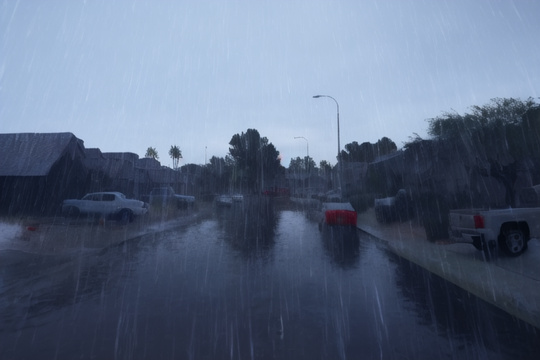}} &
    \frame{\includegraphics[width=0.4\linewidth]{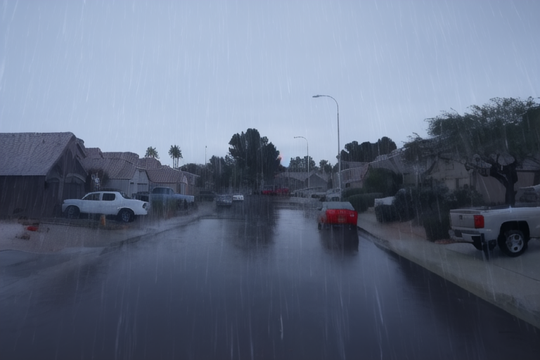}}  \\
    
    \end{tabular}%
    }
    \vspace{-3mm}
    \caption{\textbf{Ablation Study.} Our video model formulation improves the quality of transient effects and temporal consistency. Joint training with all data sources produces the best results. 
    }
    \vspace{-3mm}
    \label{fig:ablation}
\end{figure}

\begin{figure}[t]
    \centering\setlength{\tabcolsep}{0.1em}
    \resizebox{1.0\linewidth}{!}{%
    \begin{tabular}{@{}ccc@{}}

    \footnotesize{Input w/ original weather} & \footnotesize{Weather removal} & \footnotesize{Weather synthesis}  \\
     
    \frame{\includegraphics[width=0.4\linewidth]{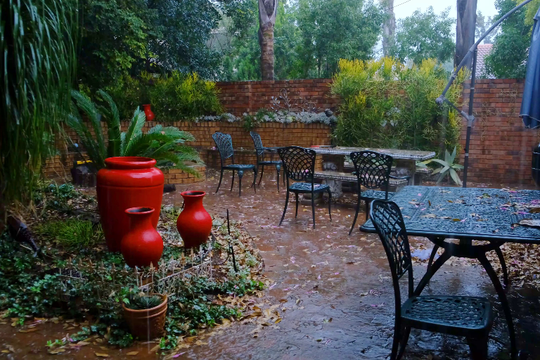}} &
    \frame{\includegraphics[width=0.4\linewidth]{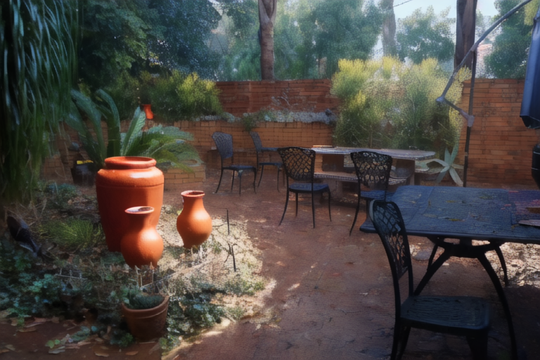}} &
    \frame{\includegraphics[width=0.4\linewidth]{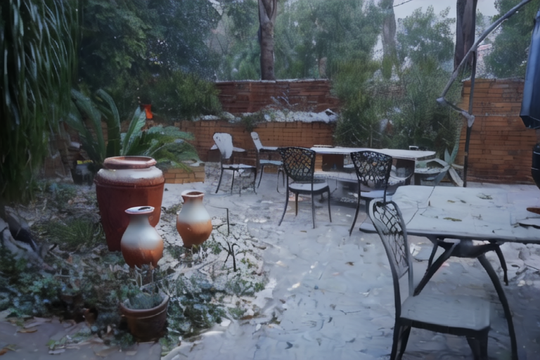}}  \\

    \frame{\includegraphics[width=0.4\linewidth]{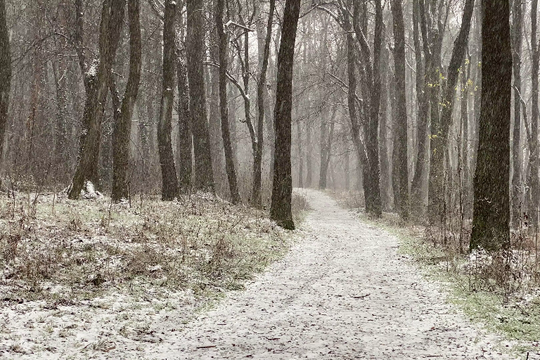}} &
    \frame{\includegraphics[width=0.4\linewidth]{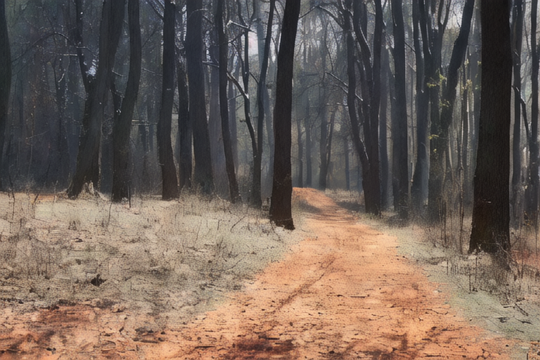}} &
    \frame{\includegraphics[width=0.4\linewidth]{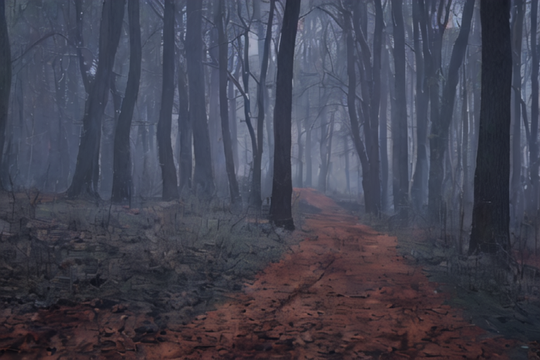}}  \\
    
    \end{tabular}%
    }
    \vspace{-3mm}
    \caption{\textbf{Weather Editing.} 
    Combined weather removal and synthesis models allow users to edit existing weather to different states. 
    }
    \vspace{-5mm}
    \label{fig:change_weather}
\end{figure}

\begin{figure}[t]
    \centering\setlength{\tabcolsep}{0.1em}
    \resizebox{1.0\linewidth}{!}{%
    \begin{tabular}{@{}lcccc@{}}
     & \footnotesize{$t = 1$} & \footnotesize{$t = 2$} & \footnotesize{$t = 3$} & \footnotesize{$t = 4$} \\

    \raisebox{1.5\normalbaselineskip}[0pt][0pt]{\rotatebox[origin=c]{90}{\footnotesize{Foggy input}}} & 
    \frame{\includegraphics[width=0.25\linewidth]{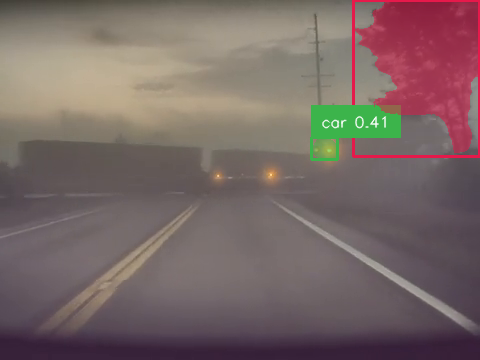}} &
    \frame{\includegraphics[width=0.25\linewidth]{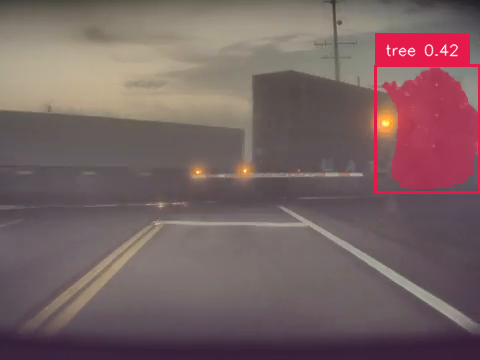}} &
    \frame{\includegraphics[width=0.25\linewidth]{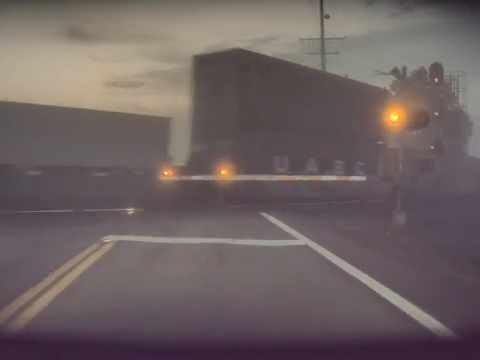}} &
    \frame{\includegraphics[width=0.25\linewidth]{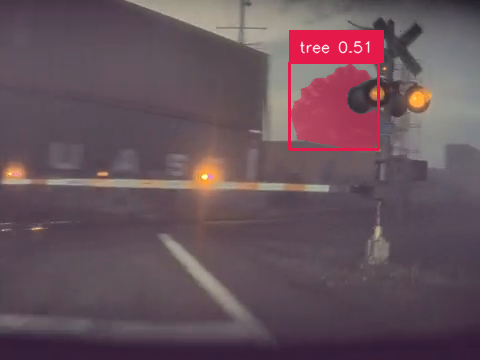}} \\

    \raisebox{1.5\normalbaselineskip}[0pt][0pt]{\rotatebox[origin=c]{90}{\footnotesize{Clear output}}} & 
    \frame{\includegraphics[width=0.25\linewidth]{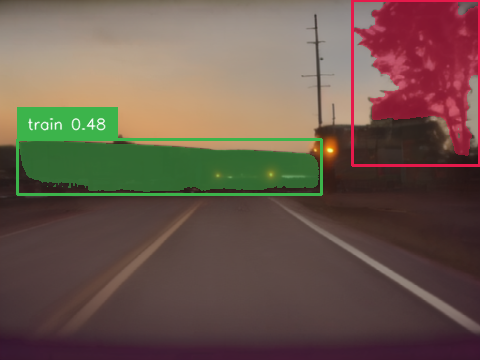}} &
    \frame{\includegraphics[width=0.25\linewidth]{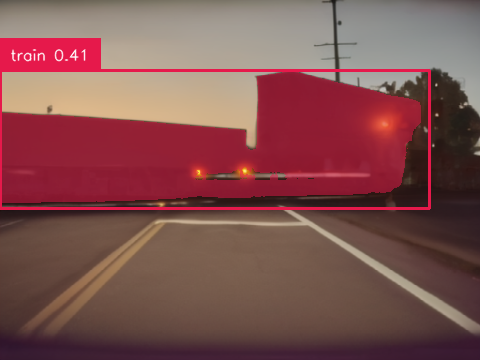}} &
    \frame{\includegraphics[width=0.25\linewidth]{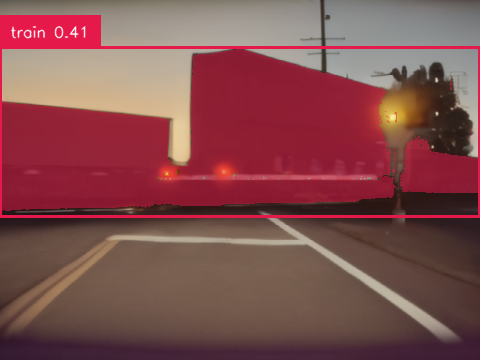}} &
    \frame{\includegraphics[width=0.25\linewidth]{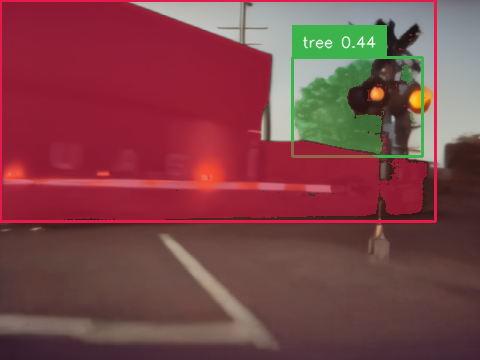}} \\
    
    \end{tabular}%
    }
    \vspace{-3mm}
    \caption{\textbf{Improved perception with weather removal.} After removing dense fog with our weather removal model, Grounded SAM~\cite{ren2024grounded} detects objects (e.g. train, tree) more accurately. %
    }
    \vspace{-3mm}
    \label{fig:perception}
\end{figure}

We compare weather removal methods in Fig.~\ref{fig:qual_inverse}. 
TokenFlow~\cite{geyer2023tokenflow} slightly changes the shading and synthesizes some background details, but struggles with strong fog, rain, puddle, and snow. 
WeatherDiffusion~\cite{ozdenizci2023} and Histoformer~\cite{sun2024restoring} are designed to remove transient snow and rain, but since they are trained only on images with synthetic patterns~\cite{valanarasu2022transweather}, they do not generalize well to diverse real-world videos and cannot handle other weather effects such as fog, puddles, and snow coverage. 
In contrast, our method is trained on diverse data sources, and effectively generalize to various weather conditions. It not only removes weather effects but also generates realistic scene content and simulates natural shading, consistently transforming videos into a clear-day appearance.  In Fig.~\ref{fig:slider}, we control the fog density and puddle reflection by changing the corresponding effect strength, demonstrating the high controllability of our method. Please refer to the supplementary for the results of all six effects.

\vspace{-1mm}
\subsection{Ablation Study}
\vspace{-1mm}
We qualitatively ablate our method in Fig.~\ref{fig:ablation}. 
Compared to our full method, the image-model variant (\ie, without temporal modules) often fails to generate transient effects such as falling raindrops and snowflakes. 

We also ablate the benefit of each data source described in Sec.~\ref{sec:data_collection}. 
When \textit{simulation} data are excluded, the model struggles to control effects and shading precisely. 
Excluding \textit{generation} data impacts the generalization of specific effects, such as rain, leading to their absence in the output. 
Without \textit{real-world} data, the generated videos often appears less realistic. 
In general, our full model combines a video-based approach with three diverse data sources, achieving the best quality and controllability.

\vspace{-1mm}
\subsection{Applications}
\vspace{-1mm}
Realistic weather editing in videos enables real-world applications. 
Combining both weather removal and synthesis models, our method enables weather editing by first applying the weather removal model, and re-generate weather effects with weather synthesis model in Fig.~\ref{fig:change_weather}. 
Furthermore, in Fig.\ref{fig:controllable}, we show that our method can be sequentially applied to the same scene to simulate ``time-lapse'' sequences with diverse weather changes. 

Effective weather removal also enhances the accuracy of perception models. In Fig.~\ref{fig:perception}, Grounded-SAM~\cite{ren2024grounded} fails to detect trains in dense fog, but succeeds after applying our weather removal model, demonstrating potential applications for self-driving and robotics.

\vspace{-2mm}
\section{Conclusion}
\vspace{-2mm}
We propose a scalable, data-driven framework for controllable weather simulation in real-world videos. Drawing inspiration from modern graphics engines, we decompose the task into \textsc{weather removal} and \textsc{weather synthesis} and train two complementary conditional video diffusion models that can be applied independently or combined. By leveraging synthetic, generated, and automatically labeled real-world data in a unified training scheme, \ourmodel consistently outperforms state-of-the-art methods.

\parahead{Limitations} While \ourmodel demonstrates realistic, controllable, and temporally consistent weather synthesis and removal, its performance is bounded by the quality of the underlying Stable Video Diffusion model. Consequently, fine details such as text and facial features are not always preserved. Our model also struggles with nighttime videos, in part due to the scarcity of such footage in our current data-curation pipeline. Finally, Stable Video Diffusion is an offline model that can only process relatively short videos. With rapid progress in video diffusion quality and efficiency, we anticipate that integrating a more robust and efficient base model will lead to even stronger performance.

\clearpage
\maketitlesupplementary
\appendix

\renewcommand{\thefigure}{S\arabic{figure}}
\setcounter{figure}{0}
\renewcommand{\thetable}{S\arabic{table}}
\setcounter{table}{0}
\renewcommand{\thesection}{\Alph{section}}
\setcounter{section}{0}

In the supplementary material, we provide additional implementation details (Sec.~\ref{sec:implementation_details}) and further results (Sec.~\ref{sec:additional_results}). Please refer to the project website for more qualitative results and comparisons.

\section{Implementation Details} \label{sec:implementation_details}

\paragraph{Training Details} 
Both weather removal and synthesis models are trained using AdamW optimizer with a learning rate of $3 \times 10^{-5}$ for 20k iterations. The models are trained on 32 A100 GPUs with fp16 mixed-precision for around 2 days.
During training, the video resolution and number of frames are randomized at multiple scales, making the model robust to various input resolutions and frame lengths. The resolutions include $384 \times 576$, $512 \times 512$, $1280 \times 1920$, and the frame lengths range from 1 to 16. 
After the full training stages, the models can precisely control six effects (benefited from simulation data), generalize to diverse content (benefited from generation data), and simulate realistic weather (benefited from real-world data), supported by the evaluation in main  Sec.~\textcolor{iccvblue}{5}.

\paragraph{Weather Strength Definition}
We adopt standard definitions from Unreal Engine, which are grounded in physically meaningful quantities, e.g., cloud coverage (ratio of the sky), fog (density), raindrop or snowflake (count per unit volume per second), ground puddle (coverage ratio), and snow cover (height). During training, their intensity values are normalized to the range $[0, 1]$. This continuous representation enables fine-grained control and smooth transitions

\section{Additional Results} \label{sec:additional_results}

\begin{figure*}[t]
    \centering\setlength{\tabcolsep}{0.1em}
    \resizebox{1.0\textwidth}{!}{%
    \begin{tabular}{@{}cccccc@{}}

    \footnotesize{Cloud: $s_{\text{cloud}} = 0.2$} & \footnotesize{Cloud: $s_{\text{cloud}} = 0.5$} & \footnotesize{Cloud: $s_{\text{cloud}} = 1.0$} & \footnotesize{Fog: $s_{\text{fog}} = 0.5$} & \footnotesize{Fog: $s_{\text{fog}} = 0.8$} & \footnotesize{Fog: $s_{\text{fog}} = 1.0$}  \\
     
    \frame{\includegraphics[width=0.2\textwidth]{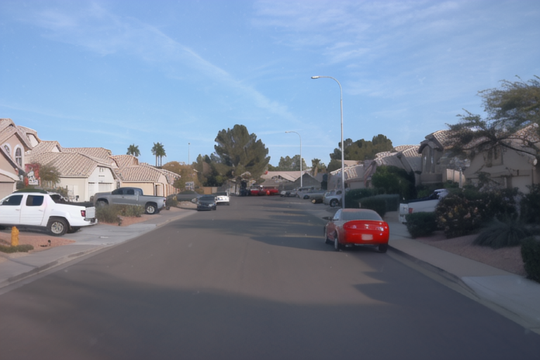}} &
    \frame{\includegraphics[width=0.2\textwidth]{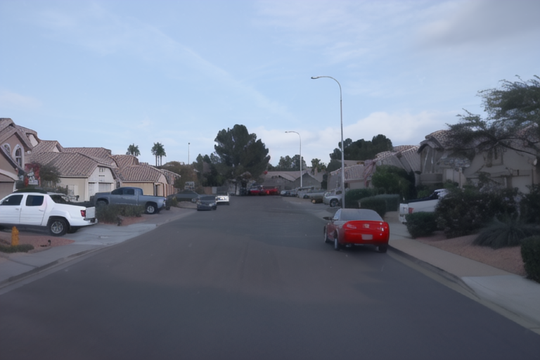}} &
    \frame{\includegraphics[width=0.2\textwidth]{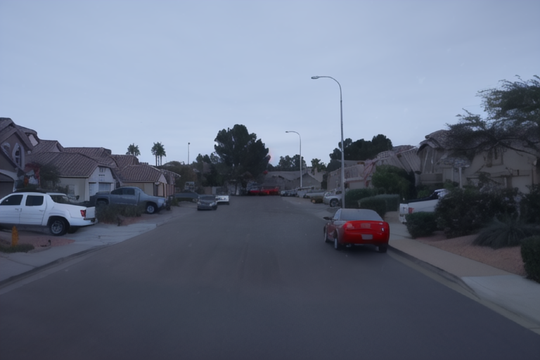}} & \frame{\includegraphics[width=0.2\textwidth]{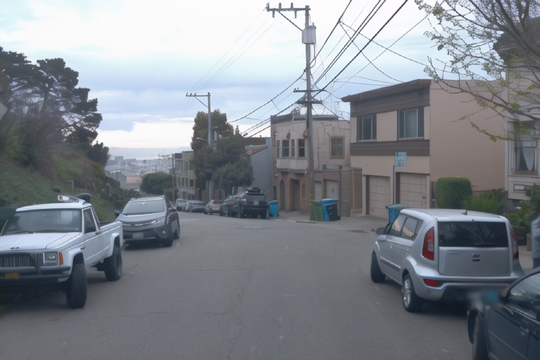}} &
    \frame{\includegraphics[width=0.2\textwidth]{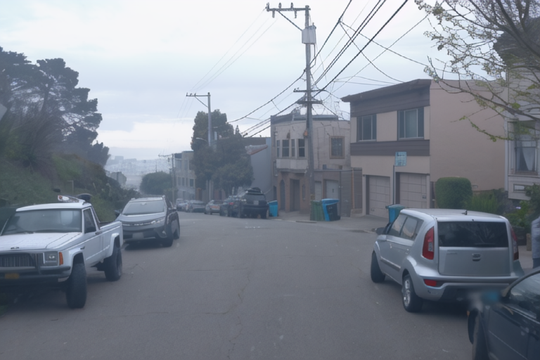}} &
    \frame{\includegraphics[width=0.2\textwidth]{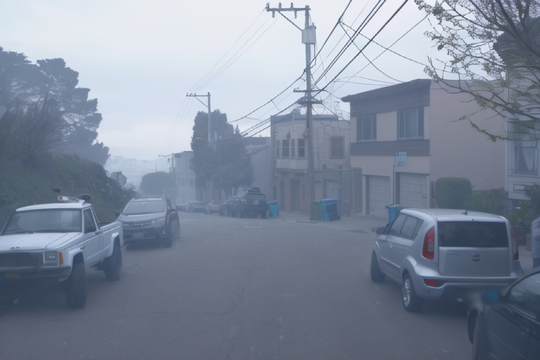}} \\

     \footnotesize{Rain: $s_{\text{rain}} = 0.2$} & \footnotesize{Rain: $s_{\text{rain}} = 0.5$} & \footnotesize{Rain: $s_{\text{rain}} = 1.0$} & \footnotesize{Puddle: $s_{\text{puddle}} = 0.2$} & \footnotesize{Puddle: $s_{\text{puddle}} = 0.5$} & \footnotesize{Puddle: $s_{\text{puddle}} = 1.0$}  \\
     
    \frame{\includegraphics[width=0.2\textwidth]{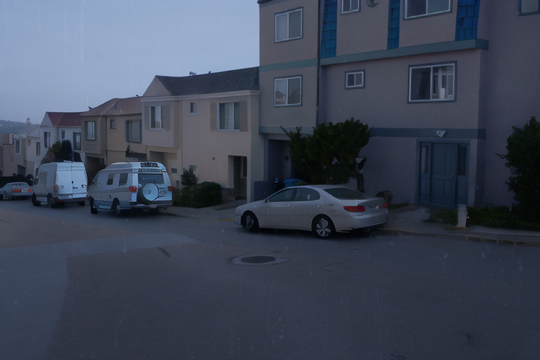}} &
    \frame{\includegraphics[width=0.2\textwidth]{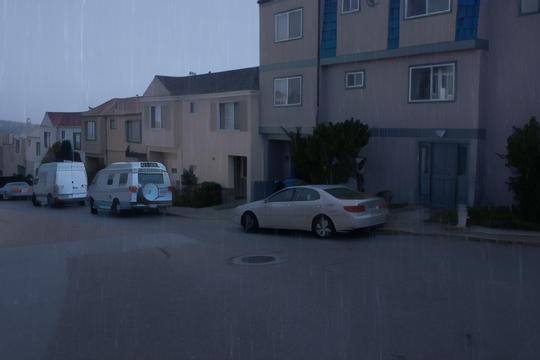}} &
    \frame{\includegraphics[width=0.2\textwidth]{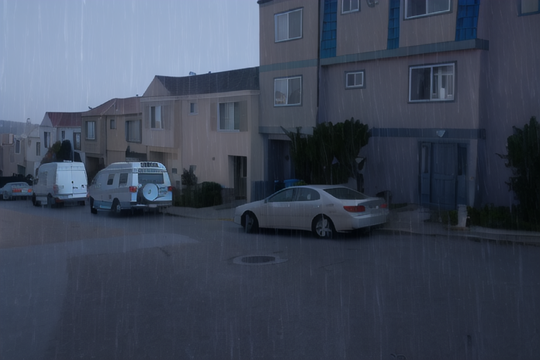}} & \frame{\includegraphics[width=0.2\textwidth]{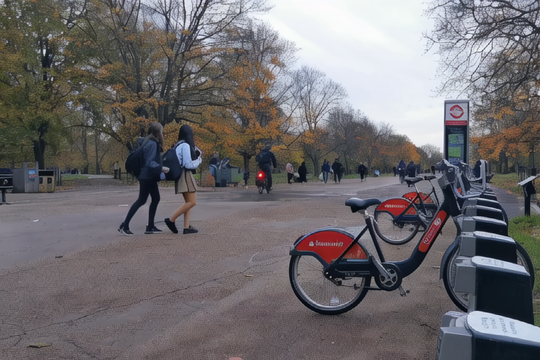}} &
    \frame{\includegraphics[width=0.2\textwidth]{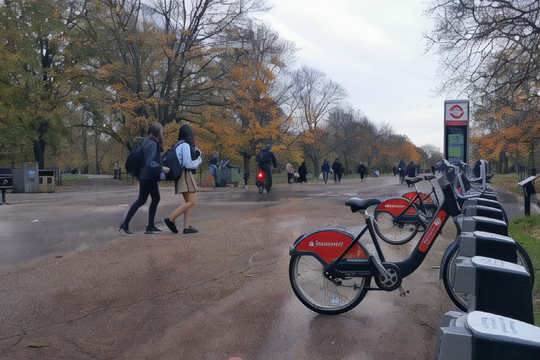}} &
    \frame{\includegraphics[width=0.2\textwidth]{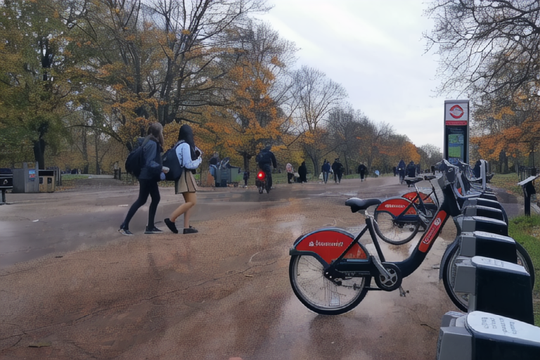}} \\

    \footnotesize{Snow: $s_{\text{snow}} = 0.2$} & \footnotesize{Snow: $s_{\text{snow}} = 0.5$} & \footnotesize{Snow: $s_{\text{snow}} = 1.0$} & \footnotesize{Snow coverage: $s_{\text{sc}} = 0.2$} & \footnotesize{Snow coverage: $s_{\text{sc}} = 0.5$} & \footnotesize{Snow coverage: $s_{\text{sc}} = 1.0$}  \\
     
    \frame{\includegraphics[width=0.2\textwidth]{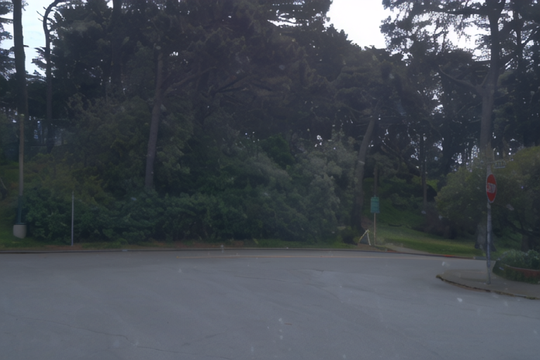}} &
    \frame{\includegraphics[width=0.2\textwidth]{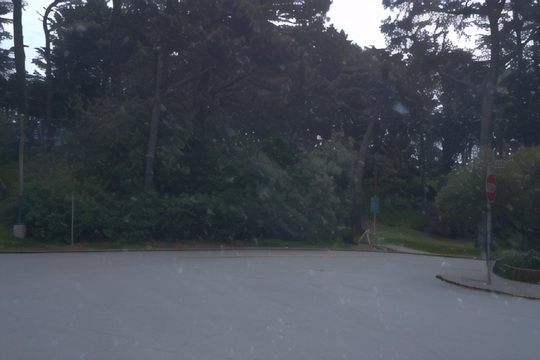}} &
    \frame{\includegraphics[width=0.2\textwidth]{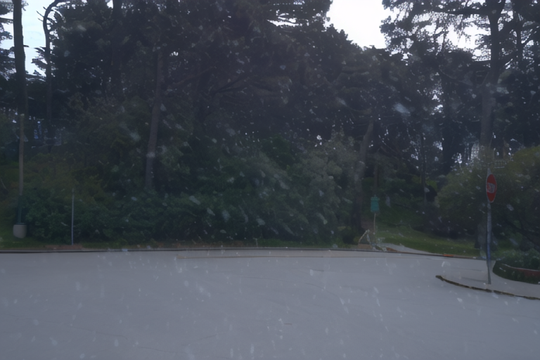}} & \frame{\includegraphics[width=0.2\textwidth]{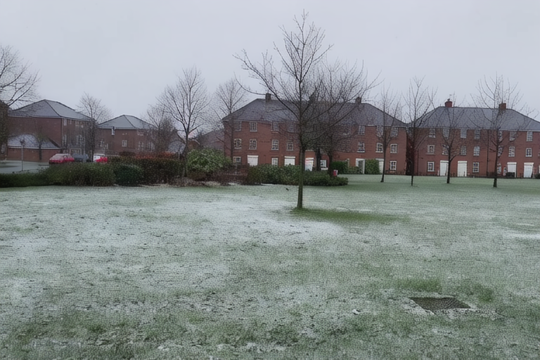}} &
    \frame{\includegraphics[width=0.2\textwidth]{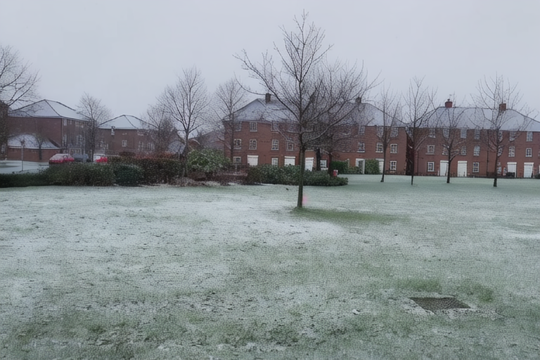}} &
    \frame{\includegraphics[width=0.2\textwidth]{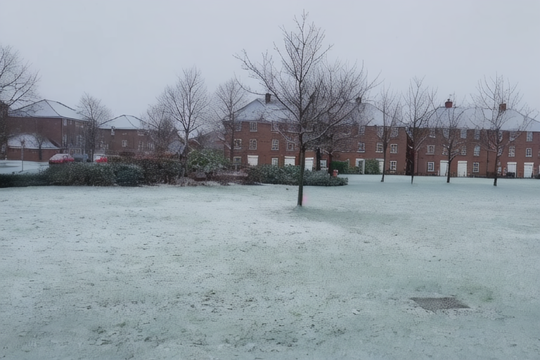}} \\
    
    \end{tabular}%
    }
    \vspace{-3mm}
    \caption{\textbf{Controlling the strength of weather effects.}
    }
    \vspace{-3mm}
    \label{fig:slider_all}
\end{figure*}

\begin{figure*}[t]
    \centering\setlength{\tabcolsep}{0.1em}
    \resizebox{1.0\textwidth}{!}{%
    \begin{tabular}{@{}lccclccc@{}}
    
    & $t=0$ & $t=1$ & $t=2$ &  & $t=0$ & $t=1$ & $t=2$ \\[0.2em]

    \raisebox{2.5\normalbaselineskip}[0pt][0pt]{\rotatebox[origin=c]{90}{Input}} & 
    \frame{\includegraphics[width=0.2\textwidth]{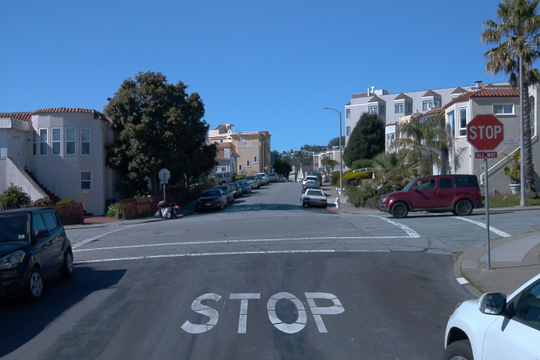}} & 
    \frame{\includegraphics[width=0.2\textwidth]{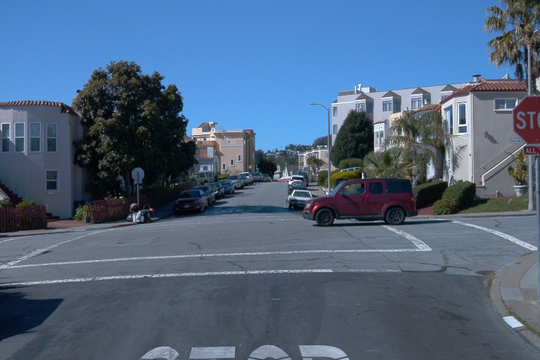}} &
    \frame{\includegraphics[width=0.2\textwidth]{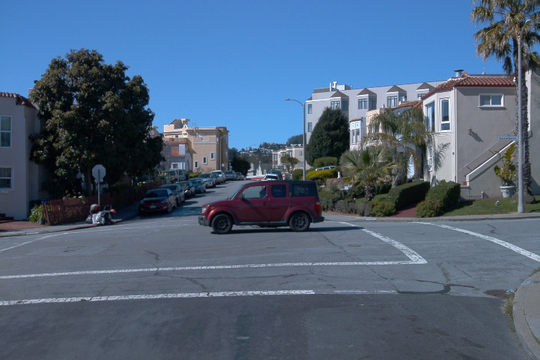}} & 
    \raisebox{2.5\normalbaselineskip}[0pt][0pt]{\rotatebox[origin=c]{90}{Input}} & 
    \frame{\includegraphics[width=0.2\textwidth]{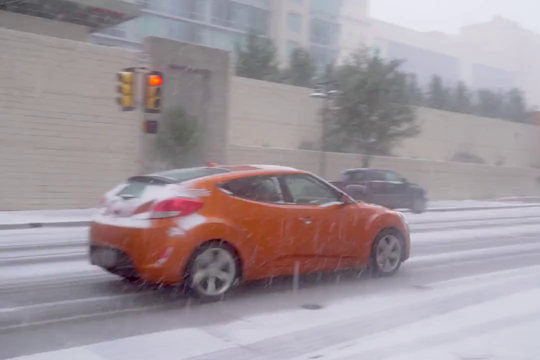}} & 
    \frame{\includegraphics[width=0.2\textwidth]{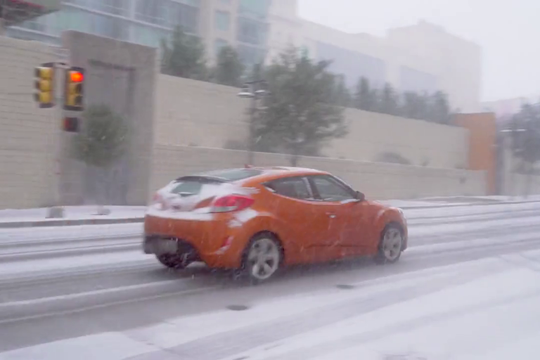}} &
    \frame{\includegraphics[width=0.2\textwidth]{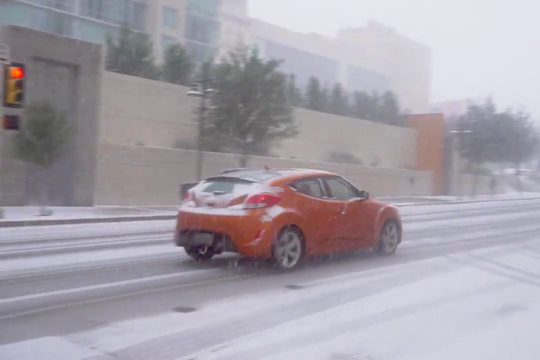}} \\

    \raisebox{2.5\normalbaselineskip}[0pt][0pt]{\rotatebox[origin=c]{90}{Ours}} & 
    \frame{\includegraphics[width=0.2\textwidth]{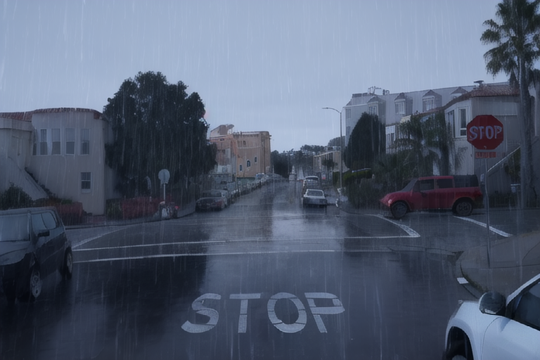}} & 
    \frame{\includegraphics[width=0.2\textwidth]{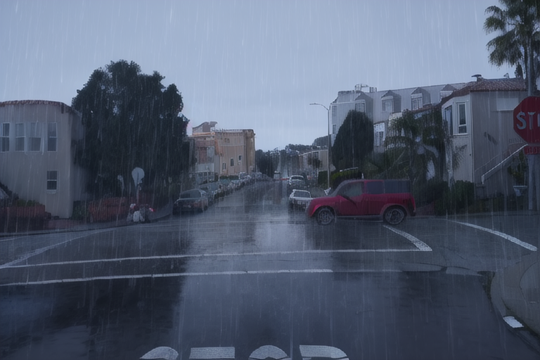}} &
    \frame{\includegraphics[width=0.2\textwidth]{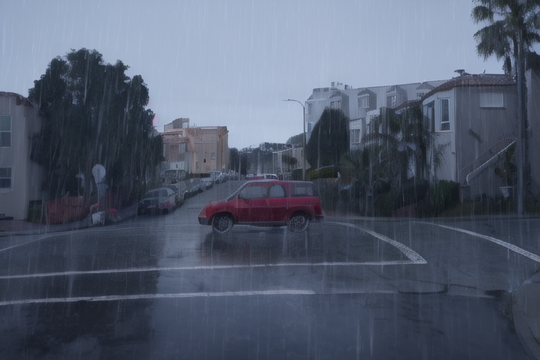}} & 
    \raisebox{2.5\normalbaselineskip}[0pt][0pt]{\rotatebox[origin=c]{90}{Ours}} & 
    \frame{\includegraphics[width=0.2\textwidth]{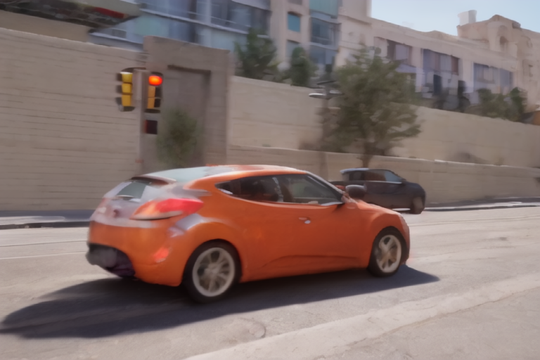}} & 
    \frame{\includegraphics[width=0.2\textwidth]{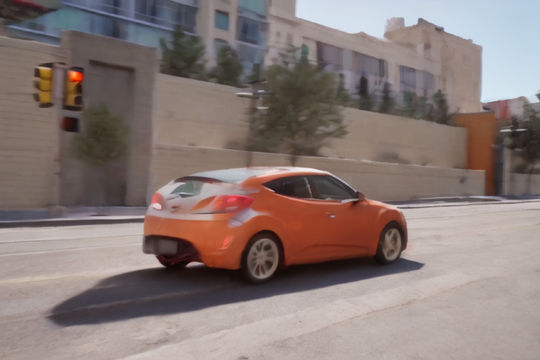}} &
    \frame{\includegraphics[width=0.2\textwidth]{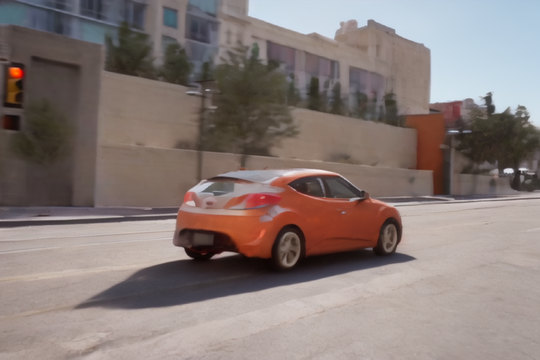}} \\

    \raisebox{2.5\normalbaselineskip}[0pt][0pt]{\rotatebox[origin=c]{90}{TokenFlow~\cite{geyer2023tokenflow}}} & 
    \frame{\includegraphics[width=0.2\textwidth]{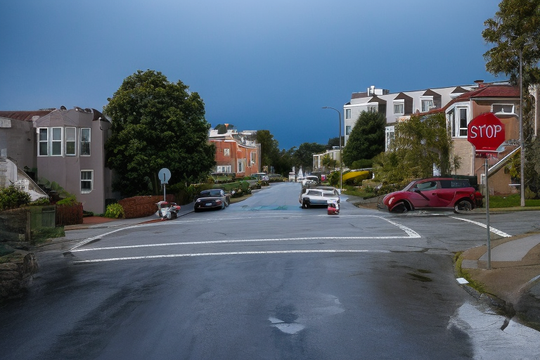}} & 
    \frame{\includegraphics[width=0.2\textwidth]{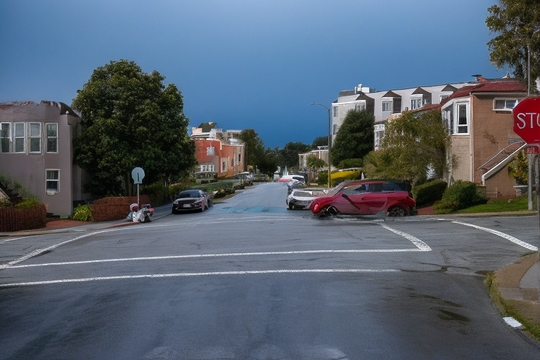}} &
    \frame{\includegraphics[width=0.2\textwidth]{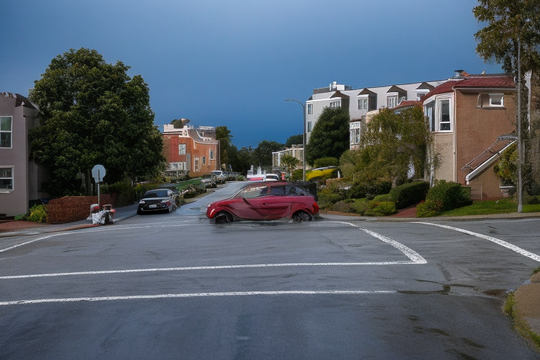}} & 
    \raisebox{2.5\normalbaselineskip}[0pt][0pt]{\rotatebox[origin=c]{90}{Histoformer~\cite{sun2024restoring}}} & 
    \frame{\includegraphics[width=0.2\textwidth]{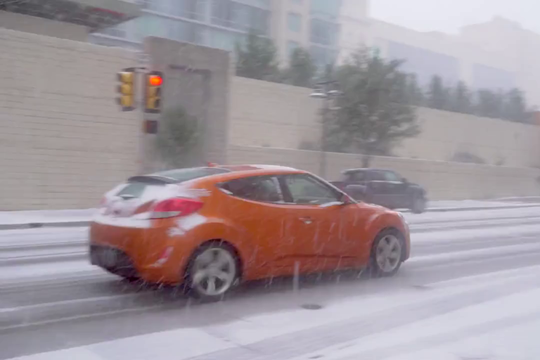}} & 
    \frame{\includegraphics[width=0.2\textwidth]{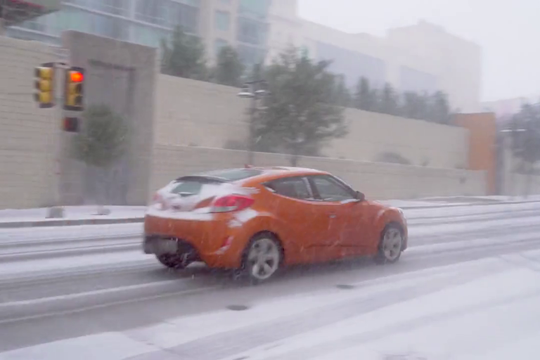}} &
    \frame{\includegraphics[width=0.2\textwidth]{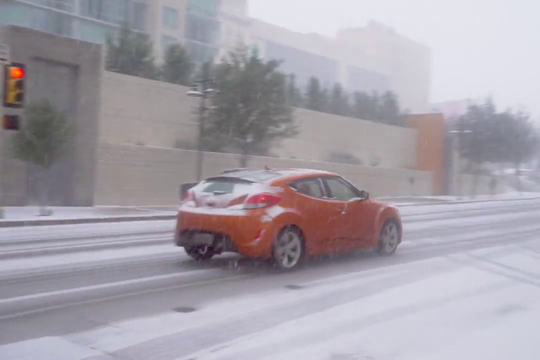}} \\

    \end{tabular}%
    }
    \caption{\textbf{Temporally-Consistent Synthesis and Removal.} Left: weather synthesis. Right: weather removal. 
    }
    \label{fig:temporal}
\end{figure*}

In Fig.~\ref{fig:temporal}, both our \textsc{weather synthesis model} and \textsc{weather removal model} effectively edit the weather, preserve details (e.g., ``STOP'' on the road), and also maintain temporal consistency. In addition, the different weather conditions can be controlled precisely by changing the strength values of each effect, shown in Fig.~\ref{fig:slider_all}.

\begin{figure}[t]
    \centering\setlength{\tabcolsep}{0.1em}
    \resizebox{1.0\linewidth}{!}{%
    \begin{tabular}{@{}ccc@{}}

    \footnotesize{Input} & \footnotesize{ClimateNeRF~\cite{Li2023ClimateNeRF}} & \footnotesize{Ours}  \\
     
    \frame{\includegraphics[width=0.4\linewidth]{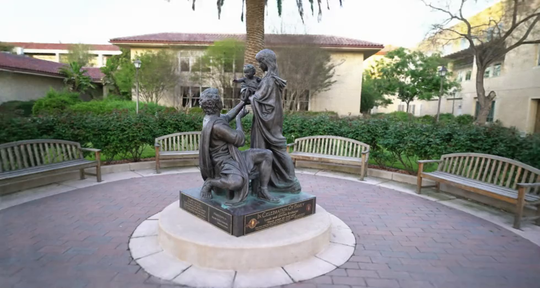}} &
    \frame{\includegraphics[width=0.4\linewidth]{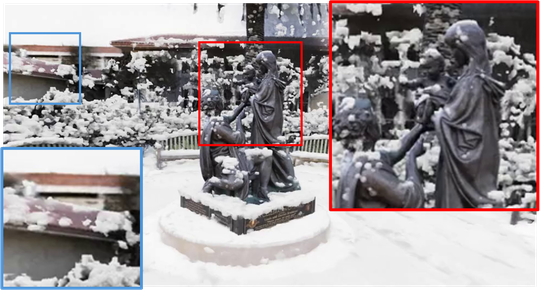}} &
    \frame{\includegraphics[width=0.4\linewidth]{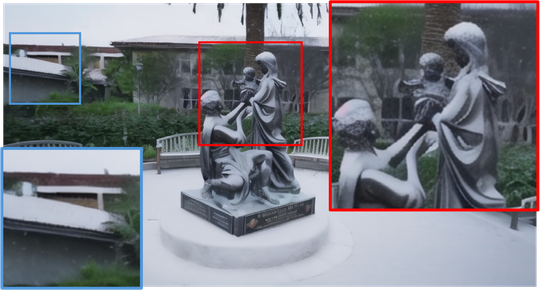}}  \\

    \end{tabular}%
    }
    \vspace{-3mm}
    \caption{\textbf{Comparison with ClimateNeRF~\cite{Li2023ClimateNeRF}.} Our video model can coat delicate snow on the statue and rooftop surfaces, and also adjust the shading, which is hard for 3D simulation approaches~\cite{Li2023ClimateNeRF}.
    }
    \vspace{-3mm}
    \label{fig:compare_climatenerf}
\end{figure}

\begin{figure*}[t]
    \centering\setlength{\tabcolsep}{0.1em}
    \resizebox{1.0\textwidth}{!}{%
    \begin{tabular}{@{}lccccc@{}}
    
    & Input & TokenFlow~\cite{geyer2023tokenflow} & WeatherDiffusion~\cite{ozdenizci2023} & Histoformer~\cite{sun2024restoring} & Ours \\[0.2em]

    \raisebox{2.5\normalbaselineskip}[0pt][0pt]{\rotatebox[origin=c]{90}{Fog}} & \frame{\includegraphics[width=0.2\textwidth]{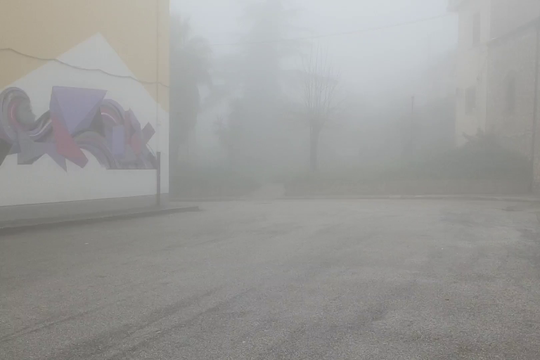}} & 
    \frame{\includegraphics[width=0.2\textwidth]{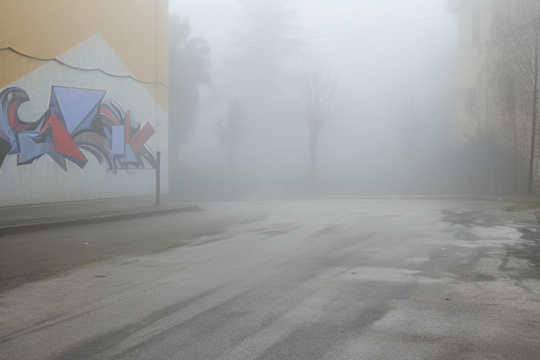}} &
    \frame{\includegraphics[width=0.2\textwidth]{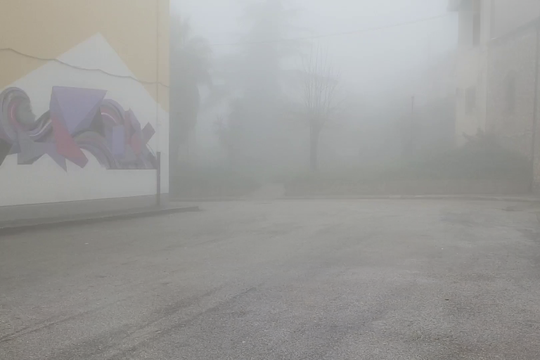}} & 
    \frame{\includegraphics[width=0.2\textwidth]{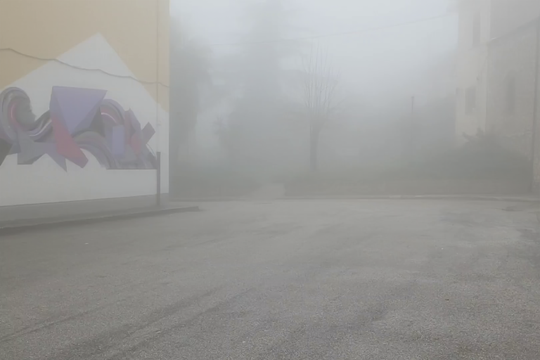}} & 
    \frame{\includegraphics[width=0.2\textwidth]{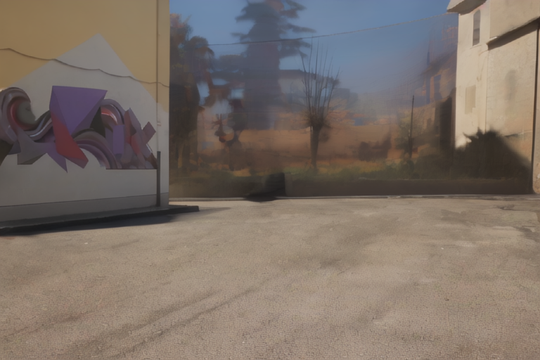}} \\

    \raisebox{2.5\normalbaselineskip}[0pt][0pt]{\rotatebox[origin=c]{90}{Rain}} & \frame{\includegraphics[width=0.2\textwidth]{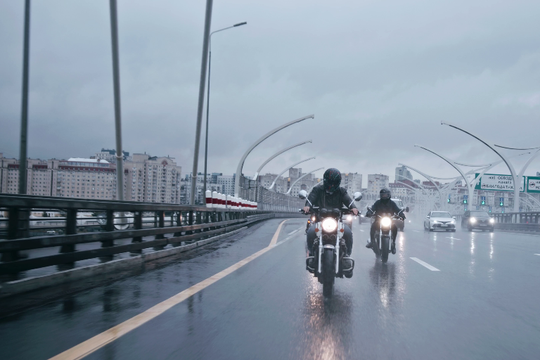}} & 
    \frame{\includegraphics[width=0.2\textwidth]{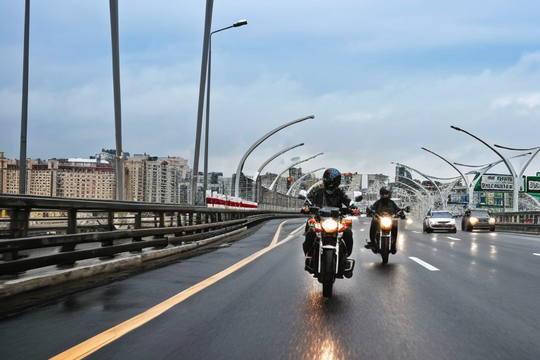}} &
    \frame{\includegraphics[width=0.2\textwidth]{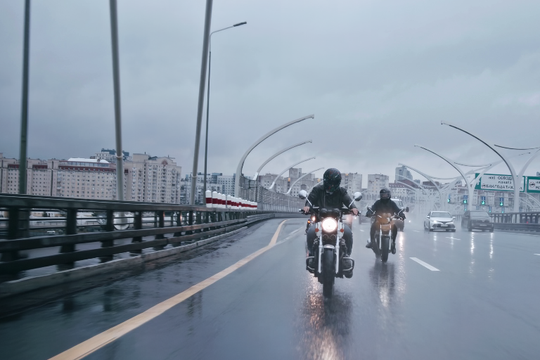}} & 
    \frame{\includegraphics[width=0.2\textwidth]{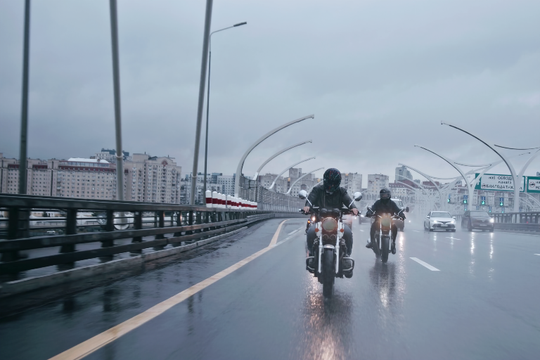}} & 
    \frame{\includegraphics[width=0.2\textwidth]{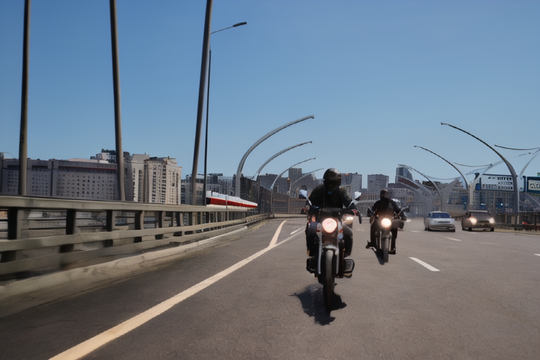}} \\

    \raisebox{2.5\normalbaselineskip}[0pt][0pt]{\rotatebox[origin=c]{90}{Snow}} & \frame{\includegraphics[width=0.2\textwidth]{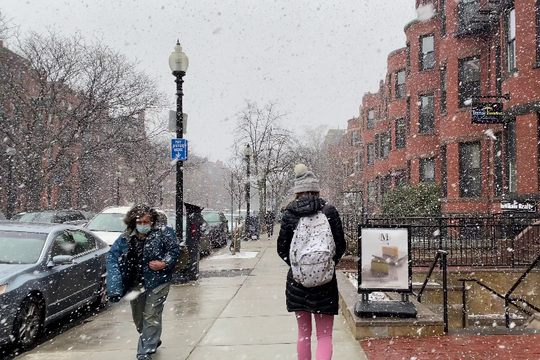}} & 
    \frame{\includegraphics[width=0.2\textwidth]{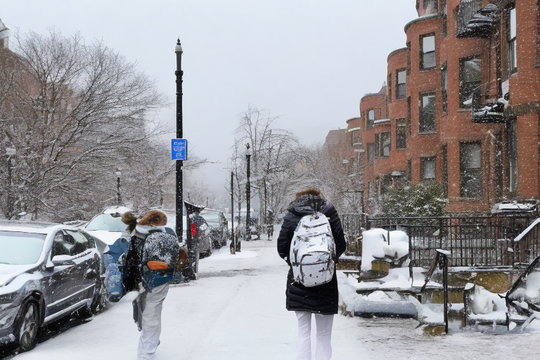}} &
    \frame{\includegraphics[width=0.2\textwidth]{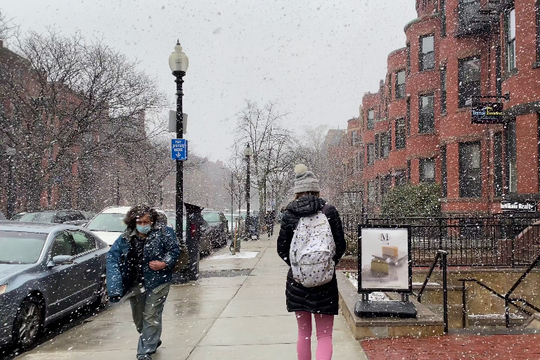}} & 
    \frame{\includegraphics[width=0.2\textwidth]{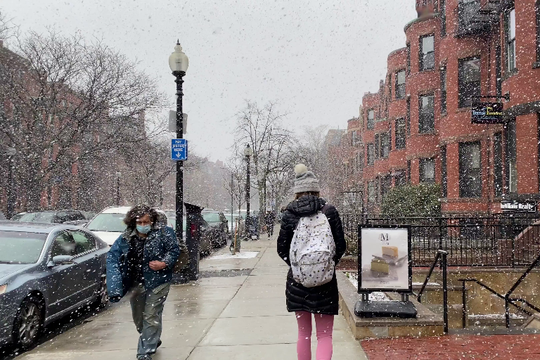}} & 
    \frame{\includegraphics[width=0.2\textwidth]{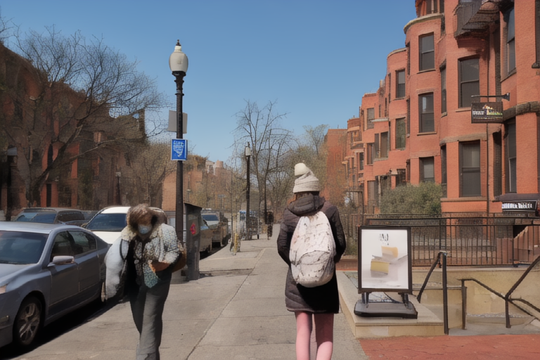}} \\
    
    \end{tabular}%
    }
    \vspace{-4mm}
    \caption{\textbf{Additional qualitative results of weather removal.}}
    \label{fig:qual_inverse_supp}
\end{figure*}

\begin{figure*}[h]
    \centering\setlength{\tabcolsep}{0.1 em}
    \resizebox{0.8\textwidth}{!}{%
    \begin{tabular}{@{}cccc@{}}
    
    Input & Rain-LHP~\cite{guo2023sky} & RainMamba~\cite{wu2024rainmamba} & Ours \\ %

     \frame{\includegraphics[width=0.25\textwidth]{figures/images/qual_inverse/pexels-rain-47554/clear/input/00000_input.png}} & \frame{\includegraphics[width=0.25\textwidth]{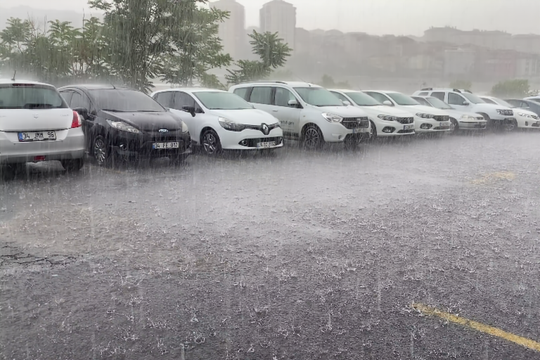}} & \frame{\includegraphics[width=0.25\textwidth]{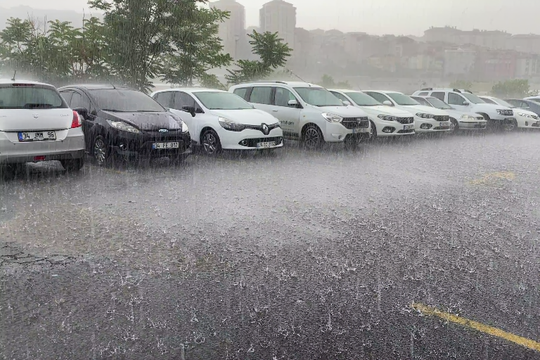}} & \frame{\includegraphics[width=0.25\textwidth]{figures/images/qual_inverse/pexels-rain-47554/clear/ours/00000_ours.png}} \\

     \frame{\includegraphics[width=0.25\textwidth]{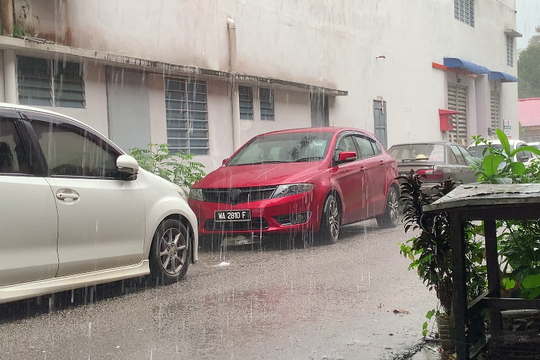}} & \frame{\includegraphics[width=0.25\textwidth]{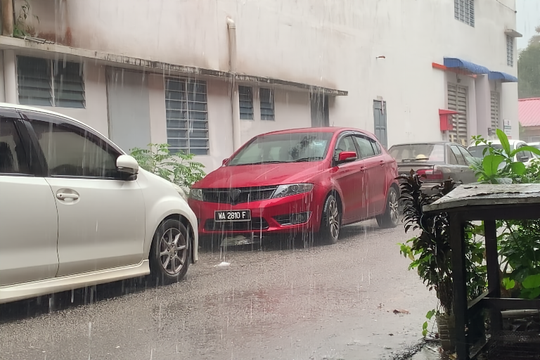}} & \frame{\includegraphics[width=0.25\textwidth]{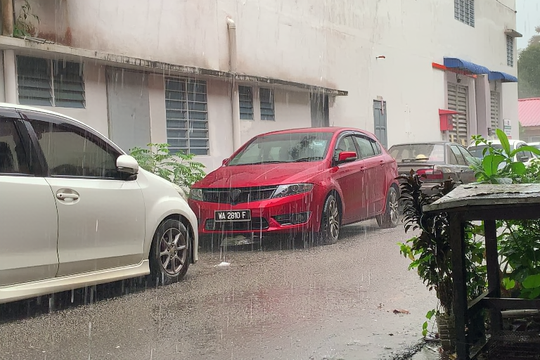}} & \frame{\includegraphics[width=0.25\textwidth]{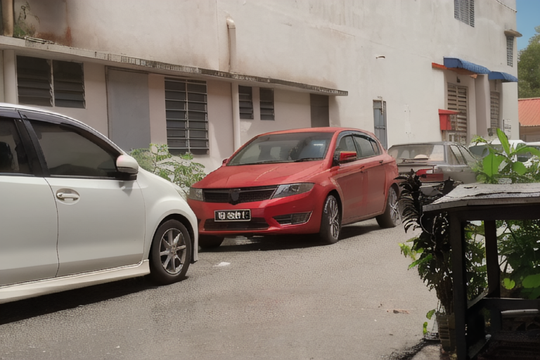}} \\

     \frame{\includegraphics[width=0.25\textwidth]{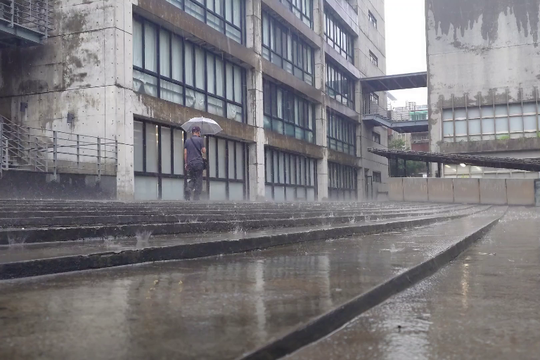}} & \frame{\includegraphics[width=0.25\textwidth]{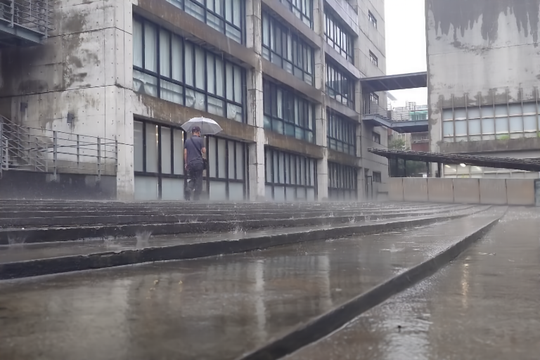}} & \frame{\includegraphics[width=0.25\textwidth]{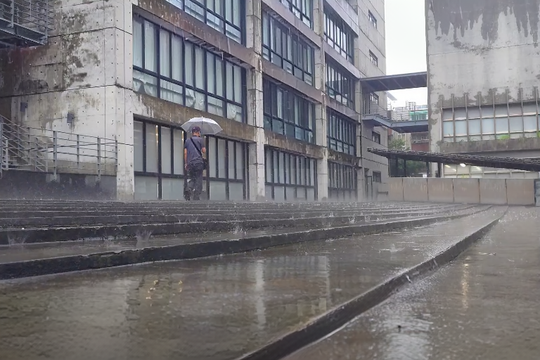}} & \frame{\includegraphics[width=0.25\textwidth]{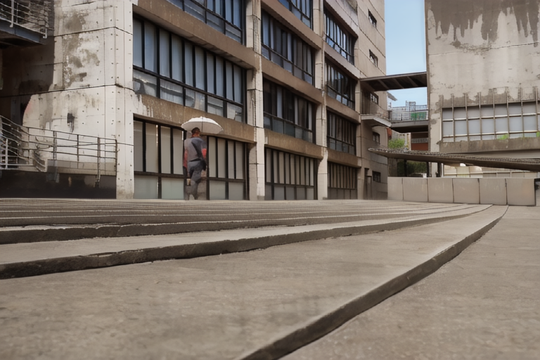}} \\

    \end{tabular}%
    }
    \caption{\textbf{Additional qualitative results of rain removal.} We compare our rain removal results with recent non-diffusion methods~\cite{guo2023sky, wu2024rainmamba}.}
    \label{fig:rain_removal}
\end{figure*}

\begin{figure*}[t]
    \centering\setlength{\tabcolsep}{0.1em}
    \resizebox{1.0\textwidth}{!}{%
    \begin{tabular}{@{}lccccc@{}}
    
    & Input & AnyV2V~\cite{ku2024anyv2v} & TokenFlow~\cite{geyer2023tokenflow} & FRESCO~\cite{yang2024fresco} & Ours \\[0.2em]

    \raisebox{2.5\normalbaselineskip}[0pt][0pt]{\rotatebox[origin=c]{90}{Fog}} & \frame{\includegraphics[width=0.2\textwidth]{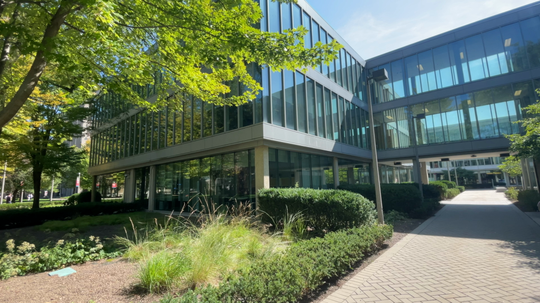}} & 
    \frame{\includegraphics[width=0.2\textwidth]{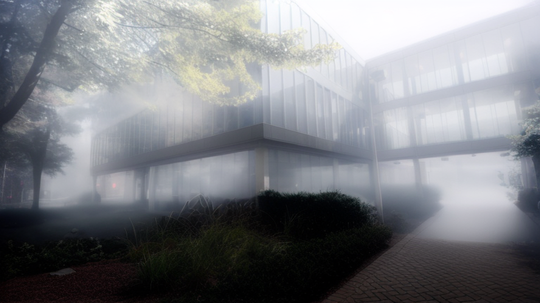}} &
    \frame{\includegraphics[width=0.2\textwidth]{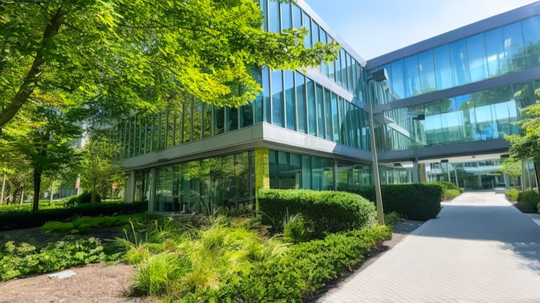}} & 
    \frame{\includegraphics[width=0.2\textwidth]{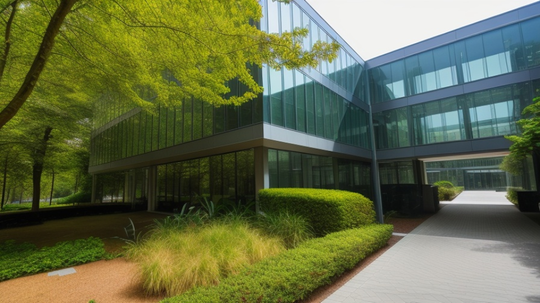}} & 
    \frame{\includegraphics[width=0.2\textwidth]{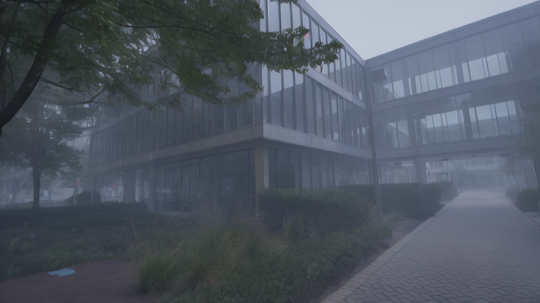}} \\

    \raisebox{2.5\normalbaselineskip}[0pt][0pt]{\rotatebox[origin=c]{90}{Rain}} & \frame{\includegraphics[width=0.2\textwidth]{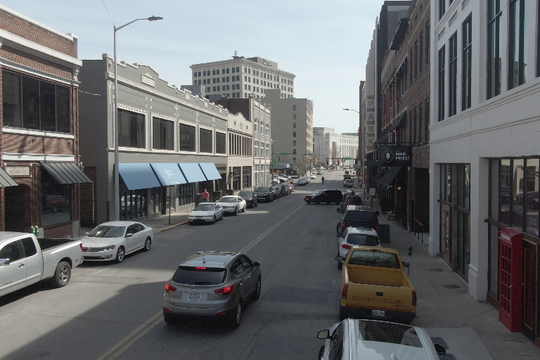}} & 
    \frame{\includegraphics[width=0.2\textwidth]{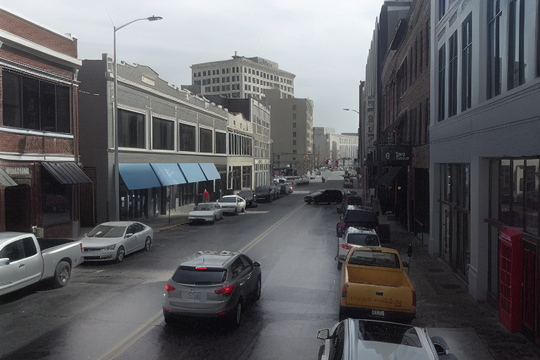}} &
    \frame{\includegraphics[width=0.2\textwidth]{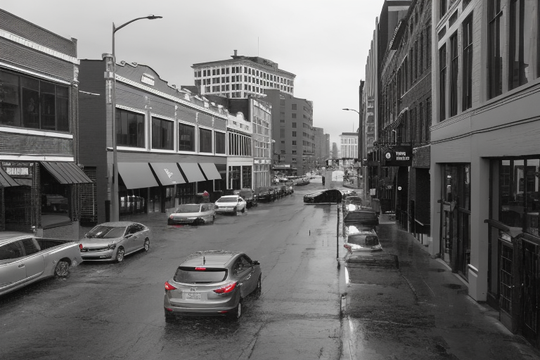}} & 
    \frame{\includegraphics[width=0.2\textwidth]{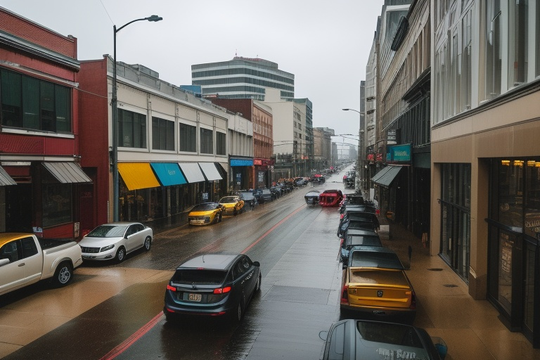}} & 
    \frame{\includegraphics[width=0.2\textwidth]{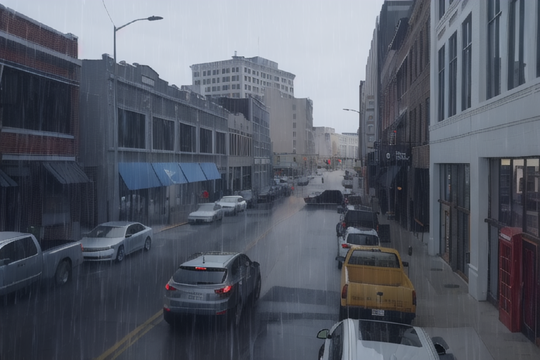}} \\

    \raisebox{2.5\normalbaselineskip}[0pt][0pt]{\rotatebox[origin=c]{90}{Snow}} & \frame{\includegraphics[width=0.2\textwidth]{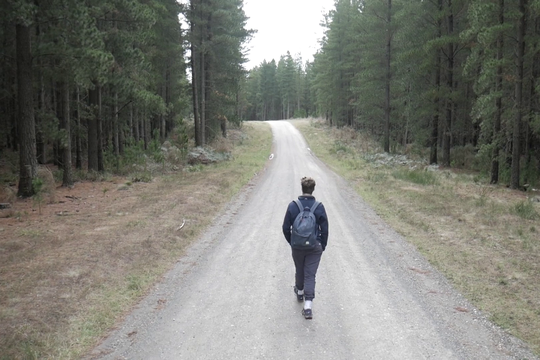}} & 
    \frame{\includegraphics[width=0.2\textwidth]{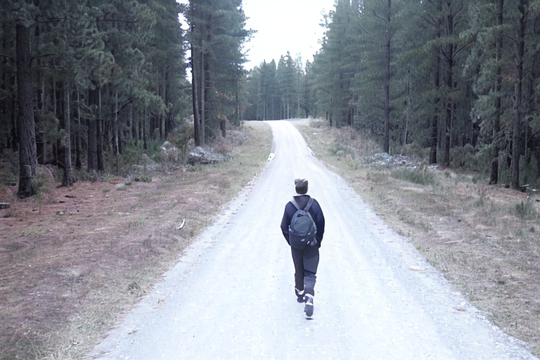}} &
    \frame{\includegraphics[width=0.2\textwidth]{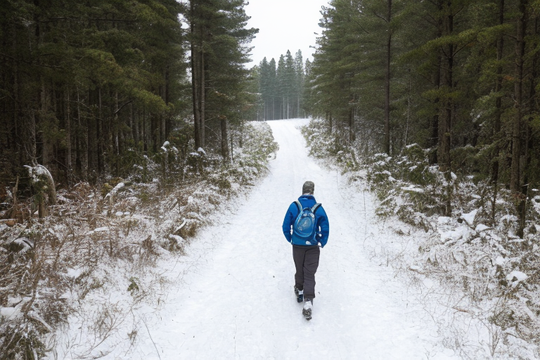}} & 
    \frame{\includegraphics[width=0.2\textwidth]{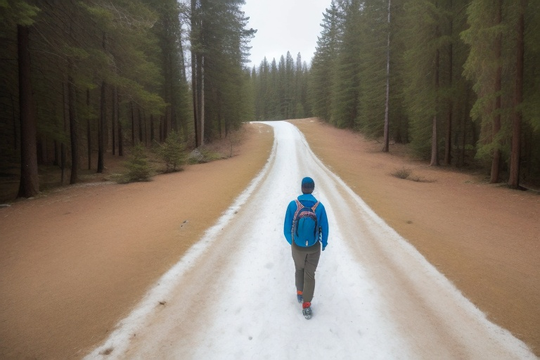}} & 
    \frame{\includegraphics[width=0.2\textwidth]{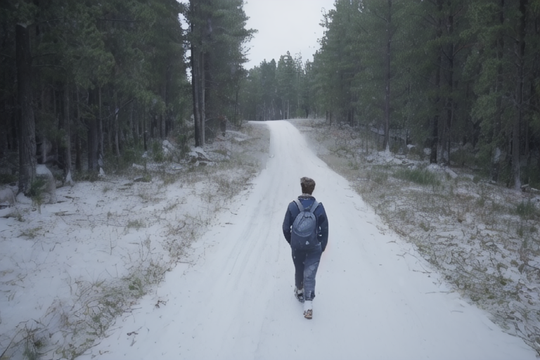}} \\
    
    \end{tabular}%
    }
    \vspace{-4mm}    
    \caption{\textbf{Additional qualitative results of weather synthesis.}
    }
    \vspace{-3mm}    
    \label{fig:qual_forward_supp}
\end{figure*}

In addition to video editing methods, we also compare the weather synthesis with 3D simulation method in Fig.~\ref{fig:compare_climatenerf}. ClimateNeRF~\cite{Li2023ClimateNeRF} relies on the high-quality geometry to integrate weather effects with the scene successfully and cannot perform well for regions that are not captured densely (e.g., rooftop). On the other hand, our weather synthesis model leverages the video diffusion model and synthesizes snowflakes, snow coverage covering the whole scene.
Furthermore, we provide additional qualitative results of weather removal and weather synthesis in Fig.~\ref{fig:qual_inverse_supp}, ~\ref{fig:rain_removal}, and ~\ref{fig:qual_forward_supp}, showing that our method generalize well to diverse video inputs.

\begin{figure*}
    \centering
    \frame{\includegraphics[width=0.9\linewidth]{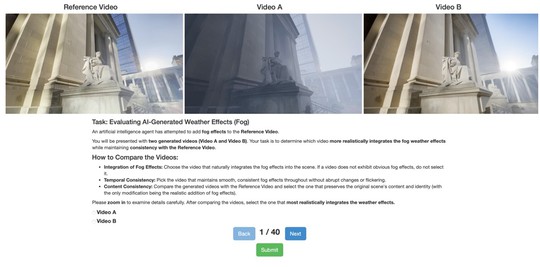}}
    \caption{\textbf{Example of user study interface for comparing two generated videos for weather synthesis.}}
    \label{fig:user_study_gui}
\end{figure*}

\parahead{User Study}
is a common approach for assessing perceptual realism. We conducted the user study mentioned in Sec.~\textcolor{iccvblue}{5.1} on Amazon Mechanical Turk (MTurk) to compare our method with other baselines. Fig.~\ref{fig:user_study_gui} visualizes the example interface used for user study on the weather synthesis task. We asked users to make perceptual decisions on the pairwise comparison with the following criteria: 1) the integration of weather effects, 2) temporal consistency, and 3) content consistency. 
For weather removal, we used a similar user interface but asked users to choose videos with the least visible weather effects.  

During the user study, we invited 11 users for each sample pair to perform binary preference selection. 
We used 40 videos for weather synthesis (4 baselines, 3 effects) and 55 for weather removal (6 baselines) evaluation. This results in $3\times 40 \times 4 \times 11 \times 3 = 15840$ and $55 \times 6 \times 11 \times 3 = 10,890$ user selections for each evaluated task. 
For each evaluated scene video, we did majority voting from 11 users to determine which method is more preferred in this scene. The majority voting can efficiently filter the effects of random users.
The full experiments are repeated 3 times to calculate the mean and standard deviation on the preference percentage.

\begin{figure*}
\centering
\begin{subfigure}{0.49\linewidth}
\centering\captionsetup{width=0.96\linewidth}%
\includegraphics[width=\linewidth]{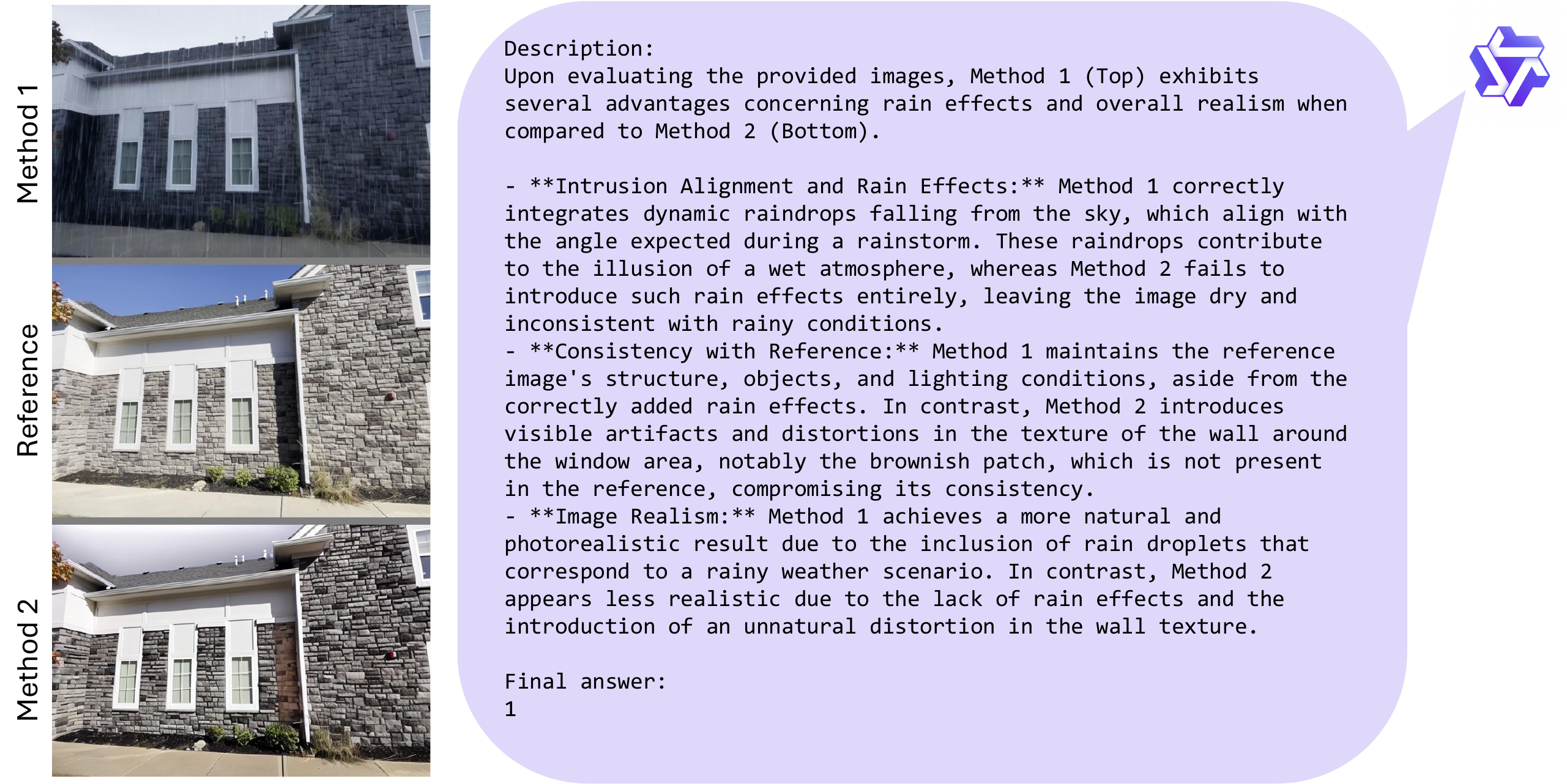}
\caption{Weather Synthesis (Rain) Example: Ours vs. AnyV2V}
\end{subfigure}
\begin{subfigure}{0.49\linewidth}
\centering\captionsetup{width=0.96\linewidth}%
\includegraphics[width=\linewidth]{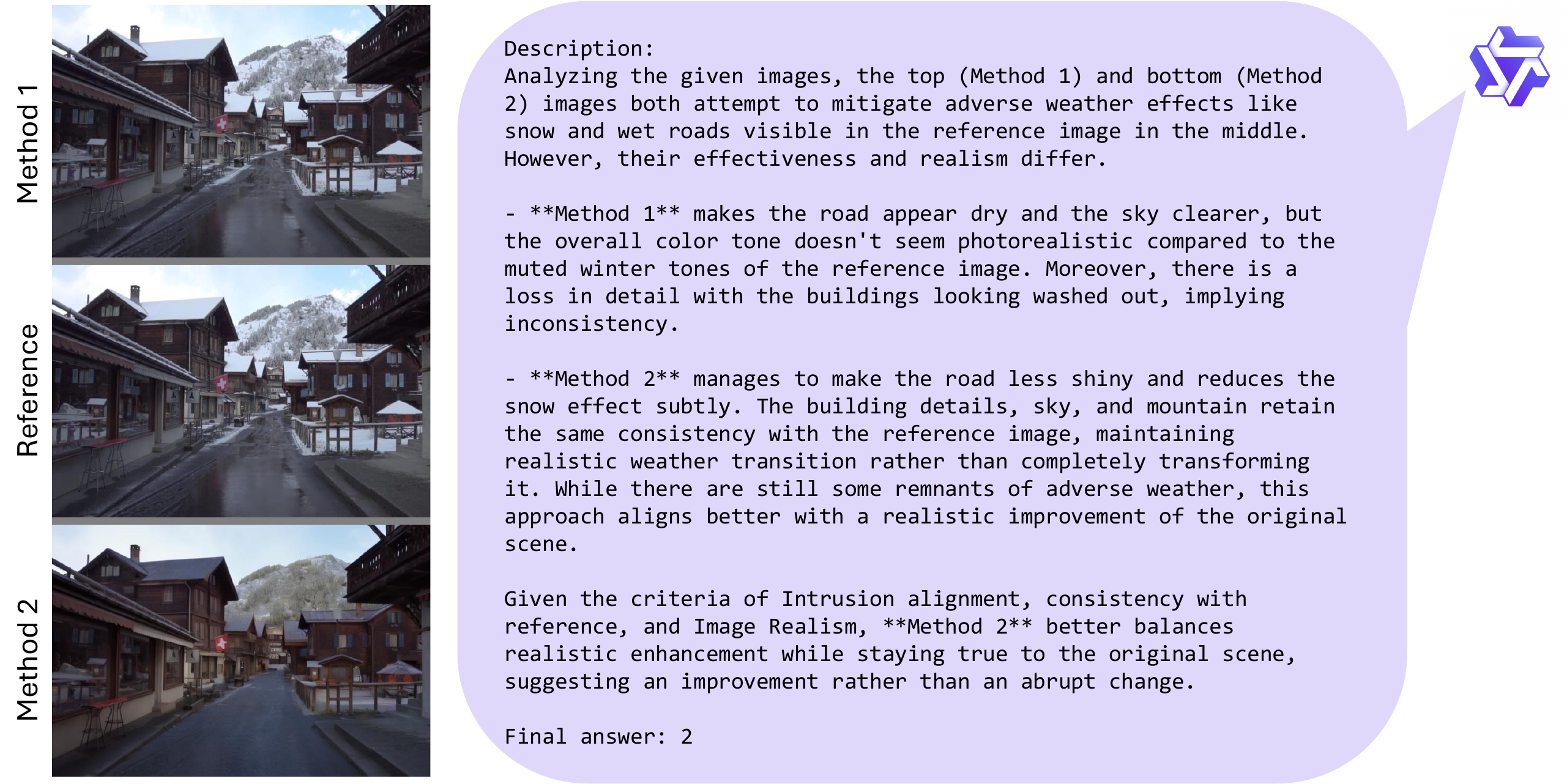}
\caption{Weather Removal Example: HistoFormer vs. Ours}
\end{subfigure}
\caption{\textbf{Examples on perceptual preference evaluation with VLM.} We instructed VLM to first briefly describe the observation, then give the reason why it makes this decision.}
\label{fig:user_study_vlm}
\end{figure*}

Inspired by \cite{wu2023gpteval3d}, we also used large vision-language models (VLM) as perceptual evaluators to perform similar perceptual preference selections. For each pair of methods to be compared, we randomly selected a frame of the video and fed these frames into VLM, then asked VLM to give a binary preference selection with the same criteria as we used in the human user study. We used Qwen2.5-VL-72B~\cite{Qwen2.5-VL} as our local VLM perceptual evaluator. For each sample pair, we run VLM 7 times with different random seeds. 
The final VLM preference of a scene video is determined by the same majority voting process. Fig. \ref{fig:user_study_vlm} demonstrates two example preference outputs from VLM.

\begin{figure*}[t]
    \centering\setlength{\tabcolsep}{0.1em}
    \resizebox{0.9\linewidth}{!}{%
    \begin{tabular}{@{}cccc@{}}

    \footnotesize{Synthesis Input} & \footnotesize{Synthesis Output} & 
    \footnotesize{Removal Input} & 
    \footnotesize{Removal Output} \\
     
    \frame{\includegraphics[width=0.25\linewidth]{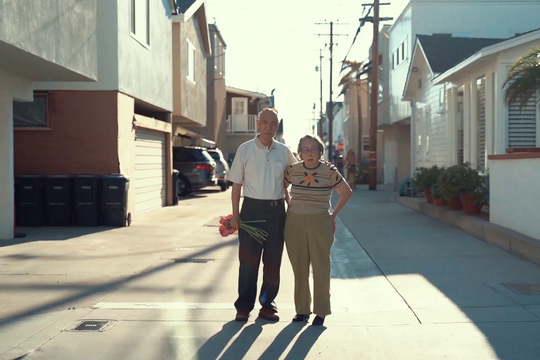}} &
    \frame{\includegraphics[width=0.25\linewidth]{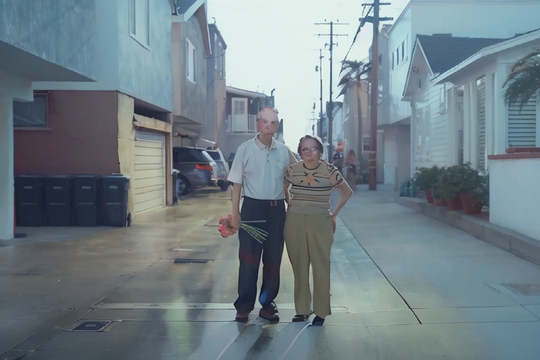}} &
    \frame{\includegraphics[width=0.25\linewidth]{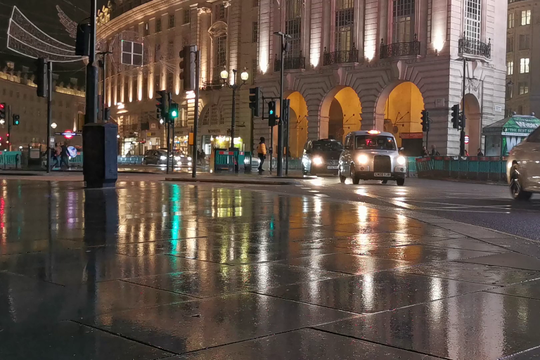}} &
    \frame{\includegraphics[width=0.25\linewidth]{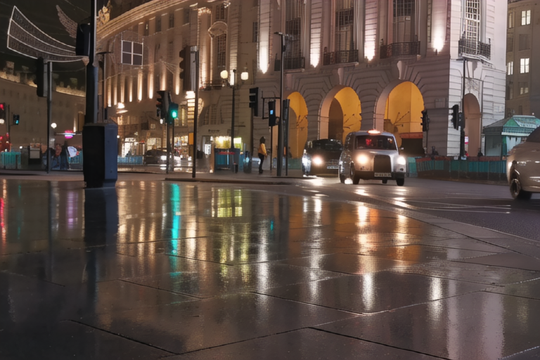}} \\
    
    \end{tabular}%
    }
    \vspace{-3mm}
    \caption{\textbf{Limitation.} Our method has a few failure cases, such as human facial details and night videos.
    }
    \vspace{-3mm}
    \label{fig:limitation}
\end{figure*}

\parahead{Failure Cases}
We show failure cases in Fig.~\ref{fig:limitation}. 
High-frequency details such as human faces are sometimes lost. This issue is primarily due to the limited capacity of our base model Stable Video Diffusion~\cite{blattmann2023stable}. 
The VAE of Stable Video Diffusion has 8x spatial compression, leading to causes significant degradation and altering of image details. 
In contrast, recent tokenizers offer significantly improved fidelity~\cite{yang2024cogvideox,agarwal2025cosmos}. 
Our results appear to have reached Stable Video Diffusion’s quality limit. Upgrading to a more powerful video model could significantly improve the overall quality. 

Our data collection includes limited night-time videos, leading to potential imperfect simulation in these scenarios. Future work could improve visual quality by collecting additional specialized data.

\vspace{-2mm}


\begin{thebibliography}{93}
\providecommand{\natexlab}[1]{#1}
\providecommand{\url}[1]{\texttt{#1}}
\expandafter\ifx\csname urlstyle\endcsname\relax
  \providecommand{\doi}[1]{doi: #1}\else
  \providecommand{\doi}{doi: \begingroup \urlstyle{rm}\Url}\fi

\bibitem[pex()]{pexels}
Pexels.com.
\newblock \url{https://www.pexels.com/}.

\bibitem[Agarwal et~al.(2025)Agarwal, Ali, Bala, Balaji, Barker, Cai, Chattopadhyay, Chen, Cui, Ding, et~al.]{agarwal2025cosmos}
Niket Agarwal, Arslan Ali, Maciej Bala, Yogesh Balaji, Erik Barker, Tiffany Cai, Prithvijit Chattopadhyay, Yongxin Chen, Yin Cui, Yifan Ding, et~al.
\newblock Cosmos world foundation model platform for physical ai.
\newblock \emph{arXiv preprint arXiv:2501.03575}, 2025.

\bibitem[Avrahami et~al.(2022)Avrahami, Lischinski, and Fried]{avrahami2022blended}
Omri Avrahami, Dani Lischinski, and Ohad Fried.
\newblock Blended diffusion for text-driven editing of natural images.
\newblock In \emph{Proceedings of the IEEE/CVF conference on computer vision and pattern recognition}, pages 18208--18218, 2022.

\bibitem[Bai et~al.(2025)Bai, Chen, Liu, Wang, Ge, Song, Dang, Wang, Wang, Tang, Zhong, Zhu, Yang, Li, Wan, Wang, Ding, Fu, Xu, Ye, Zhang, Xie, Cheng, Zhang, Yang, Xu, and Lin]{Qwen2.5-VL}
Shuai Bai, Keqin Chen, Xuejing Liu, Jialin Wang, Wenbin Ge, Sibo Song, Kai Dang, Peng Wang, Shijie Wang, Jun Tang, Humen Zhong, Yuanzhi Zhu, Mingkun Yang, Zhaohai Li, Jianqiang Wan, Pengfei Wang, Wei Ding, Zheren Fu, Yiheng Xu, Jiabo Ye, Xi Zhang, Tianbao Xie, Zesen Cheng, Hang Zhang, Zhibo Yang, Haiyang Xu, and Junyang Lin.
\newblock Qwen2.5-vl technical report.
\newblock \emph{arXiv preprint arXiv:2502.13923}, 2025.

\bibitem[Bain et~al.(2021)Bain, Nagrani, Varol, and Zisserman]{bain2021frozen}
Max Bain, Arsha Nagrani, G{\"u}l Varol, and Andrew Zisserman.
\newblock Frozen in time: A joint video and image encoder for end-to-end retrieval.
\newblock In \emph{ICCV}, 2021.

\bibitem[Bar-Tal et~al.(2022)Bar-Tal, Ofri-Amar, Fridman, Kasten, and Dekel]{bar2022text2live}
Omer Bar-Tal, Dolev Ofri-Amar, Rafail Fridman, Yoni Kasten, and Tali Dekel.
\newblock Text2live: Text-driven layered image and video editing.
\newblock In \emph{ECCV}. Springer, 2022.

\bibitem[Blattmann et~al.(2023)Blattmann, Dockhorn, Kulal, Mendelevitch, Kilian, Lorenz, Levi, English, Voleti, Letts, et~al.]{blattmann2023stable}
Andreas Blattmann, Tim Dockhorn, Sumith Kulal, Daniel Mendelevitch, Maciej Kilian, Dominik Lorenz, Yam Levi, Zion English, Vikram Voleti, Adam Letts, et~al.
\newblock Stable video diffusion: Scaling latent video diffusion models to large datasets.
\newblock \emph{arXiv preprint arXiv:2311.15127}, 2023.

\bibitem[Brooks et~al.(2023)Brooks, Holynski, and Efros]{brooks2023instructpix2pix}
Tim Brooks, Aleksander Holynski, and Alexei~A Efros.
\newblock Instructpix2pix: Learning to follow image editing instructions.
\newblock In \emph{CVPR}, 2023.

\bibitem[Brown et~al.(2020)Brown, Mann, Ryder, Subbiah, Kaplan, Dhariwal, Neelakantan, Shyam, Sastry, Askell, et~al.]{brown2020language}
Tom Brown, Benjamin Mann, Nick Ryder, Melanie Subbiah, Jared~D Kaplan, Prafulla Dhariwal, Arvind Neelakantan, Pranav Shyam, Girish Sastry, Amanda Askell, et~al.
\newblock Language models are few-shot learners.
\newblock \emph{NeurIPS}, 2020.

\bibitem[Cai et~al.(2016)Cai, Xu, Jia, Qing, and Tao]{cai2016dehazenet}
Bolun Cai, Xiangmin Xu, Kui Jia, Chunmei Qing, and Dacheng Tao.
\newblock Dehazenet: An end-to-end system for single image haze removal.
\newblock \emph{IEEE transactions on image processing}, 25\penalty0 (11):\penalty0 5187--5198, 2016.

\bibitem[Ceylan et~al.(2023)Ceylan, Huang, and Mitra]{ceylan2023pix2video}
Duygu Ceylan, Chun-Hao~P Huang, and Niloy~J Mitra.
\newblock Pix2video: Video editing using image diffusion.
\newblock In \emph{Proceedings of the IEEE/CVF International Conference on Computer Vision}, pages 23206--23217, 2023.

\bibitem[Chen et~al.(2023)Chen, Ye, Liu, Liao, Jiang, Chen, and Chen]{chen2023msp}
Sixiang Chen, Tian Ye, Yun Liu, Taodong Liao, Jingxia Jiang, Erkang Chen, and Peng Chen.
\newblock Msp-former: Multi-scale projection transformer for single image desnowing.
\newblock In \emph{ICASSP 2023-2023 IEEE International Conference on Acoustics, Speech and Signal Processing (ICASSP)}. IEEE, 2023.

\bibitem[Chen et~al.(2024)Chen, Ye, Zhang, Xing, Lin, and Zhu]{chen2024teaching}
Sixiang Chen, Tian Ye, Kai Zhang, Zhaohu Xing, Yunlong Lin, and Lei Zhu.
\newblock Teaching tailored to talent: Adverse weather restoration via prompt pool and depth-anything constraint.
\newblock In \emph{European Conference on Computer Vision}, pages 95--115. Springer, 2024.

\bibitem[Chen et~al.(2020)Chen, Fang, Ding, Tsai, and Kuo]{chen2020jstasr}
Wei-Ting Chen, Hao-Yu Fang, Jian-Jiun Ding, Cheng-Che Tsai, and Sy-Yen Kuo.
\newblock Jstasr: Joint size and transparency-aware snow removal algorithm based on modified partial convolution and veiling effect removal.
\newblock In \emph{ECCV}. Springer, 2020.

\bibitem[Cong et~al.(2024)Cong, Xu, Simon, Chen, Ren, Xie, Perez-Rua, Rosenhahn, Xiang, and He]{cong2023flatten}
Yuren Cong, Mengmeng Xu, Christian Simon, Shoufa Chen, Jiawei Ren, Yanping Xie, Juan-Manuel Perez-Rua, Bodo Rosenhahn, Tao Xiang, and Sen He.
\newblock Flatten: optical flow-guided attention for consistent text-to-video editing.
\newblock \emph{ICLR}, 2024.

\bibitem[Cosne et~al.(2020)Cosne, Juraver, Teng, Schmidt, Vardanyan, Luccioni, and Bengio]{cosne2020using}
Gautier Cosne, Adrien Juraver, M{\'e}lisande Teng, Victor Schmidt, Vahe Vardanyan, Alexandra Luccioni, and Yoshua Bengio.
\newblock Using simulated data to generate images of climate change.
\newblock \emph{ICLR Workshop}, 2020.

\bibitem[Dai et~al.(2025)Dai, Ni, Shen, Chen, Chen, and Chu]{dai2025rainygs}
Qiyu Dai, Xingyu Ni, Qianfan Shen, Wenzheng Chen, Baoquan Chen, and Mengyu Chu.
\newblock Rainygs: Efficient rain synthesis with physically-based gaussian splatting.
\newblock \emph{arXiv preprint arXiv:2503.21442}, 2025.

\bibitem[Dhariwal and Nichol(2021)]{dhariwal2021diffusion}
Prafulla Dhariwal and Alexander~Quinn Nichol.
\newblock Diffusion models beat {GAN}s on image synthesis.
\newblock In \emph{NeurIPS}, 2021.

\bibitem[{Epic Games}(2019)]{unrealengine}
{Epic Games}.
\newblock Unreal engine, 2019.

\bibitem[Esser et~al.(2023)Esser, Chiu, Atighehchian, Granskog, and Germanidis]{esser2023structure}
Patrick Esser, Johnathan Chiu, Parmida Atighehchian, Jonathan Granskog, and Anastasis Germanidis.
\newblock Structure and content-guided video synthesis with diffusion models.
\newblock In \emph{Proceedings of the IEEE/CVF international conference on computer vision}, pages 7346--7356, 2023.

\bibitem[Feldman and O'Brien(2002)]{feldman2002modeling}
Bryan~E Feldman and James~F O'Brien.
\newblock Modeling the accumulation of wind-driven snow.
\newblock In \emph{ACM SIGGRAPH 2002 conference abstracts and applications}, 2002.

\bibitem[Feng et~al.(2024)Feng, Weng, Wang, Yuan, Bao, Luo, Chen, and Guo]{feng2024ccedit}
Ruoyu Feng, Wenming Weng, Yanhui Wang, Yuhui Yuan, Jianmin Bao, Chong Luo, Zhibo Chen, and Baining Guo.
\newblock Ccedit: Creative and controllable video editing via diffusion models.
\newblock In \emph{Proceedings of the IEEE/CVF Conference on Computer Vision and Pattern Recognition}, pages 6712--6722, 2024.

\bibitem[Fiebelman et~al.(2025)Fiebelman, Averbuch-Elor, and Benaim]{fiebelman2025let}
Gal Fiebelman, Hadar Averbuch-Elor, and Sagie Benaim.
\newblock Let it snow! animating static gaussian scenes with dynamic weather effects.
\newblock \emph{arXiv preprint arXiv:2504.05296}, 2025.

\bibitem[Gal et~al.(2022)Gal, Patashnik, Maron, Bermano, Chechik, and Cohen-Or]{gal2022stylegan}
Rinon Gal, Or Patashnik, Haggai Maron, Amit~H Bermano, Gal Chechik, and Daniel Cohen-Or.
\newblock Stylegan-nada: Clip-guided domain adaptation of image generators.
\newblock \emph{ACM Transactions on Graphics (TOG)}, 2022.

\bibitem[Garg and Nayar(2006)]{garg2006photorealistic}
Kshitiz Garg and Shree~K Nayar.
\newblock Photorealistic rendering of rain streaks.
\newblock \emph{ACM Transactions on Graphics (TOG)}, 2006.

\bibitem[Geyer et~al.(2024)Geyer, Bar-Tal, Bagon, and Dekel]{geyer2023tokenflow}
Michal Geyer, Omer Bar-Tal, Shai Bagon, and Tali Dekel.
\newblock Tokenflow: Consistent diffusion features for consistent video editing.
\newblock \emph{ICLR}, 2024.

\bibitem[Gissler et~al.(2020)Gissler, Henne, Band, Peer, and Teschner]{gissler2020implicit}
Christoph Gissler, Andreas Henne, Stefan Band, Andreas Peer, and Matthias Teschner.
\newblock An implicit compressible sph solver for snow simulation.
\newblock \emph{ACM Transactions on Graphics (TOG)}, 2020.

\bibitem[Guo~et al.(2023)]{guo2023sky}
Yun Guo~et al.
\newblock From sky to the ground: A large-scale benchmark and simple baseline towards real rain removal.
\newblock In \emph{ICCV}, 2023.

\bibitem[Hahner et~al.(2019)Hahner, Dai, Sakaridis, Zaech, and Van~Gool]{hahner2019semantic}
Martin Hahner, Dengxin Dai, Christos Sakaridis, Jan-Nico Zaech, and Luc Van~Gool.
\newblock Semantic understanding of foggy scenes with purely synthetic data.
\newblock In \emph{IEEE Intelligent Transportation Systems Conference (ITSC)}, 2019.

\bibitem[Haque et~al.(2023)Haque, Tancik, Efros, Holynski, and Kanazawa]{haque2023instruct}
Ayaan Haque, Matthew Tancik, Alexei~A Efros, Aleksander Holynski, and Angjoo Kanazawa.
\newblock Instruct-nerf2nerf: Editing 3d scenes with instructions.
\newblock In \emph{Proceedings of the IEEE/CVF International Conference on Computer Vision}, pages 19740--19750, 2023.

\bibitem[Hertz et~al.(2022)Hertz, Mokady, Tenenbaum, Aberman, Pritch, and Cohen-Or]{hertz2022prompt}
Amir Hertz, Ron Mokady, Jay Tenenbaum, Kfir Aberman, Yael Pritch, and Daniel Cohen-Or.
\newblock Prompt-to-prompt image editing with cross attention control.
\newblock \emph{arXiv preprint arXiv:2208.01626}, 2022.

\bibitem[Ho et~al.(2020)Ho, Jain, and Abbeel]{ho2020denoising}
Jonathan Ho, Ajay Jain, and Pieter Abbeel.
\newblock Denoising diffusion probabilistic models.
\newblock \emph{NeurIPS}, 2020.

\bibitem[Hong et~al.(2022)Hong, Ding, Zheng, Liu, and Tang]{hong2022cogvideo}
Wenyi Hong, Ming Ding, Wendi Zheng, Xinghan Liu, and Jie Tang.
\newblock Cogvideo: Large-scale pretraining for text-to-video generation via transformers.
\newblock \emph{arXiv preprint arXiv:2205.15868}, 2022.

\bibitem[Hsu et~al.(2024)Hsu, Lin, Zhai, Xia, and Wang]{hsu2024autovfx}
Hao-Yu Hsu, Zhi-Hao Lin, Albert Zhai, Hongchi Xia, and Shenlong Wang.
\newblock Autovfx: Physically realistic video editing from natural language instructions.
\newblock \emph{arXiv preprint arXiv:2411.02394}, 2024.

\bibitem[Huang et~al.(2024{\natexlab{a}})Huang, He, Yu, Zhang, Si, Jiang, Zhang, Wu, Jin, Chanpaisit, Wang, Chen, Wang, Lin, Qiao, and Liu]{huang2023vbench}
Ziqi Huang, Yinan He, Jiashuo Yu, Fan Zhang, Chenyang Si, Yuming Jiang, Yuanhan Zhang, Tianxing Wu, Qingyang Jin, Nattapol Chanpaisit, Yaohui Wang, Xinyuan Chen, Limin Wang, Dahua Lin, Yu Qiao, and Ziwei Liu.
\newblock {VBench}: Comprehensive benchmark suite for video generative models.
\newblock In \emph{CVPR}, 2024{\natexlab{a}}.

\bibitem[Huang et~al.(2024{\natexlab{b}})Huang, Zhang, Xu, He, Yu, Dong, Ma, Chanpaisit, Si, Jiang, Wang, Chen, Chen, Wang, Lin, Qiao, and Liu]{huang2024vbench++}
Ziqi Huang, Fan Zhang, Xiaojie Xu, Yinan He, Jiashuo Yu, Ziyue Dong, Qianli Ma, Nattapol Chanpaisit, Chenyang Si, Yuming Jiang, Yaohui Wang, Xinyuan Chen, Ying-Cong Chen, Limin Wang, Dahua Lin, Yu Qiao, and Ziwei Liu.
\newblock Vbench++: Comprehensive and versatile benchmark suite for video generative models.
\newblock \emph{arXiv preprint arXiv:2411.13503}, 2024{\natexlab{b}}.

\bibitem[Karras et~al.(2022)Karras, Aittala, Aila, and Laine]{Karras2022edm}
Tero Karras, Miika Aittala, Timo Aila, and Samuli Laine.
\newblock Elucidating the design space of diffusion-based generative models.
\newblock In \emph{NeurIPS}, 2022.

\bibitem[Kerbl et~al.(2023)Kerbl, Kopanas, Leimk{\"u}hler, and Drettakis]{kerbl3Dgaussians}
Bernhard Kerbl, Georgios Kopanas, Thomas Leimk{\"u}hler, and George Drettakis.
\newblock 3d gaussian splatting for real-time radiance field rendering.
\newblock \emph{ACM Transactions on Graphics}, 2023.

\bibitem[Khachatryan et~al.(2023)Khachatryan, Movsisyan, Tadevosyan, Henschel, Wang, Navasardyan, and Shi]{khachatryan2023text2video}
Levon Khachatryan, Andranik Movsisyan, Vahram Tadevosyan, Roberto Henschel, Zhangyang Wang, Shant Navasardyan, and Humphrey Shi.
\newblock Text2video-zero: Text-to-image diffusion models are zero-shot video generators.
\newblock In \emph{Proceedings of the IEEE/CVF International Conference on Computer Vision}, pages 15954--15964, 2023.

\bibitem[Kirstain et~al.(2023)Kirstain, Polyak, Singer, Matiana, Penna, and Levy]{kirstain2023pick}
Yuval Kirstain, Adam Polyak, Uriel Singer, Shahbuland Matiana, Joe Penna, and Omer Levy.
\newblock Pick-a-pic: An open dataset of user preferences for text-to-image generation.
\newblock \emph{NeurIPS}, 2023.

\bibitem[Ku et~al.(2024)Ku, Wei, Ren, Yang, and Chen]{ku2024anyv2v}
Max Ku, Cong Wei, Weiming Ren, Harry Yang, and Wenhu Chen.
\newblock Anyv2v: A tuning-free framework for any video-to-video editing tasks.
\newblock \emph{arXiv preprint arXiv:2403.14468}, 2024.

\bibitem[Li et~al.(2017)Li, Peng, Wang, Xu, and Feng]{li2017aod}
Boyi Li, Xiulian Peng, Zhangyang Wang, Jizheng Xu, and Dan Feng.
\newblock Aod-net: All-in-one dehazing network.
\newblock In \emph{Proceedings of the IEEE international conference on computer vision}, pages 4770--4778, 2017.

\bibitem[Li et~al.(2020)Li, Tan, and Cheong]{li2020all}
Ruoteng Li, Robby~T Tan, and Loong-Fah Cheong.
\newblock All in one bad weather removal using architectural search.
\newblock In \emph{CVPR}, 2020.

\bibitem[Li et~al.(2023{\natexlab{a}})Li, Lin, Forsyth, Huang, and Wang]{Li2023ClimateNeRF}
Yuan Li, Zhi-Hao Lin, David Forsyth, Jia-Bin Huang, and Shenlong Wang.
\newblock Climatenerf: Extreme weather synthesis in neural radiance field.
\newblock In \emph{ICCV}, 2023{\natexlab{a}}.

\bibitem[Li et~al.(2023{\natexlab{b}})Li, Zhu, Han, Hou, Guo, and Cheng]{li2023amt}
Zhen Li, Zuo-Liang Zhu, Ling-Hao Han, Qibin Hou, Chun-Le Guo, and Ming-Ming Cheng.
\newblock Amt: All-pairs multi-field transforms for efficient frame interpolation.
\newblock In \emph{CVPR}, 2023{\natexlab{b}}.

\bibitem[Liang et~al.(2024)Liang, Wu, Wang, Yu, Li, Zhao, Misra, Huang, Zhang, Vajda, et~al.]{liang2024flowvid}
Feng Liang, Bichen Wu, Jialiang Wang, Licheng Yu, Kunpeng Li, Yinan Zhao, Ishan Misra, Jia-Bin Huang, Peizhao Zhang, Peter Vajda, et~al.
\newblock Flowvid: Taming imperfect optical flows for consistent video-to-video synthesis.
\newblock In \emph{Proceedings of the IEEE/CVF Conference on Computer Vision and Pattern Recognition}, pages 8207--8216, 2024.

\bibitem[Liang et~al.(2025)Liang, Gojcic, Ling, Munkberg, Hasselgren, Lin, Gao, Keller, Vijaykumar, Fidler, and Wang]{DiffusionRenderer}
Ruofan Liang, Zan Gojcic, Huan Ling, Jacob Munkberg, Jon Hasselgren, Zhi-Hao Lin, Jun Gao, Alexander Keller, Nandita Vijaykumar, Sanja Fidler, and Zian Wang.
\newblock Diffusionrenderer: Neural inverse and forward rendering with video diffusion models.
\newblock \emph{arXiv preprint arXiv: 2501.18590}, 2025.

\bibitem[Ling et~al.(2024)Ling, Sheng, Tu, Zhao, Xin, Wan, Yu, Guo, Yu, Lu, et~al.]{ling2024dl3dv}
Lu Ling, Yichen Sheng, Zhi Tu, Wentian Zhao, Cheng Xin, Kun Wan, Lantao Yu, Qianyu Guo, Zixun Yu, Yawen Lu, et~al.
\newblock Dl3dv-10k: A large-scale scene dataset for deep learning-based 3d vision.
\newblock In \emph{CVPR}, 2024.

\bibitem[Liu et~al.(2019)Liu, Ma, Shi, and Chen]{liu2019griddehazenet}
Xiaohong Liu, Yongrui Ma, Zhihao Shi, and Jun Chen.
\newblock Griddehazenet: Attention-based multi-scale network for image dehazing.
\newblock In \emph{Proceedings of the IEEE/CVF international conference on computer vision}, pages 7314--7323, 2019.

\bibitem[Liu et~al.(2018)Liu, Jaw, Huang, and Hwang]{liu2018desnownet}
Yun-Fu Liu, Da-Wei Jaw, Shih-Chia Huang, and Jenq-Neng Hwang.
\newblock Desnownet: Context-aware deep network for snow removal.
\newblock \emph{IEEE TIP}, 2018.

\bibitem[Meng et~al.(2021)Meng, Song, Song, Wu, Zhu, and Ermon]{meng2021sdedit}
Chenlin Meng, Yang Song, Jiaming Song, Jiajun Wu, Jun-Yan Zhu, and Stefano Ermon.
\newblock Sdedit: Image synthesis and editing with stochastic differential equations.
\newblock \emph{arXiv preprint arXiv:2108.01073}, 2021.

\bibitem[Mildenhall et~al.(2020)Mildenhall, Srinivasan, Tancik, Barron, Ramamoorthi, and Ng]{mildenhall2020nerf}
Ben Mildenhall, Pratul~P. Srinivasan, Matthew Tancik, Jonathan~T. Barron, Ravi Ramamoorthi, and Ren Ng.
\newblock Nerf: Representing scenes as neural radiance fields for view synthesis.
\newblock In \emph{ECCV}, 2020.

\bibitem[Molad et~al.(2023)Molad, Horwitz, Valevski, Acha, Matias, Pritch, Leviathan, and Hoshen]{molad2023dreamix}
Eyal Molad, Eliahu Horwitz, Dani Valevski, Alex~Rav Acha, Yossi Matias, Yael Pritch, Yaniv Leviathan, and Yedid Hoshen.
\newblock Dreamix: Video diffusion models are general video editors.
\newblock \emph{arXiv preprint arXiv:2302.01329}, 2023.

\bibitem[M{\"u}ller et~al.(2022)M{\"u}ller, Evans, Schied, and Keller]{muller2022instant}
Thomas M{\"u}ller, Alex Evans, Christoph Schied, and Alexander Keller.
\newblock Instant neural graphics primitives with a multiresolution hash encoding.
\newblock \emph{ACM TOG}, 2022.

\bibitem[OpenAI(2024)]{OpenAI_ChatGPT}
OpenAI.
\newblock Chatgpt: A conversational ai model, 2024.
\newblock Accessed: 2025-03-04.

\bibitem[\"{O}zdenizci and Legenstein(2023)]{ozdenizci2023}
Ozan \"{O}zdenizci and Robert Legenstein.
\newblock Restoring vision in adverse weather conditions with patch-based denoising diffusion models.
\newblock \emph{IEEE TPAMI}, 2023.

\bibitem[Parmar et~al.(2024)Parmar, Park, Narasimhan, and Zhu]{parmar2024one}
Gaurav Parmar, Taesung Park, Srinivasa Narasimhan, and Jun-Yan Zhu.
\newblock One-step image translation with text-to-image models.
\newblock \emph{arXiv preprint arXiv:2403.12036}, 2024.

\bibitem[Podell et~al.(2023)Podell, English, Lacey, Blattmann, Dockhorn, M{\"u}ller, Penna, and Rombach]{podell2023sdxl}
Dustin Podell, Zion English, Kyle Lacey, Andreas Blattmann, Tim Dockhorn, Jonas M{\"u}ller, Joe Penna, and Robin Rombach.
\newblock Sdxl: Improving latent diffusion models for high-resolution image synthesis.
\newblock \emph{arXiv preprint arXiv:2307.01952}, 2023.

\bibitem[Qi et~al.(2023)Qi, Cun, Zhang, Lei, Wang, Shan, and Chen]{qi2023fatezero}
Chenyang Qi, Xiaodong Cun, Yong Zhang, Chenyang Lei, Xintao Wang, Ying Shan, and Qifeng Chen.
\newblock Fatezero: Fusing attentions for zero-shot text-based video editing.
\newblock In \emph{Proceedings of the IEEE/CVF International Conference on Computer Vision}, pages 15932--15942, 2023.

\bibitem[Qian et~al.(2018)Qian, Tan, Yang, Su, and Liu]{qian2018attentive}
Rui Qian, Robby~T Tan, Wenhan Yang, Jiajun Su, and Jiaying Liu.
\newblock Attentive generative adversarial network for raindrop removal from a single image.
\newblock In \emph{Proceedings of the IEEE conference on computer vision and pattern recognition}, pages 2482--2491, 2018.

\bibitem[Quan et~al.(2021)Quan, Yu, Liang, and Yang]{quan2021removing}
Ruijie Quan, Xin Yu, Yuanzhi Liang, and Yi Yang.
\newblock Removing raindrops and rain streaks in one go.
\newblock In \emph{CVPR}, 2021.

\bibitem[Radford et~al.(2021)Radford, Kim, Hallacy, Ramesh, Goh, Agarwal, Sastry, Askell, Mishkin, Clark, et~al.]{radford2021learning}
Alec Radford, Jong~Wook Kim, Chris Hallacy, Aditya Ramesh, Gabriel Goh, Sandhini Agarwal, Girish Sastry, Amanda Askell, Pamela Mishkin, Jack Clark, et~al.
\newblock Learning transferable visual models from natural language supervision.
\newblock In \emph{ICML}, 2021.

\bibitem[Ren et~al.(2024)Ren, Liu, Zeng, Lin, Li, Cao, Chen, Huang, Chen, Yan, Zeng, Zhang, Li, Yang, Li, Jiang, and Zhang]{ren2024grounded}
Tianhe Ren, Shilong Liu, Ailing Zeng, Jing Lin, Kunchang Li, He Cao, Jiayu Chen, Xinyu Huang, Yukang Chen, Feng Yan, Zhaoyang Zeng, Hao Zhang, Feng Li, Jie Yang, Hongyang Li, Qing Jiang, and Lei Zhang.
\newblock Grounded sam: Assembling open-world models for diverse visual tasks, 2024.

\bibitem[Schmalfuss et~al.(2023)Schmalfuss, Mehl, and Bruhn]{schmalfuss2023distracting}
Jenny Schmalfuss, Lukas Mehl, and Andr{\'e}s Bruhn.
\newblock Distracting downpour: Adversarial weather attacks for motion estimation.
\newblock In \emph{ICCV}, 2023.

\bibitem[Schmidt et~al.(2019)Schmidt, Luccioni, Mukkavilli, Balasooriya, Sankaran, Chayes, and Bengio]{schmidt2019visualizing}
Victor Schmidt, Alexandra Luccioni, S~Karthik Mukkavilli, Narmada Balasooriya, Kris Sankaran, Jennifer Chayes, and Yoshua Bengio.
\newblock Visualizing the consequences of climate change using cycle-consistent adversarial networks.
\newblock \emph{ICLR}, 2019.

\bibitem[Schmidt et~al.(2022)Schmidt, Luccioni, Teng, Zhang, Reynaud, Raghupathi, Cosne, Juraver, Vardanyan, Hernandez-Garcia, et~al.]{schmidt2021climategan}
Victor Schmidt, Alexandra~Sasha Luccioni, M{\'e}lisande Teng, Tianyu Zhang, Alexia Reynaud, Sunand Raghupathi, Gautier Cosne, Adrien Juraver, Vahe Vardanyan, Alex Hernandez-Garcia, et~al.
\newblock Climategan: Raising climate change awareness by generating images of floods.
\newblock \emph{ICLR}, 2022.

\bibitem[Schuhmann et~al.(2022)Schuhmann, Beaumont, Vencu, Gordon, Wightman, Cherti, Coombes, Katta, Mullis, Wortsman, et~al.]{schuhmann2022laion}
Christoph Schuhmann, Romain Beaumont, Richard Vencu, Cade Gordon, Ross Wightman, Mehdi Cherti, Theo Coombes, Aarush Katta, Clayton Mullis, Mitchell Wortsman, et~al.
\newblock Laion-5b: An open large-scale dataset for training next generation image-text models.
\newblock \emph{NeurIPS}, 2022.

\bibitem[Shin et~al.(2024)Shin, Kim, Lee, Lee, and Yoon]{shin2024edit}
Chaehun Shin, Heeseung Kim, Che~Hyun Lee, Sang-gil Lee, and Sungroh Yoon.
\newblock Edit-a-video: Single video editing with object-aware consistency.
\newblock In \emph{Asian Conference on Machine Learning}, pages 1215--1230. PMLR, 2024.

\bibitem[{Sohl-Dickstein} et~al.(2015){Sohl-Dickstein}, Weiss, Maheswaranathan, and Ganguli]{sohl2015deep}
Jascha {Sohl-Dickstein}, Eric Weiss, Niru Maheswaranathan, and Surya Ganguli.
\newblock Deep unsupervised learning using nonequilibrium thermodynamics.
\newblock In \emph{International Conference on Machine Learning}, 2015.

\bibitem[Stomakhin et~al.(2013)Stomakhin, Schroeder, Chai, Teran, and Selle]{stomakhin2013material}
Alexey Stomakhin, Craig Schroeder, Lawrence Chai, Joseph Teran, and Andrew Selle.
\newblock A material point method for snow simulation.
\newblock \emph{ACM Transactions on Graphics (TOG)}, 2013.

\bibitem[Sulsky et~al.(1995)Sulsky, Zhou, and Schreyer]{sulsky1995application}
Deborah Sulsky, Shi-Jian Zhou, and Howard~L Schreyer.
\newblock Application of a particle-in-cell method to solid mechanics.
\newblock \emph{Computer physics communications}, 1995.

\bibitem[Sun et~al.(2020)Sun, Kretzschmar, Dotiwalla, Chouard, Patnaik, Tsui, Guo, Zhou, Chai, Caine, et~al.]{sun2020scalability}
Pei Sun, Henrik Kretzschmar, Xerxes Dotiwalla, Aurelien Chouard, Vijaysai Patnaik, Paul Tsui, James Guo, Yin Zhou, Yuning Chai, Benjamin Caine, et~al.
\newblock Scalability in perception for autonomous driving: Waymo open dataset.
\newblock In \emph{CVPR}, 2020.

\bibitem[Sun et~al.(2024)Sun, Ren, Gao, Wang, and Cao]{sun2024restoring}
Shangquan Sun, Wenqi Ren, Xinwei Gao, Rui Wang, and Xiaochun Cao.
\newblock Restoring images in adverse weather conditions via histogram transformer.
\newblock \emph{ECCV}, 2024.

\bibitem[Tremblay et~al.(2021)Tremblay, Halder, De~Charette, and Lalonde]{tremblay2021rain}
Maxime Tremblay, Shirsendu~Sukanta Halder, Raoul De~Charette, and Jean-Fran{\c{c}}ois Lalonde.
\newblock Rain rendering for evaluating and improving robustness to bad weather.
\newblock \emph{IJCV}, 2021.

\bibitem[Tumanyan et~al.(2022)Tumanyan, Bar-Tal, Bagon, and Dekel]{tumanyan2022splicing}
Narek Tumanyan, Omer Bar-Tal, Shai Bagon, and Tali Dekel.
\newblock Splicing vit features for semantic appearance transfer.
\newblock In \emph{CVPR}, 2022.

\bibitem[Tumanyan et~al.(2023)Tumanyan, Geyer, Bagon, and Dekel]{tumanyan2023plug}
Narek Tumanyan, Michal Geyer, Shai Bagon, and Tali Dekel.
\newblock Plug-and-play diffusion features for text-driven image-to-image translation.
\newblock In \emph{Proceedings of the IEEE/CVF Conference on Computer Vision and Pattern Recognition}, pages 1921--1930, 2023.

\bibitem[Valanarasu et~al.(2022)Valanarasu, Yasarla, and Patel]{valanarasu2022transweather}
Jeya Maria~Jose Valanarasu, Rajeev Yasarla, and Vishal~M Patel.
\newblock Transweather: Transformer-based restoration of images degraded by adverse weather conditions.
\newblock In \emph{CVPR}, 2022.

\bibitem[Volk et~al.(2019)Volk, M{\"u}ller, Von~Bernuth, Hospach, and Bringmann]{volk2019towards}
Georg Volk, Stefan M{\"u}ller, Alexander Von~Bernuth, Dennis Hospach, and Oliver Bringmann.
\newblock Towards robust cnn-based object detection through augmentation with synthetic rain variations.
\newblock In \emph{2019 IEEE intelligent transportation systems conference (ITSC)}. IEEE, 2019.

\bibitem[Von~Bernuth et~al.(2019)Von~Bernuth, Volk, and Bringmann]{von2019simulating}
Alexander Von~Bernuth, Georg Volk, and Oliver Bringmann.
\newblock Simulating photo-realistic snow and fog on existing images for enhanced cnn training and evaluation.
\newblock In \emph{2019 IEEE Intelligent Transportation Systems Conference (ITSC)}. IEEE, 2019.

\bibitem[Wang et~al.(2024)Wang, Bai, Tan, Wang, Fan, Bai, Chen, Liu, Wang, Ge, Fan, Dang, Du, Ren, Men, Liu, Zhou, Zhou, and Lin]{Qwen2-VL}
Peng Wang, Shuai Bai, Sinan Tan, Shijie Wang, Zhihao Fan, Jinze Bai, Keqin Chen, Xuejing Liu, Jialin Wang, Wenbin Ge, Yang Fan, Kai Dang, Mengfei Du, Xuancheng Ren, Rui Men, Dayiheng Liu, Chang Zhou, Jingren Zhou, and Junyang Lin.
\newblock Qwen2-vl: Enhancing vision-language model's perception of the world at any resolution.
\newblock \emph{arXiv preprint arXiv:2409.12191}, 2024.

\bibitem[Wu et~al.(2021)Wu, Qu, Lin, Zhou, Qiao, Zhang, Xie, and Ma]{wu2021contrastive}
Haiyan Wu, Yanyun Qu, Shaohui Lin, Jian Zhou, Ruizhi Qiao, Zhizhong Zhang, Yuan Xie, and Lizhuang Ma.
\newblock Contrastive learning for compact single image dehazing.
\newblock In \emph{Proceedings of the IEEE/CVF conference on computer vision and pattern recognition}, pages 10551--10560, 2021.

\bibitem[Wu et~al.(2023)Wu, Ge, Wang, Lei, Gu, Shi, Hsu, Shan, Qie, and Shou]{wu2023tune}
Jay~Zhangjie Wu, Yixiao Ge, Xintao Wang, Stan~Weixian Lei, Yuchao Gu, Yufei Shi, Wynne Hsu, Ying Shan, Xiaohu Qie, and Mike~Zheng Shou.
\newblock Tune-a-video: One-shot tuning of image diffusion models for text-to-video generation.
\newblock In \emph{ICCV}, 2023.

\bibitem[Wu et~al.(2024)Wu, Yang, Li, Zhang, Liu, Guibas, Lin, and Wetzstein]{wu2023gpteval3d}
Tong Wu, Guandao Yang, Zhibing Li, Kai Zhang, Ziwei Liu, Leonidas Guibas, Dahua Lin, and Gordon Wetzstein.
\newblock Gpt-4v(ision) is a human-aligned evaluator for text-to-3d generation.
\newblock In \emph{CVPR}, 2024.

\bibitem[Wu~et al.(2024)]{wu2024rainmamba}
Hongtao Wu~et al.
\newblock Rainmamba: Enhanced locality learning with state space models for video deraining.
\newblock In \emph{ACM MM}, 2024.

\bibitem[Xiao et~al.(2022)Xiao, Fu, Liu, Wu, and Zha]{xiao2022image}
Jie Xiao, Xueyang Fu, Aiping Liu, Feng Wu, and Zheng-Jun Zha.
\newblock Image de-raining transformer.
\newblock \emph{IEEE transactions on pattern analysis and machine intelligence}, 45\penalty0 (11):\penalty0 12978--12995, 2022.

\bibitem[Yang et~al.(2024{\natexlab{a}})Yang, Zhou, Liu, and Loy]{yang2024fresco}
Shuai Yang, Yifan Zhou, Ziwei Liu, and Chen~Change Loy.
\newblock Fresco: Spatial-temporal correspondence for zero-shot video translation.
\newblock In \emph{CVPR}, 2024{\natexlab{a}}.

\bibitem[Yang et~al.(2017)Yang, Tan, Feng, Liu, Guo, and Yan]{yang2017deep}
Wenhan Yang, Robby~T Tan, Jiashi Feng, Jiaying Liu, Zongming Guo, and Shuicheng Yan.
\newblock Deep joint rain detection and removal from a single image.
\newblock In \emph{Proceedings of the IEEE conference on computer vision and pattern recognition}, pages 1357--1366, 2017.

\bibitem[Yang et~al.(2023)Yang, Aviles-Rivero, Fu, Liu, Wang, and Zhu]{yang2023video}
Yijun Yang, Angelica~I Aviles-Rivero, Huazhu Fu, Ye Liu, Weiming Wang, and Lei Zhu.
\newblock Video adverse-weather-component suppression network via weather messenger and adversarial backpropagation.
\newblock In \emph{Proceedings of the IEEE/CVF International Conference on Computer Vision}, pages 13200--13210, 2023.

\bibitem[Yang et~al.(2024{\natexlab{b}})Yang, Teng, Zheng, Ding, Huang, Xu, Yang, Hong, Zhang, Feng, et~al.]{yang2024cogvideox}
Zhuoyi Yang, Jiayan Teng, Wendi Zheng, Ming Ding, Shiyu Huang, Jiazheng Xu, Yuanming Yang, Wenyi Hong, Xiaohan Zhang, Guanyu Feng, et~al.
\newblock Cogvideox: Text-to-video diffusion models with an expert transformer.
\newblock \emph{arXiv preprint arXiv:2408.06072}, 2024{\natexlab{b}}.

\bibitem[Ye et~al.(2023)Ye, Chen, Bai, Shi, Xue, Jiang, Yin, Chen, and Liu]{ye2023adverse}
Tian Ye, Sixiang Chen, Jinbin Bai, Jun Shi, Chenghao Xue, Jingxia Jiang, Junjie Yin, Erkang Chen, and Yun Liu.
\newblock Adverse weather removal with codebook priors.
\newblock In \emph{Proceedings of the IEEE/CVF international conference on computer vision}, pages 12653--12664, 2023.

\bibitem[Zhang et~al.(2023)Zhang, Wei, Jiang, Zhang, Zuo, and Tian]{zhang2023controlvideo}
Yabo Zhang, Yuxiang Wei, Dongsheng Jiang, Xiaopeng Zhang, Wangmeng Zuo, and Qi Tian.
\newblock Controlvideo: Training-free controllable text-to-video generation.
\newblock \emph{arXiv preprint arXiv:2305.13077}, 2023.

\bibitem[Zhu et~al.(2017)Zhu, Park, Isola, and Efros]{zhu2017unpaired}
Jun-Yan Zhu, Taesung Park, Phillip Isola, and Alexei~A Efros.
\newblock Unpaired image-to-image translation using cycle-consistent adversarial networks.
\newblock In \emph{ICCV}, 2017.

\bibitem[Zhu et~al.(2024)Zhu, Tu, Liu, Bovik, and Fan]{zhu2024mwformer}
Ruoxi Zhu, Zhengzhong Tu, Jiaming Liu, Alan~C Bovik, and Yibo Fan.
\newblock Mwformer: Multi-weather image restoration using degradation-aware transformers.
\newblock \emph{IEEE TIP}, 2024.

\end{thebibliography}
\end{document}